\documentclass[letterpaper,twocolumn,10pt]{article}
\usepackage{usenix}

\usepackage{amsmath,amsfonts}
 
\usepackage{amssymb}

\usepackage{pifont} %use the command \ding

\usepackage{tikz}
\usepackage{tkz-euclide}
\tikzset{%
    pics/sema/.style args={#1/#2/#3}{code={%
        \ifstrequal{#2}{0}{%
            \node[circle,minimum width=1mm,draw,fill=#1] {};
        }{%
            \tkzDefPoint(0,0){O}
            \tkzDrawSector[R,fill=#1](O,1mm)(90,90-#2)
            \tkzDrawSector[R,fill=#3](O,1mm)(90-#2,90-360)
    }
    }},
}

\usepackage{xcolor}
\definecolor{armygreen}{rgb}{0.29, 0.33, 0.13}
\definecolor{electricpurple}{rgb}{0.75, 0.0, 1.0}
\definecolor{limegreen}{rgb}{0.75, 1.0, 0.0}
\definecolor{skyblue}{rgb}{0.53, 0.81, 0.98}
\definecolor{goldenrod}{rgb}{0.98, 0.80, 0.20}

\usepackage{amssymb}
\usepackage{bbding} %对勾叉号

\usepackage{makecell}
\usepackage{colortbl} % table color
\usepackage{adjustbox}

\usepackage{pgf-pie}
\usepackage{graphicx}
\usepackage{subfigure}

\usepackage{xcolor}
\definecolor{babypink}{rgb}{0.96, 0.76, 0.76}
\definecolor{flamingopink}{rgb}{0.99, 0.56, 0.67}
\definecolor{bananamania}{rgb}{0.98, 0.91, 0.71}
\definecolor{cambridgeblue}{rgb}{0.64, 0.76, 0.68}
\definecolor{asparagus}{rgb}{0.53, 0.66, 0.42}
\definecolor{desertsand}{rgb}{0.93, 0.79, 0.69}
\definecolor{tropicalholiday}{HTML}{8ECFC9}
\definecolor{ao(english)}{rgb}{0.0, 0.5, 0.0}

\usepackage{lipsum}
\usepackage{tikz}
\usetikzlibrary{arrows.meta}
\usetikzlibrary{automata,positioning}

\usepackage{threeparttable}

\renewcommand*{\arraystretch}{1.5}%
\definecolor{tabred}{RGB}{230,36,0}%
\definecolor{tabgreen}{RGB}{0,116,21}%
\definecolor{taborange}{RGB}{250,124,30}%
\definecolor{tabbrown}{RGB}{171,70,0}%
\definecolor{tabyellow}{RGB}{251,253,169}%
\newcommand*{\vcorr}{%
  \vadjust{\vspace{-\dp\csname @arstrutbox\endcsname}}%
  \global\let\vcorr\relax
}% 

\usepackage{mathtools}

\usepackage{lscape}
\usepackage[figuresright]{rotating}
\usepackage[graphicx]{realboxes}
\usepackage{adjustbox}
\usepackage{caption}

\usepackage{hyperref}
\newcommand{\hlhref}[2]{\href{#1}{\textcolor{blue}{{#2}}}}

\usepackage{colortbl} 
\usepackage{xfrac}

\usepackage{appendix}
\usepackage{float}

\usepackage{fbox} %use box
\usepackage{fancybox}%use other types of boxes
\usepackage{framed}

\usepackage{multirow}

\def\BibTeX{{\rm B\kern-.05em{\sc i\kern-.025em b}\kern-.08em
    T\kern-.1667em\lower.7ex\hbox{E}\kern-.125emX}}
    
\usepackage{CJKutf8}  
\usepackage{pgfplots}
\usepackage{filecontents}

\usepackage{stackengine}
\usepackage{scalerel}
\newcommand\dangersign[1][2ex]{%
  \renewcommand\stacktype{L}%
  \scaleto{\stackon[1.3pt]{\color{red}$\triangle$}{\tiny !}}{#1}%
}

\usepackage{tabularx,makecell, caption}
\usepackage{enumitem} % to control Itemization spacing

\newcolumntype{L}{>{\arraybackslash}X}

\usepackage{multicol}
\usepackage{blindtext}
\usepackage{etoolbox,xstring,mfirstuc,textcase}

\usepackage{subfiles}
\usepackage[most]{tcolorbox}
\usepackage{arydshln}
\usepackage{tikz}
\usepackage{relsize}

\usepackage{xcolor}
\usepackage{siunitx}
\usepackage{comment}
\usepackage{indentfirst} 
\usepackage{framed}

\usepackage[font=small,labelfont=bf,tableposition=top]{caption}
\usepackage{booktabs}
\usepackage{dashbox}

\usepackage{listings}
\usepackage{url}

\newenvironment{packeditemize}{
	\begin{list}{$\bullet$}{
			\setlength{\labelwidth}{4pt}
			\setlength{\itemsep}{0pt}
			\setlength{\leftmargin}{\labelwidth}
			\addtolength{\leftmargin}{\labelsep}
			\setlength{\parindent}{0pt}
			\setlength{\listparindent}{\parindent}
			\setlength{\parsep}{0pt}
			\setlength{\topsep}{1pt}}}{\end{list}}

\newenvironment{circitemize}{
	\begin{list}{$\circ$}{
			\setlength{\labelwidth}{4pt}
			\setlength{\itemsep}{0pt}
			\setlength{\leftmargin}{\labelwidth}
			\addtolength{\leftmargin}{\labelsep}
			\setlength{\parindent}{0pt}
			\setlength{\listparindent}{\parindent}
			\setlength{\parsep}{0pt}
			\setlength{\topsep}{1pt}}}{\end{list}}

\usepackage{color}
\usepackage{algorithm,algpseudocode} %% select one of them

\begin{document}
\title{SoK: Bitcoin Layer Two (L2)}

%\begin{comment}
\author{
{\rm Minfeng Qi$^{1,}$\thanks{These authors contributed equally to the work.}\, , Qin Wang$^{2,\textcolor{green}{*}}$,  Zhipeng Wang$^{3}$, Manvir Schneider$^4$,} \\
{\rm Tianqing Zhu$^{1}$, Shiping Chen$^{2}$, William Knottenbelt$^3$, Thomas Hardjono$^{5}$}  \\
\textit{$^1$City University of Macau, China} | 
\textit{$^2$CSIRO Data61, Australia} | 
\textit{$^3$Imperial College London, UK }  \\ 
\textit{$^4$Cardano Foundation, Switzerland} | 
\textit{$^5$Massachusetts Institute of Technology, US} 
}

%\thanks{ These authors contributed equally to the work. }

%\and
%{\rm Second Name}\\
%Second Institution

%\author{Minfeng Qi$^{1}$$^{\dag}$, Qin Wang$^{2}$$^{\dag}$, Manvir Schneider$^3$, Zhipeng Wang$^{4}$, Lin Zhong$^{5}$, \\  Shiping Chen$^{2}$, William Knottenbelt$^4$, Thomas Hardjono$^{6}$} \thanks{$^{\dag}$ These authors contributed equally to the work. }

%\affiliation{ \textit{$^1$City University of Macau, China} | \textit{$^2$CSIRO Data61, Australia} |  \textit{$^3$Cardano Foundation, Switzerland} \\ \textit{$^4$Imperial College London, UK }   | \textit{$^5$New Huo Digital Limited, Singapore} | \textit{$^6$MIT, US} \\ }

%\end{comment}

%=================================================
\maketitle

\begin{abstract}
We present the first Systematization of Knowledge (SoK) on constructing Layer Two (L2) solutions for Bitcoin.

We carefully examine a representative subset of ongoing Bitcoin L2 solutions (40 out of 335 extensively investigated cases) and provide a concise yet impactful identification of six classic design patterns through two approaches (i.e., modifying transactions \& creating proofs). Notably, we are the first to incorporate the inscription technology (emerged in mid-2023), along with a series of related innovations. We further establish a reference framework that serves as a baseline criterion ideally suited for evaluating the security aspects of Bitcoin L2 solutions, and which can also be extended to broader L2 applications. We apply this framework to evaluate each of the projects we investigated.

We find that the inscription-based approaches introduce new \textit{functionality} (i.e., programability) to Bitcoin systems, whereas existing proof-based solutions primarily address scalability challenges. Our security analysis reveals new attack vectors targeting data/state (availability, verification), assets (withdrawal, recovery), and users (disputes, censorship).

\end{abstract}

%\keywords{Bitcoin, SoK, Layer Two, Blockchain}

%=================================================   
\section{Introduction}
%=================================================   

Bitcoin~\cite{nakamoto2008bitcoin,bonneau2015sok,narayanan2016bitcoin} has surged in the crypto market since mid-2023~\cite{binanceHY}. Several indicators demonstrate the growth of its ecosystem~\cite{binance1,binance2,binance3,binance4,binance5,binance6}: (i) The Bitcoin price (and marketcap) have reached a historical record price of US\$73,079  (Mar. 14, 2024, \#$\mathsf{CoinMarketCap}$); (ii) The average block size of Bitcoin has increased significantly, from 1.2MB to over 2MB; (iii) The transaction volume in the memory pool has consistently risen, reaching nearly 24,000 transactions, compared to the stable level of around 5,000 transactions in 2022; (iv) Statistical platforms such as Dune Analytics~\cite{duneanalytic}\cite{duneanalytic1} and UniSat~\cite{unisat} confirm the upward trend in Bitcoin transactions; (v) Numerous Bitcoin new projects~\cite{binance6} secure investment and been actively launched in the market; (vi) Governments soften policies toward Bitcoin assets, as evidenced by the approval of Bitcoin ETFs by SEC~\cite{news1} and HK authorities~\cite{news2}.

In this paper, we focus on the technical underpinnings, specifically the new developments designed and implemented on Bitcoin, collectively referred to as \textit{Bitcoin Layer Two} (L2).

\smallskip
\noindent\textbf{L2 protocols.} Layer two in general is a mirrored concept to layer one (L1) protocols. While L1 protocols are designed to be self-sufficient and operate independently, typically referring to the blockchain base that consists of a series of core protocols (consensus mechanism, data structure, transaction/block processing) and network layer (P2P networks), L2 protocols often refer to secondary frameworks or technologies that are built on top of L1. The concept of L2 at the early stage (which we called ``old-fashioned" to salute its classiness) majorly discusses off-chain computing techniques such as sidechain, channels (e.g., payment \& lighting channels). Those solutions primarily improve the system's scalability while partially (i.e., byproduct) enhancing its interoperability.

\begin{packeditemize}
    \item \textit{Scalability}\footnote{For clarity and quick reference, we use ``$\bullet$'' for general itemized classifications, ``$\circ$'' for workflow-related steps, and ``$\triangleright$'' for metric-related items.} refers to the ability to efficiently (also implies performance) manage an increasing volume of transactions, commonly measured by transaction speed (quicker confirmation) and throughput (transactions per second, or TPS). 
    L2 protocols are one of the main solutions for improving scalability compared to concurrent L1 techniques (e.g., block size~\cite{xu2018cub, dai2019jidar, wang2022txilm}, consensus mechanisms~\cite{bentov2016snow, kiayias2017ouroboros, eyal2016bitcoin, gilad2017algorand}, sharding~\cite{luu2016secure, kokoris2018omniledger, zamani2018rapidchain, wang2019monoxide}, and chain structure~\cite{wang2023sok, lewenberg2015inclusive, sompolinsky2016spectre, sompolinsky2018phantom, li2018scaling}). 

    \item \textit{Interoperability} refers to the capability of heterogeneous blockchain systems to exchange data efficiently. Numerous interoperability solutions~\cite{zamyatin2021sok,belchior2021survey,wang2023exploring} leverage similar L2 underlying techniques (e.g., sidechains~\cite{li2022zerocross}, hash-locks~\cite{thyagarajan2022universal,manevich2022cross}) to facilitate cross-chain communications.

\end{packeditemize}

\noindent\textbf{Concept refinement: Bitcoin L2.} When exploring the specific concept of Bitcoin L2, we find that the intrinsic nature of L2 introduces new dimensions, particularly concerning Bitcoin's functionality (i.e., programmability).

\begin{packeditemize}
\item \textit{Functionality} refers to the capacity of a blockchain to support and execute customized code, enabling state transitions directly on-chain. This term is predominantly associated with the Ethereum ecosystem (and other EVM-compatible platforms), where state-of-the-art solutions mostly utilize smart contracts to facilitate the creation of decentralized applications (dApps~\cite{antonopoulos2018mastering}), automated workflows~\cite{wang2019artchain}, and complex financial instruments (e.g., DeFi~\cite{werner2022sok,jiang2023decentralized}).
\end{packeditemize}

However, due to the structural differences between the UTXO model (e.g.,  Bitcoin)~\cite{delgado2019analysis} and the account model (e.g., EVM-compatible blockchains), discussions of Bitcoin's functionality were historically limited. This is largely because of the inherent constraints~\cite{brakmic2019bitcoin} in UTXO which is primarily designed for simple and stateless transactions. Unlike EVM-compatible platforms, Bitcoin and similar UTXO designs are not well-suited for hosting complex functional applications beyond their original purpose. To date, Bitcoin's primary role remains as a store of value (peaking at US\$1.44T, Mar. 2024), serving as the anchor of crypto markets~(occupying 69.2\%\footnote{The global crypto market cap was US\$2.08T(\#$\mathsf{CoinMarketCap}$).}).

%Fortunately, a series of sustained technical efforts (detailed below) have collectively demonstrated the potential for enhancing the functionality of the Bitcoin network.

\smallskip
\noindent\textbf{\textcolor{red}{\dangersign} Gaps.} Despite the growing importance of Bitcoin L2 solutions, awareness of this concept remains surprisingly low. Our preliminary surveys (i.e., random sampling of small groups) indicate that even experienced blockchain researchers and developers (31 out of 39) are not familiar with it. Therefore, we pose the following research questions:

\begin{packeditemize}
\item \textbf{RQ1:} What is the current status of existing projects?
\item \textbf{RQ2:} What are the common patterns and architectural differences among these solutions?
\end{packeditemize}

Moreover, introducing an additional layer to the mainchain may expose the system to unexpected attacking vectors, posing significant security risks. For instance, a challenge-response pattern between the mainchain and sidechains could face risks such as falsified proofs or witnesses~\cite{chaliasos2024sok}; unreliable bridging services between two chains could lead to asset corruption~\cite{zhang2023sok}; and direct inscription on a transaction field might lead to transaction malleability issues~\cite{decker2014bitcoin}. Existing L2 studies center around classifying operational mechanisms, while security-focused research tends to address specific technical issues, often overlooking the general applicability of various constructions. This leads us to ask: %third research question:

\begin{packeditemize}
\item \textbf{RQ3:} What distinct security threats (either new or existing) arise from those different design patterns?
\end{packeditemize}

%\mfcomment{TODO: I will discuss the details of each criterion against security threats more}

\noindent\textbf{Our approach.} We show how we address these gaps.

\begin{packeditemize}

\item[\ding{172}] \noindent\textbf{Collecting projects} (Sec.\ref{subsec-data}). We start by conducting a comprehensive investigation of the projects in the wild. We then carefully examine the selected projects (40 out of 335) that can represent the most state-of-the-art developments.

\item[\ding{173}] \noindent\textbf{Identifying \& assessing design patterns.} Our classification is based on the modification areas within Bitcoin (cf. Fig.\ref{fig-design}) where we identified two major approaches (Sec.\ref{ref-designspace}): \textit{embedding inscribed scripts into transactions} and \textit{incorporating contract management for off-chain proof verification}. 

% Each of these approaches is further divided into five specific design patterns. 

We then analyze architectural differences among various solutions (Sec.\ref{sec-archi}) (i.e., inscription, BitVM, rollups, sidechains, state channels, client-side verification) and examine how they integrate with the Bitcoin mainchain. To our knowledge, this is the first work to include and discuss inscription-based solutions that have not been explored previously.

\begin{figure}[!hbt]
\vspace{-0.3em}
    \centering
    \includegraphics[width=0.8\linewidth]{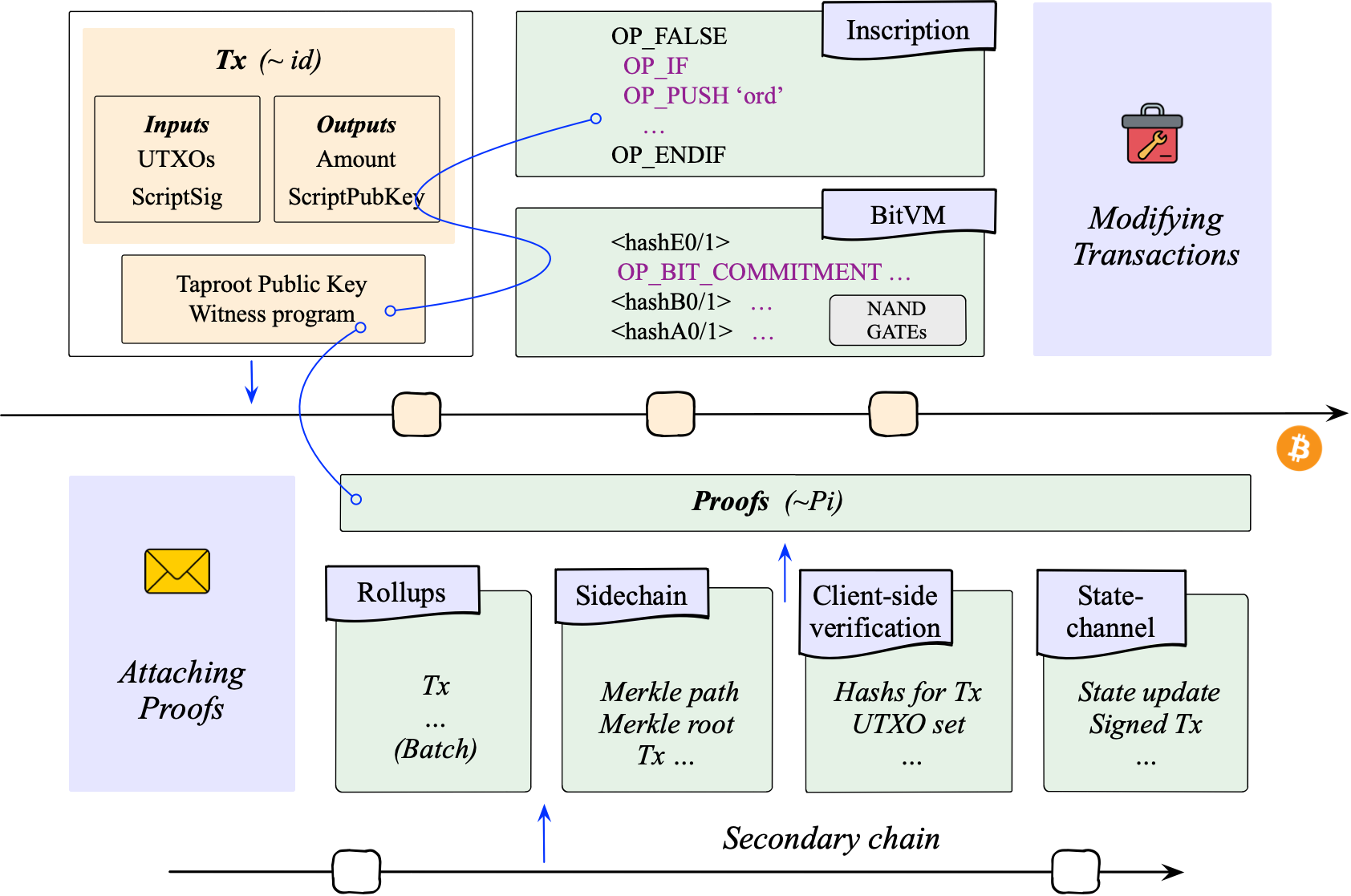}
   \caption{Our Identified Bitcoin L2 Designs}
   \label{fig-design}
   \vspace{-0.1in}
\end{figure}

\item[\ding{174}] \noindent\textbf{Establishing new security frame} (Sec.\ref{sec-security}). We also, for the first time, identify a series of security threats aligned with the transaction execution workflow, specifically tailored to Bitcoin L2. We consider the security of native assets (e.g., withdrawal), the integrity of data during processes (e.g., proof/state verification, data availability), and incorporate the user’s perspective (e.g., dispute resolution).
%, while also applicable to general L2 solutions
\end{packeditemize}

\noindent\textbf{Key findings.} Bitcoin L2 introduces new dimensions (i.e., functionality) compared to earlier L2 conceptions (majorly scalability) and new methods (inscription, BitVM), while negatively also bringing more complexity and attacking vectors.  

\begin{packeditemize}

\item Embedding scripts directly on-chain allows for implementing NFTs by uniquely numbering each satoshi and enabling state transitions via logic gates through commitments. 

\item Proof-based solutions such as rollups and sidechains enable trustless validation of off-chain transactions, ensuring that only valid state transitions are recorded on the mainchain. 

\item Complexity increases due to complicated commitment gates in BitVM, and cryptographic computations (e.g., generation, verification) in proof-based solutions.

\item Our identified attacking vectors include potential vulnerabilities in the cryptographic primitives used for transaction verification, the risk of data manipulation or censorship in the secondary chain, and the exploitation of weaknesses in withdrawal mechanisms and dispute resolution processes.

%However, these solutions encounter significant challenges, such as increased computational complexity, reduced flexibility in asset recovery, and limitations in handling multi-party interactions.

%However, the reliance on complex cryptographic proofs introduces challenges such as high computational overhead, the need for efficient proof generation. Additionally, proof-based approaches face difficulties in optimizing for latency and scalability while maintaining decentralization.

%\item We also note that attacking vectors in Bitcoin L2 solutions arise due to the introduction of off-chain components, complex cryptographic proofs, and multi-party protocols. These vectors include potential vulnerabilities in the cryptographic primitives used for transaction verification, the risk of data manipulation or censorship in the secondary chain, and the exploitation of weaknesses in withdrawal mechanisms and dispute resolution processes. The increased reliance on off-chain computation also opens up attack surfaces where adversaries might disrupt the availability of critical transaction data, posing significant security challenges across different L2 architectures.

\end{packeditemize}

\smallskip
\noindent\textbf{Our credits.} We highlight several SoKs that are partially related to ours and credit (shaping our deep understanding) their discussions on Bitcoin~\cite{bonneau2015sok}, ``old-fashioned'' L2 (off-chain) protocols~\cite{gudgeon2020sok, jourenko2019sok}, SNARKs~\cite{chaliasos2024sok}, rollups~\cite{kotzer2024sok}, interoperability~\cite{zamyatin2021sok,belchior2021survey,wang2023exploring,augusto2024sok,zhang2023sok}, and light client~\cite{chatzigiannis2022sok}.

Many recent parallel studies have made great contributions to exploring new techniques within Bitcoin~\cite{li2024bitcoin,wang2023understanding,yu2023bridging,messias2024writing,wen2024modular,bertucci2024bitcoin}  and enhancing the scalability and robustness of Bitcoin systems~\cite{tas2023bitcoin,tas2023interchain,feng2023scalable}. Despite these topics being slightly beyond ours, we also acknowledge these important advancements.

%=======================================
\section{Our Methodology}
\label{sec-prelimi}
%=======================================

%===============================
\subsection{Project Collection}\label{subsec-data}
%===============================

%data source 
%(1) https://www.rootdata.com/EcosystemMap/list/271?n=Bitcoin
%(2) https://www.btceden.org/
%(3) https://L2.watch/
%(4) https://edgein.io/

\noindent\textbf{Data sources.} We primarily collect Bitcoin L2 projects from four statistical websites: Rootdata~\cite{rootdata2023}, BTCEden~\cite{btceden2023}, L2.watch~\cite{l2watch2023}, and EdgeIn.io~\cite{edgein2023} (more in Appendix~\ref{sec-web}). These platforms have already amassed large-scale datasets and specialize in blockchain projects. Following this, we dive into each project's official website, GitHub repository, and open forums to gather additional information.

\smallskip
\noindent\textbf{Selection strategy.} Based on gathering initial lists of L2 projects, we apply a filtering process: (i) we conducted a deduplication process to eliminate redundancies; (ii) we primarily selected projects that had either achieved the mainnet stage or were on the cusp of doing so; (iii) we included a few projects in the testnet and pre-testnet phases with notable developmental advancements. For some borderline projects, we manually and subjectively evaluated their standing within the community, considering factors including the project team's background, media coverage, and industry recognition. 

\smallskip
\noindent\textbf{Statistical results.} Our initial results report \textbf{335} projects (detailed results in Table~\ref{tab-distribution} and Fig.\ref{fig:projects} in Appendix~\ref{sec-web}). By applying the filtering process, we screened out \textbf{40} eligible Bitcoin L2 projects (cf. the \textit{second column} in Table~\ref{tab-summary}).

\smallskip
\noindent\textbf{Project details.} We further provide a summary of those 40 projects, which is included in Appendix~\ref{sec-project}.

\begin{center}
\fbox{%
    \begin{minipage}{0.95\linewidth}

    % \textbf{(RQ1) Finding 1:} Bitcoin L2 projects increased significantly (\textbf{335}) within a short period (15 months), while only a small portion of them (\textbf{40}) are ready for examination.
    \textbf{(RQ1) Finding 1:} Bitcoin L2 projects have experienced significant growth, with \textbf{335} new projects emerging within a short period (15 months). Of these, \textbf{40} have reached a stage where they are ready for examination.
    \end{minipage}
}
\end{center}

%------------------------------------ 
\subsection{Design Spaces}\label{ref-designspace}
%------------------------------------ 

\noindent\textbf{Skeleton protocol.} The Bitcoin mainchain operates through a series of fundamental protocols that ensure the secure generation of addresses and keys. Address generation involves creating a public-private key pair, where the private key \( k \) is a randomly generated 256-bit number, and the public key \( K \) is derived using the elliptic curve multiplication \( K = k \cdot G \), with \( G \) being the generator point on the secp256k1 curve. The Bitcoin address \( A \) is then generated by hashing the public key using SHA-256 and RIPEMD-160: \( A = \mathsf{RIPEMD\text{-}160(SHA\text{-}256(}K\mathsf{))} \). 

% Transaction creation in Bitcoin involves constructing inputs and outputs referencing unspent transaction outputs (UTXOs). Each transaction \( T \) can be represented as \( T = \{ \mathsf{inputs}, \mathsf{outputs} \} \), where inputs include references to previous UTXOs and signatures that prove ownership. The transaction is broadcast to the network and, once validated, included in a block.

\smallskip
\noindent\textbf{Transaction details.} In Bitcoin, a transaction \( T \) is the fundamental unit for transferring assets between addresses. Each transaction is composed of several key fields that define its structure. Specifically, a transaction \( T \) can be represented as:
\(
T = \{ \mathsf{inputs}, \mathsf{outputs}, \mathsf{locktime}, \mathsf{version}, \mathsf{witness} \},
\)
where,
\begin{packeditemize}
    \item \textcolor{teal}{$\mathsf{inputs}$} are references to previous unspent transaction outputs (UTXOs) that the current transaction is spending. Each input contains a reference to a previous output, an unlocking script (\( \mathsf{ScriptSig} \)), and a sequence number.
    \item \textcolor{teal}{$\mathsf{outputs}$} define the destination of the transferred Bitcoin, specifying the amount and a locking script ($\mathsf{ ScriptPubKey}$) that controls who can spend the output in the future.
    \item \(\mathsf{locktime}\) is an optional field that specifies the earliest time or block height at which the transaction can be included.
    \item \(\mathsf{version}\) indicates the transaction format version and can signal the use of new transaction types or features.
    \item \textcolor{teal}{\(\mathsf{witness}\)} is used in SegWit transactions that contain the Taproot public key, witness program, and other witness data required to validate the transaction. 
    
    % There are some opcodes related to witness processing, such as \( \mathsf{OP\_DUP} \), \( \mathsf{OP\_HASH160} \), and \( \mathsf{OP\_EQUALVERIFY} \).
    
\end{packeditemize}

% Transactions are then included in blocks, where a block \( B \) is defined as:
% \(
% B = \{ \mathsf{block\_header}, \mathsf{tx\_count}, \mathsf{txs} \},
% \)
% with the block header \(\mathsf{block\_header}\) containing fields such as the previous block hash \( \mathsf{prev\_block\_hash} \), the Merkle root \( \mathsf{merkle\_root} \) summarizing the included transactions, the timestamp \( \mathsf{time} \), the difficulty target \( \mathsf{bits} \), and the nonce \( \mathsf{nonce} \).

% These fields serve as the foundation for several Layer 2 improvements:
% \begin{itemize}
%     \item \textbf{Modifying transaction data:} Techniques like Inscriptions introduce new data types by modifying specific fields in transactions, such as embedding metadata within \(\mathsf{outputs}\) or \(\mathsf{witness}\) fields, allowing for additional functionalities like digital assets or complex scripts.
%     \item \textbf{Verifying proofs:} Solutions that rely on cryptographic proofs typically include these proofs in the \(\mathsf{witness}\) field or as part of \(\mathsf{inputs}\) or \(\mathsf{outputs}\). Off-chain mechanisms like State Channels interact with the Bitcoin network by periodically updating the state reflected in these transactions, particularly affecting the UTXOs referenced in \(\mathsf{inputs}\) and \(\mathsf{outputs}\).
% \end{itemize}

\smallskip
\noindent\textbf{Supported techniques.} BIP141~\cite{bip141} introduced SegWit which separates signature data from the transaction data. BIP341~\cite{bip341} allows complex scripts to be represented as single keys through the Taproot public key and Taptree structures, revealing only the executed path during transaction validation. We defer more technical details in Appendix~\ref{sec-technical}.

%\url{https://pure.tudelft.nl/ws/files/98438118/mainSP.pdf} \qw{SecII for formal model}

\smallskip
\noindent\textbf{Category by modification fields.} Fields in Bitcoin transactions serve as the foundation for several L2 improvements. We categorized these improvements into two primary categories.

\smallskip
\noindent\textbf{\ding{172} Modifying transaction data} (Sec.\ref{subsec-inscript}). Techniques like inscriptions introduce new data types by modifying specific fields in transactions (i.e., embedding metadata within \textcolor{teal}{$\mathsf{outputs}$} or \textcolor{teal}{$\mathsf{witness}$} fields), allowing for additional functionalities like digital assets or complex scripts.

\begin{packeditemize}
\item \textit{Inscriptions:} The inscription technique, enabled by the Ordinals protocol~\cite{ordinalbitcoinwhitepaper}, allows embedding arbitrary data into Bitcoin transaction. This process involves encoding data \( D \), constructing a witness script \( S_w = \mathsf{OP\_RETURN \|} D \). A transaction \( T \) is then created with an output script \( (\mathsf{ScriptPubKey}) \) that references \( S_w \) as part of the witness data. The transaction is then broadcast and included in a block, embedding the data on-chain.

% \item \textit{BitVM:} BitVM~\cite{bitvmofficial2023} leverages the $\mathsf{Taproot}$ and $\mathsf{Taptree}$ structures to embed multiple conditions within the $\mathsf{Taproot}$ public key \(\mathsf{K_{\text{Taproot}}}\), allowing transactions to represent more sophisticated logic. In BitVM, the $\mathsf{Taptree}$ structure is utilized to organize different possible execution paths, where each leaf in the tree ($\mathsf{Tapleaf}$) can represent a specific logic condition or script. The root of those $\mathsf{Taptree}$s as part of \(\mathsf{K_{\text{Taproot}}}\) is included in $\mathsf{ScriptPubKey}$. This setup allows for the representation of a complex logic circuit that only reveals the executed branch during transaction validation.

\item \textit{BitVM:} 
BitVM~\cite{bitvmofficial2023} leverages the $\mathsf{Taproot}$ and $\mathsf{Taptree}$ structures to embed multiple conditions within the witness data, specifically within the Taproot public key \(\mathsf{K_{\text{Taproot}}}\). This setup allows transactions to represent more sophisticated logic by utilizing the $\mathsf{Taptree}$ structure to organize different possible execution paths. Each leaf in the Tapleaf can represent a specific logic condition or script. The root of those $\mathsf{Taptree}$s is included in the witness field as part of \(\mathsf{K_{\text{Taproot}}}\) and referenced in $\mathsf{ScriptPubKey}$.

% For example, BitVM can embed multiple $\mathsf{Taproot}$ public keys \{$\mathsf{K_1, K_2, \ldots, K_n}$\} into the $\mathsf{ScriptPubKey}$, which are connected through a $\mathsf{Taptree}$ structure. Each $\mathsf{Taproot}$ public key \(\mathsf{K_i}\) can represent one or a seris of logic gate? \(\mathsf{G_i}\) (such as \(\mathsf{AND}\), \(\mathsf{OR}\), or \(\mathsf{NAND}\)), forming a complex logic circuit.

% \item \textit{BitVM:} In BitVM~\cite{bitvmofficial2023}, multiple spending conditions are organized in a $\mathsf{Taptree}$, where each leaf node corresponds to a different script or condition. These scripts are connected through the $\mathsf{Taptree}$ structure, and the root of this tree is combined with a single public key to create the $\mathsf{Taproot}$. Each script within the $\mathsf{Taptree}$ can represent one or a series of logic gates \(\mathsf{G_i}\) (such as \(\mathsf{AND}\), \(\mathsf{OR}\), or \(\mathsf{NAND}\)), creating a complex logic circuit that only reveals the executed branch during transaction validation.

% BitVM~\cite{bitvmofficial2023} leverages the $\mathsf{Taproot}$ mechanism, which uses a $\mathsf{Taptree}$ structure to embed multiple conditions and complex logic within a single public key. 

\end{packeditemize}

\noindent\textbf{\ding{173} Verifying proofs}. Solutions that rely on cryptographic proofs typically include these proofs in the \textcolor{teal}{$\mathsf{witness}$} field or as part of \textcolor{teal}{$\mathsf{inputs}$} or \textcolor{teal}{$\mathsf{outputs}$}, enabling the validation of off-chain computations or state transitions. Customized smart contracts often act as an automated verifier that checks the validity of the proof and manages the subsequent state changes.

\begin{packeditemize}
\item  \textit{Rollups} (Sec.\ref{subsec-proof}): Rollups utilize proofs to validate transactions off-chain and post summarized data on-chain. The validity of a batch of transactions is proved by a proof \( \pi \):
\(
\pi = \mathsf{Proof}(T_1, T_2, \ldots, T_n),
\)
where \( T_i \) represents individual transactions. The rollup contract on the mainchain verifies \( \pi \), ensuring that all transactions in the batch are valid.

\item \textit{Sidechains} (Sec.\ref{subsec-sidechain}): Sidechains use Simplified Payment Verification (SPV) proofs \cite{bitcoinSVwiki2024} to provide a cryptographic link between a transaction on the Bitcoin mainchain and the corresponding operation on the sidechain. This proof includes the Merkle path \( \mathsf{P_{\text{main}}} \) from the transaction \( T_{\text{lock}} \) to the block's Merkle root \( \mathsf{R_{\text{main}} }\), as well as the block header \( \mathsf{H_{\text{main}}} \):
\(
\mathsf{\pi_{SPV}} = \left\{\mathsf{P_{\text{main}}}, \mathsf{H_{\text{main}}}, \mathsf{R_{\text{main}}}, T_{\text{lock}}\right\}.
\)

% enables the creation of independent blockchains that operate alongside the Bitcoin mainchain, allowing assets and data to move between the two chains. The proof verification process in sidechains typically involves a two-way peg mechanism:
% \(
% T_{\text{lock}} = \mathsf{Tx}(A \rightarrow \mathsf{lock\_address}),
% \)
% where \( T_{\text{lock}} \) is the transaction that locks Bitcoin on the mainchain. An SPV proof is then used to unlock equivalent assets on the sidechain.

\item  \textit{Client-side verification} (Sec.\ref{subsec-csv}): It leverages client-side validation~\cite{chatzigiannis2022sok} and the UTXO set to ensure data integrity. The provided proof \( \pi(T_x) \) for verification includes the following components:
\(
\pi(T_x) = \left\{ C(T_x), H(U_i) \right\},
\)
where \( C(T_x) \) is a commitment for a transaction \( T_x \), \( U_i \) is the UTXO associated with \( T_x \), and \( H(U_i) \) is its hash. Verification is performed by comparing \( \pi(T_x) \) against the known state of UTXOs to ensure the commitment is valid.

\item \textit{State channels} (Sec.\ref{subsec-sidechannel}): State channels enable off-chain transactions between participants by exchanging signed transactions \( T_i = \text{SignedTx}(S_i)\). These transactions result in state updates \( S_i \), with the signatures serving as cryptographic proof of each participant’s consent to the new state. Participants can update the state without broadcasting them to the Bitcoin network. After \( n \) transactions, the updated state is:
\(
S_n = \{\mathsf{(A, v_A')}, \mathsf{(B, v_B')}\},
\)
where \( \mathsf{v_A'} \) and \( \mathsf{v_B'} \) are new balances of \( \mathsf{A} \) and \( \mathsf{B} \) after the \( n \)-th transaction.

% Participants deposit Bitcoin into a multi-signature address, creating an opening transaction:
% \(
% T_{\text{open}} = \mathsf{Tx}(A_{1}, A_{2} \rightarrow A_{\mathsf{multi}}),
% \)
% where \( A_{1} \) and \( A_{2} \) are the participants' addresses.

\end{packeditemize}

% \noindent\textbf{\ding{174} Facilitating off-chain processing:} This technique allows for transactions to be processed off-chain by exchanging signed transactions\qw{like inherently a tx or signature on tx...}

% \begin{packeditemize}

% \item \textit{State channels} (Sec.\ref{subsec-sidechannel}) enable frequent, off-chain transactions between participants. Participants deposit Bitcoin into a multi-signature address, creating an opening transaction:
% \(
% T_{\text{open}} = \mathsf{Tx}(A_{1}, A_{2} \rightarrow A_{\mathsf{multi}}),
% \)
% where \( A_{1} \) and \( A_{2} \) are the participants' addresses. Off-chain transactions update the state of the channel, and the final state is recorded on-chain when the channel is closed.

% \end{packeditemize}

\subsection{Security Reference Frame} 

For the first time, we establish a comprehensive security reference framework specifically designed to evaluate a wide range of L2 solutions, including those previously overlooked in early-stage studies~\cite{gudgeon2020sok, zamyatin2021sok}. Our framework is rooted in an analysis of asset flow throughout its entire lifecycle, systematically categorized into three distinct phases:

\smallskip
\noindent\textbf{\ding{172} Pre-execution.} This phase focuses on security measures that need to be in place before transactions are initiated, ensuring that the system is prepared to handle transactions securely.

\begin{packeditemize}
\item \textit{Data availability}: Ensures that all necessary transaction data is fully available and accessible for verification prior to the finalization of the transaction.
\end{packeditemize}

\smallskip
\noindent\textbf{\ding{173} Transaction execution.}
This phase focuses on the integrity of transactions as they are being processed, including the transfer of assets and the recording of transactions.

\begin{packeditemize}
\item \textit{State verification}: Ensures that the L2 solution employs robust validation techniques to accurately mirror off-chain transactions on the mainchain.
\item \textit{Withdrawal mechanism}: Safeguards the process of withdrawing assets, ensuring that funds can be securely transferred back to the mainchain without unauthorized access.
\item \textit{Anti-censorship measures}: Prevents any manipulation or censorship during the transaction process, ensuring that all transactions are processed fairly and without interference.
\end{packeditemize}

\smallskip
\noindent\textbf{\ding{173} Post-transaction.} 
This phase addresses the measures that ensure the continued security and integrity of the system after transactions have been completed, including conflict resolution and recovery mechanisms.

\begin{packeditemize}
\item \textit{Dispute resolution}: Provide mechanisms for resolving conflicts that arise after transactions, such as fraudulent activities or disputes, to maintain network stability.
\item \textit{Emergency asset recovery}: Ensures that users can reclaim assets in the case of emergencies (e.g., network failures), protecting their deposits after the transaction completion.
\end{packeditemize}

We defer the detailed definitions to Sec.\ref{sec-security}.

%===============================
\section{Evaluating Bitcoin L2 Protocols}\label{sec-archi}
%===============================

% \qw{Pikachu: Securing PoS Blockchains from Long-Range Attacks by Checkpointing into Bitcoin PoW using Taproot} 

% The scalability of Bitcoin refers to its capacity to handle a growing amount of transactions over time~\cite{sanka2021systematic}. As the adoption of Bitcoin increases, so does the volume of transactions that need to be processed on the network. However, Bitcoin faces a significant scalability issue due to its design: the size of each block in the blockchain is limited (traditionally 1MB), and blocks are added approximately every ten minutes~\cite{karame2016security}. This design limits the network to process an average of up to 7 tps, a rate significantly lower than traditional payment systems like VISA, which can handle over 65,000 tps~\cite{qi2021blockchain}. To address this challenge, the Bitcoin community has explored various L2 solutions.

% \hi{limit: the size of each block is limited (traditionally 1MB)}

% \hi{limit: an average of up to 7 transactions per second (tps)} 

% equal to  \hi{limit: scalability issue}
%This section delved into the technical foundations of various L2 solutions based on improving points.
% \hi{limit: scripting capability limits in 100 opcodes}

%------------------------------------ 
\subsection{Bitcoin's Inscription}\label{subsec-inscript}
%------------------------------------ 

Bitcoin’s scripting capabilities are inherently limited due to the constraints of its scripting language~\cite{brakmic2019bitcoin}, which supports only a limited set of operations to ensure network security. %This limitation hinders the development of computational logic on the Bitcoin network. 
Bitcoin inscription extends the functionality of Bitcoin by allowing string data to be embedded directly within transactions, enabling a weaker version of customized operations.

\smallskip
\noindent\textbf{Working mechanism.} The inscription process involves encoding the data (e.g., text, image, video) into a byte string or JSON format, which is then embedded in a Bitcoin transaction. The data are encapsulated within a $\mathsf{Taproot}$ script using opcodes such as \{\( \mathsf{OP\_FALSE} \), \( \mathsf{OP\_IF} \), \ldots, \( \mathsf{OP\_ENDIF} \)\}, creating a structure known as an ``envelope''. This envelope is embedded in a $\mathsf{Taproot}$ output, leveraging the witness discount provided by SegWit~\cite{segwit2017} to reduce storage costs. The process is executed in two phases:

\begin{circitemize}
    \item \textit{\underline{Com}mit Transaction:} A $\mathsf{Taproot}$  output \( T_\text{com} \) is created, committing to a script \( \mathsf{S_{com}} \) that references the inscription metadata \( \mathsf{D} \) without revealing it. This can be mathematically represented as:
    \(
    T_\text{com} = \mathsf{Taproot}( \mathsf{S_{com}}(H(\mathsf{D})) )
    \)
    where \( H(\mathsf{D}) \) is the cryptographic hash of the inscription data. This transaction is broadcast and included in a block \( \mathsf{B_{com}} \).

    \item \textit{\underline{Rev}eal Transaction:} The output from \( T_\text{com} \) is spent in a subsequent transaction \( T_\text{rev} \), which includes the actual inscription data \( \mathsf{D} \) within its script \( \mathsf{S_{rev}} \). The reveal transaction is mathematically represented as:
    \(
    T_\text{rev} = \mathsf{Taproot}( \mathsf{S_{rev}}(\mathsf{D}) )
    \)
    When \( T_\text{rev} \) is confirmed, the inscription metadata \( \mathsf{D} \) is permanently recorded on the blockchain.
\end{circitemize}

The inscription technique can be used to create NFT on Bitcoin~(e.g., BRC20~\cite{wang2023understanding}) or inscription tokens on Ethereum (e.g., Ethscriptions~\cite{Ethscriptions}). For the case of BRC20, 
each satoshi during the minting process (cf.Listing~\ref{list-mint}) is inscribed with four pieces of metadata to represent the token's attributes. The ordinal/sequential numbers ($\mathsf{Ord}$~\cite{ordinalbitcoinwhitepaper}) associated with these satoshis are tracked from the initial commitment to the final inscription~\cite{li2024bitcoin,wang2023understanding,binance3}, ensuring that metadata is permanently linked to those specific satoshis on-chain. 

\begin{lstlisting}[caption={Operations (\textcolor{teal}{$\mathsf{Mint}$)} for Bitcoin Inscription}, label={list-mint},basicstyle=\ttfamily\scriptsize]
# On-chain Inscription
"p" : "brc-20", # protocol name
"op": "mint", # operation
"tick": "ordi", # token name
"amt": "1000" # the amount of token being minted

# Off-chain update
if state[tick] NOT exists OR 
                 "amt" > "lim" OR sum("amt") > "max":
    raise errors
else 
    account_state[tick]["balance"][minter] += amt
\end{lstlisting}

%------------------------------------ 
\subsection{BitVM \& BitVM2}\label{subsec-inscript}
%------------------------------------ 

\smallskip
\noindent{$\bullet$ \textit{BitVM.} BitVM extends the design principles of inscriptions by enabling Turing-complete computation through Discreet Log Contracts (DLCs~\cite{mitdci2024}) and cryptographic proofs.

% \hi{idea: Taptree created in Taproot and simulated circuits}

% \hi{comput > Binanry circuit > opcode script > taproot commitment}

% \hi{bit value commitment for copy constraints (link)}

% \hi{logic gate commitment for gate constraints (gate)}

% \hi{interactive proving: challenge response by presign offchain tx}

% \hi{limit: high complexity} 

% \qw{@minfeng, could also give a itemized working mechanism for BitVm/Inscirption?} done

% \smallskip
% \noindent\textbf{Bit value commitment.}
% The process of compiling code into machine-readable binary, which is fundamentally a sequence of $0$s and $1$s, enables a processor to execute instructions using binary logic circuits. These circuits are composed of logical gates such as $AND$, $OR$, $NOT$, and $NAND$, with the $NAND$ gate being particularly notable as it can be used to construct all other types of gates~\cite{michael2000implications}. 

% \mfcomment{TODO}

\smallskip
\noindent\textbf{NAND gate.} The $NAND$ gate is an elementary component in BitVM. It results in a "$1$" only when both inputs are "$0$". For all other combinations of inputs, the result is "$0$". Within the Bitcoin scripting environment, the opcodes $\mathsf{OP\_BOOLAND}$ and $\mathsf{OP\_NOT}$ can be strategically combined to emulate the behavior of a $NAND$ gate. This composite operation, essential for constructing more complex logic, is termed "$\mathsf{OP\_NAND}$" in BitVM. For example, Listing~\ref{lst:hash-commitment-verification} illustrates (also see Fig.\ref{fig:nand}) how a prover constructs the logic gate \( NAND \)  by pushing three bit values (\( E \), \( B \), and \( A \)) onto the stack, corresponding to the preimages of \( \mathsf{hashE} \), \( \mathsf{hashB} \), and \( \mathsf{hashA} \) using \( \mathsf{OP\_BIT\_COMMITMENT} \). The verifier then checks the correctness of \( NAND \) by applying \( \mathsf{OP\_EQUALVERIFY} \) operation to compare the consistency of outputs. 

% Furthermore, BitVM can express any circuit by composing gate commitments known as binary circuit commitment, such as Figure~\ref{fig:nand}, which has 8 different \( NAND \) gates and 4 inputs \( A \), \( B \), \( C \), and \( D \).

\begin{lstlisting}[caption={Hash Commitments and Verification in NAND Gate}, label={lst:hash-commitment-verification}, basicstyle=\ttfamily\scriptsize, breaklines=true]
# Hash Commitments
<hashE0/1>  # Commitment to hashE
OP_BIT_COMMITMENT
OP_TOALTSTACK
<hashB0/1>  # Commitment to hashB
OP_BIT_COMMITMENT
OP_TOALTSTACK
<hashA0/1>  # Commitment to hashA
OP_BIT_COMMITMENT
OP_TOALTSTACK

# Verification
OP_FROMALTSTACK
OP_NAND
OP_EQUALVERIFY  # Verify A NAND B == E
\end{lstlisting}

% The process of compiling code into machine-readable binary, which is fundamentally a sequence of $0$s and $1$s, enables a processor to execute instructions using binary logic circuits. 

% A prover can further set the bit’s value of either to "$0$" or "$1$" through an Opcode known as \( \mathsf{OP\_BITCOMMITMENT} \)~\cite{linus2023bitvm}.

% \smallskip
% \noindent\textbf{Binary circuit commitment.} 

% \begin{center}
% \fbox{%
%     \begin{minipage}{0.95\linewidth}

%     \textbf{(RQ2) Finding 2:} BitVM emulates NAND gates using Bitcoin Script’s opcodes $\mathsf{OP\_BOOLAND}$ and $\mathsf{OP\_NOT}$, enabling complex logic circuit construction. Provers commit to binary values using "$\mathsf{OP\_BIT\_COMMITMENT}$", and verifiers ensure computational accuracy through a challenge-response protocol.
    
%     \end{minipage}
% }
% \end{center}

\smallskip
\noindent\textbf{Challenge-response protocol.}
The challenge-response mechanism~\cite{lerner2024bitvmx} is crucial for ensuring the accuracy of BitVM computations. Similar to Optimistic rollups (discussed later), it provides a method for one party (i.e., the verifier) to test the veracity of another party's (i.e., the prover's) claims through a series of challenges. Both the prover and the verifier stake a certain amount of Bitcoin assets as collateral and pre-sign a series of transactions to prepare for potential disputes. The verifier, acting as a skeptic, randomly selects a logical gate from the circuit and issues a challenge to the prover. This challenge requires the prover to disclose the inputs and outputs associated with the selected gate. In response to the challenge, the prover reveals the specified gate's inputs and outputs. Notably, the process of challenging and disclosing is repeated multiple times. The verifier may challenge different gates to ensure the prover's claims are consistent across the circuit. If at any point the prover's claims about the same gate contradict each other, the verifier detects this inconsistency. Then the verifier uses this inconsistency as proof of fraud and can take punitive action, such as confiscating the prover's funds. 

%------------figure 4 is in the appendex
% For example, in Fig.~\ref{fig:nand}, the Verifier claims an output \( E \) equal to 1 for gate \( \mathsf{NAND1} \) and 0 for gate \( \mathsf{NAND4} \), which logically cannot both be true, and thus the Verifier can seize the funds.

% \hi{idea: adopting Optimistic rollups for checking inconsistency}

\smallskip
\noindent\textbf{Working mechanism.} BitVM utilizes $\mathsf{Taproot}$ addresses to create a matrix, which functions similarly to a binary circuit. Each instruction within a Script corresponds to the smallest unit of a program, producing binary outcomes (true or false). These are organized into a matrix, resulting in a sequence of binary outputs (e.g., \textit{01100101}). The complexity of the program is directly proportional to the number of $\mathsf{Taproot}$ addresses combined, with traversal and execution time estimated at $ \mathcal{O}(n) $ for a $\mathsf{Taptree}$ with $n$ nodes.

\begin{circitemize}
    \item \textit{Key pair generation:} Participants generate their respective key pairs \( (k_A, K_A) \) and \( (k_B, K_B) \).
    
    \item \textit{Funding transaction:} A funding transaction is created with the inputs from \( k_A \) and \( k_B \). The output script of this transaction utilizes a Taproot public key \( K_{\text{Taproot}} \), which is a combination of the participants' keys and possibly other conditions:
    \(
    T_f = \mathsf{\{inputs}(k_A, k_B), \mathsf{outputs(Taproot}(K_{\text{Taproot}})\mathsf{)\}}.
    \)
    
    \item \textit{Oracle setup:} An oracle $\mathcal{O}$ publishes a public key \( K_{\mathcal{O}} \) and a signed message \( \sigma_{\mathcal{O}} \) indicating the outcome of the event.
    
    \item \textit{Contract execution transactions (CETs):} Pre-signed CETs are created for all possible outcomes. These CETs incorporate the Taproot public key and are executed based on the oracle's signature, distributing funds accordingly.
    
    \item \textit{State channels and off-chain state transitions:} The BitVM execution process involves setting up state channels and performing off-chain state transitions \( s_{i+1} = \delta(s_i) \), where these transitions can be optionally validated using zero-knowledge proofs (ZKP):
    \(
    \pi_i = \mathsf{zkp}(s_i, s_{i+1}).
    \)
    
    \item \textit{On-chain verification:} The final state \( s_n \) and proof \( \pi_n \), and Taproot public key, are submitted on-chain for verification.
\end{circitemize}

\smallskip
\noindent\textbf{Limitations.} While BitVM presents an approach to expanding the capabilities of the Bitcoin network for smart contract functionality, it also faces significant challenges. These include its current limitation to two-party interactions, the costs associated with scripting, and a narrow scope of ideal use cases~\cite{bitfinex2023bitvm, hirobitvm2023}. Specifically, BitVM's architecture is limited to interactions between two pre-defined entities. Extending this to support $N$-to-$N$ interactions, where multiple parties are involved in a single contract, would require a more complex logical design. In addition, the execution of preset unlock conditions for $\mathsf{Taproot}$ addresses in BitVM incurs miner fees, and as the number of addresses involved increases, so does the cost. Furthermore, BitVM is best suited for applications that rely on heavy off-chain computations. However, for applications that require frequent on-chain interactions, BitVM may not be the most efficient solution.

% \hi{limit: only two-party involved}

% \hi{limit: high cost for scripts}

% \hi{limit: frequent on-chain interactions}

% \qw{based on paragraph BitVM2, is there a limit on permissioned Verifier?}

%https://bitvm.org/bitvm2.html

\smallskip
\noindent{$\bullet$ \textit{BitVM2.} BitVM2 is an advanced iteration of the original BitVM, designed to enable permissionless verification~\cite{bitvm22023,bitvm22024}. This iteration overcomes the limitations of the initial design, which was confined to predefined two-party setups, by allowing any participant to act as a verifier during runtime.
%We highlight their differences. 

\smallskip
\noindent\textbf{SNARK verifier.} In BitVM2, the system employs a one-time setup with a \(1\text{-}n \) honesty assumption, where anyone can challenge an invalid assertion without being part of the initial verifier group. The verification process in BitVM2 utilizes the Groth16 proof system~\cite{baghery2020simulation} to verify assertions efficiently. A key technical enhancement is the division of the SNARK verifier program into smaller, manageable sub-programs, each verified through a sequence of steps using Lamport signatures~\cite{lamport1979signatures}. Specifically, let's denote the initial state as \( s_0 \) and the final state as \( s_n \). The state transitions are managed by an optimized state transition function \( \delta \):
\(
s_{i+1} = \delta(s_i).
\)
Each transition generates a proof \( \pi_i \) that verifies the transition without revealing any confidential information:
\(
\pi_i = \mathsf{Groth16}(s_i, s_{i+1}).
\)
To handle the large size of the SNARK verifier programs, the computation is split into multiple steps. Each step \( j \) is signed using Lamport signatures \( \sigma_j \):
\(
\sigma_j = \mathsf{LamportSign}(H_j, K_j),
\)
where \( H_j \) represents the hash of the computation step, and \( K_j \) is the corresponding key.

% Additionally, BitVM2 strengthens the role of the oracle \( O' \)~\cite{beniiche2020study} by using distributed oracles and enhanced cryptographic commitments. The oracle publishes a more secure public key \( K'_O \) and a signed message \( \sigma'_O \), ensuring the integrity of the contract's outcome. Pre-signed CETs in BitVM2 are executed based on the oracle's enhanced signature, mitigating the risks associated with a single point of failure. The final state and proof are submitted on-chain for verification: $\mathsf{(s'_n, \pi'_n) \rightarrow on-chain verification}$.

% \hi{pro: permissionless verifier (1-of-n)}

% \hi{pro: weak assumption on 1 honesty}

% \hi{pro: reduced complexity by SNARK}

\smallskip
\noindent\textbf{Limitations.} One challenges is the need to emulate covenants due to their absence in Bitcoin, which requires the involvement of a signer committee, thereby increasing off-chain coordination efforts. This adds complexity to the system and heightens the risk of operational delays or inefficiencies. Moreover, the design depends on the presence of at least one honest operator to prevent funds from becoming unspendable, creating the risk of ransom attacks, where malicious operators could freeze assets and hold them hostage~\cite{bitvm22023}. Additionally, the system generates large, non-standard transactions that may not be relayed by default Bitcoin nodes, necessitating alternative propagation methods and increasing reliance on custom network configurations~\cite{bitvm22024}. Lastly, the use of fixed deposit amounts restricts flexibility, which may force users to engage with professional services for peg-ins and peg-outs. This reliance on external services could reduce accessibility, particularly for smaller or casual users, and may create entry barriers for a broader range of participants.

%One of the primary challenges is the need to emulate covenants due to their absence in Bitcoin, requiring a signer committee and increasing off-chain coordination efforts. Additionally, the design necessitates the presence of at least one honest operator to avoid funds becoming unspendable, posing a risk of ransom attacks where dishonest operators could freeze funds~\cite{bitvm22023}. Furthermore, the system's design results in large, non-standard transactions that may not be relayed by default Bitcoin nodes, necessitating alternative propagation methods~\cite{bitvm22024}. Lastly, the fixed deposit amounts limit flexibility, potentially requiring users to interact with professional services for peg-ins and peg-outs, which could limit accessibility for some users.

% \qw{this should be pros or still cons?} 
% \mfcomment{should be a con, referring to \url{https://bitvm.org/bitvm2.html}}

% \hi{idea: permissionless verifier (1-of-n)}

% \hi{limit: presettings for operators: strong assumption, complexity}

% \begin{tcolorbox}[colback=gray!10!white, colframe=gray!80!black, arc=2mm, boxrule=0.5mm, left=1mm, right=1mm, top=1mm, bottom=1mm]
%     \textbf{\textit{(RQ2-b) Finding 3:}} \textit{Bitcoin L2 projects integrate with BitVM by utilizing Taproot addresses to create Taptrees, enabling Turing-complete smart contracts. They usually employ ZKP for secure state validation and use a UTXO model to ensure compatibility with Bitcoin’s transaction structure.}
% \end{tcolorbox}

\smallskip
\noindent\textbf{Discussion.} The surveyed projects~\cite{bitlayer2024, zbyte2024, bitstake2024, citrea2024, satoshivm2024}, each building on BitVM in different ways, collectively advance Bitcoin’s capabilities of off-chain computation. For instance, BitLayer builds on BitVM by integrating a more sophisticated state management system. ZKByte focuses on incorporating ZKP into Bitcoin’s L2, utilizing a ZKP validator while managing cross-chain state transitions through the UTXO model. Bitstake leverages the BitVM framework to implement a PoS mechanism, enabling dynamic participation in transaction management. On the other hand, Citrea implements a ZK Rollup architecture, introducing the concept of execution slices that batch process thousands of transactions. SatoshiVM simplifies circuit complexity by adopting the Bristol format for logical gate circuit structures and introduces single-round non-interactive on-chain verification.

%------------------------------------
\subsection{Rollups}\label{subsec-proof}
%------------------------------------

Data availability requires all necessary data for verifying to be readily accessible to all participants~\cite{weber2017availability}. However, as Bitcoin networks grow in size, ensuring that every participant has access to all transaction data becomes increasingly challenging. 
Rollups move most computational work off-chain, transmitting only small size data (e.g., transaction outputs, state updates, proof data) to the mainchain~\cite{chaliasos2024towards}. By processing transactions off-chain and periodically posting summarized data on-chain, rollups can improve transaction throughput.

\smallskip
\noindent\textbf{Components.} The mechanism relies on three  components to ensure the off-chain transaction processing and validation.

\begin{packeditemize}

\item\textit{Sequencer:} A sequencer plays a crucial role in collecting and ordering transactions before they are aggregated and processed. Essentially, the sequencer's primary job is to receive transactions from users, order them, and then batch them into a single rollup block~\cite{motepalli2023sok}. The way transactions are ordered can affect the state of the rollup and potentially the fees associated with those transactions.

\item\textit{Operator:} An operator executes transactions in an off-chain environment and manages the resulting state transitions~\cite{gopalan2020stability}. This off-chain execution and management facilitate reduced load and less computational effort on the mainnet. Operators are responsible for submitting proof of the rollup blocks to the blockchain mainnet. 

\item\textit{Challenger:} The role of challengers is to act as external verifiers who check the validity of the operator's commitments to the mainchain. They continuously monitor the state transitions and proofs. If challengers detect inconsistencies or fraudulent activity in the submitted data, they can submit a fraud-proof to the mainchain.

\end{packeditemize}

\smallskip
\noindent\textbf{Working mechanism.} Bitcoin rollups work by processing a large volume of transactions off-chain on an L2 network and then summarizing the results to submit to the main Bitcoin blockchain. The process involves five main steps:

\begin{circitemize}
    \item \textit{Transaction initialization on L2:} Users initiate transactions on the L2 Rollup. The sequencer and operator are responsible for sorting these transactions, aggregating them into batches, executing the transaction logic, and computing the post-transaction state.
    
    \item \textit{State representation:} The state of the blockchain, including account balances and state variables, is represented in a specific data structure such as a Merkle tree. The Pre-state Root \( R_{\text{pre}} \) represents the state before the transactions are executed, while the Post-state Root \( R_{\text{post}} \) represents the state after the transactions are executed:
    $R_{\text{pre}  = \mathsf{MerkleRoot}(S_{\text{pre}})} $ and $
    R_{\text{post}  = \mathsf{MerkleRoot}(S_{\text{post}})}$.

    \item \textit{Transaction proof generation:} The Operator generates a transaction proof \( \pi \) along with the Post-state Root \( R_{\text{post}} \). For zk-rollups, this proof is a succinct zk-proof:
    \(
    \pi = \mathsf{Groth16}(R_{\text{pre}}, R_{\text{post}}, T).
    \)
    For Optimistic rollups, the proof involves a fraud-proof mechanism where participants can challenge incorrect state transitions:
    \(
    \pi_{\text{fraud}} = \mathsf{FraudProof}(R_{\text{pre}}, R_{\text{post}}, T).
    \)
    
    \item \textit{Submission to the mainchain:} The Operator submits the Post-state Root \( R_{\text{post}} \) and the transaction proof \( \pi \) to the Rollup smart contract on the mainchain.
    
    \item \textit{Verification on the mainchain:} The Rollup contract verifies whether the state transition from \( R_{\text{pre}} \) to \( R_{\text{post}} \) is correct using the submitted proof:
    \(
    \mathsf{Verify}(R_{\text{pre}}, R_{\text{post}}, \pi) \rightarrow \mathsf{True}.
    \)
    If the verification passes, the Rollup contract updates the state on the mainchain, confirming the transactions that took place on L2. If an incorrect \( R_{\text{post}} \) is submitted, other participants can challenge it using \( R_{\text{pre}} \) and the transaction batch data:
    \(
    \mathsf{Verify}(R_{\text{pre}}, R_{\text{post}}, \pi_{\text{fraud}}) \rightarrow \mathsf{False}.
    \)
\end{circitemize}

\smallskip
\noindent\textbf{Types of rollups.} There are two primary types of rollups: Zero-knowledge Rollups (zk-rollups) that utilize ZKPs to validate transaction efficacy, and Optimistic rollups that employ fraud-proof mechanisms to ascertain transaction validity.

\begin{packeditemize}

\item \textit{Zk-rollups:} Operators in zk-rollups generate and submit a ZKP, specifically zk-SNARks and zk-STARKs~\cite{guan2020blockmaze}, to the mainchain. This validity proof confirms the correctness of all transactions in the batch and the resultant new state. zk-SNARKks, while fast due to their small proof sizes and quick verification times, require a trusted setup that can be a security concern if compromised~\cite{parno2016pinocchio}. Conversely, zk-STARKs avoid this issue by not needing a common reference string, offering better resistance against quantum computing, albeit at the cost of larger proof sizes~\cite{ozdemir2022experimenting, huang2022zkchain}.

\item \textit{Optimistic rollups:} The fundamental assumption in Optimistic rollups is that the transactions and state changes are valid by default. Different from using a ZKP, Optimistic rollups use a fraud-proof mechanism~\cite{al2018fraud}. When an illegal transaction is identified, challengers can challenge it by submitting a \textit{fraud proof} to the network. The person who submitted the fraudulent transaction, faces penalties, typically losing their bond. A portion of this bond is awarded to the challenger as a reward. While both types of rollups increase throughput, implementing zk-rollups is generally more complex than Optimistic rollups because of the intricacies involved in generating ZKPs. Conversely, Optimistic rollups typically have a longer withdrawal time due to the challenge period necessary for potential fraud proofs.

\end{packeditemize}

\smallskip
\noindent \textbf{Limitations.}
Implementing rollups on Bitcoin presents a unique set of challenges due to Bitcoin's underlying design, which is quite different from platforms like Ethereum that support on-chain programmability. Rollups require a rich set of functionalities to manage state transitions, process fraud or validity proofs, and handle complex dispute resolutions, which are beyond the capabilities of Bitcoin's current scripting language. In addition, to implement rollups efficiently, Bitcoin would likely need adjustments at the base-layer protocol to support essential features, such as covenants. Such changes require broad consensus within the community, which has been slow to progress given the Bitcoin community's wariness of any agreement. Furthermore, the absence of a native smart contract layer in Bitcoin complicates the deployment of rollups. Technologies like zk-rollups depend on complex smart contracts for generating and verifying zpks. Similarly, Optimistic rollups need smart contracts for managing the challenge-response mechanisms essential for their security model. The deployment of such systems would require a foundational shift in how Bitcoin operates.

\smallskip
\noindent \textbf{Discussion.} The rollup-based projects~\cite{bsquared2024, bisonlabs2024, tunachain2024, bl22024, sovryn2024, merlinchain2024, lumibit2024, bioplabs2024, rollux2024, gobob2024, bel22024, hacash2024, rollkit2024} on Bitcoin demonstrate a range of strategies to enhance scalability. In the zk-rollups category, Projects like B² Network and Bison leverage ZKP for secure transaction validation, with B² Network incorporating a hybrid zkEVM and fraud-proof challenge mechanism, while Bison emphasizes computational efficiency with zk-STARKs. Tuna Chain and BL2 use zk-rollups for batch processing and integrating Taproot and decentralized data availability layers, respectively, with BL2 introducing a dual-layer architecture for distributed verification. Sovryn and Merlin Chain further extend this category by combining zk-rollups with decentralized oracle networks and bi-directional bridges to enhance cross-chain interactions. In the Optimistic rollups category, projects like Biop and Rollux utilize the Optimistic Rollup protocol to batch transactions off-chain, with Biop integrating PoS consensus and fault-proof mechanisms, and Rollux leveraging the Optimism Bedrock for EVM equivalence. BeL2 adopts Elastos SmartWeb and a relayer mechanism for fraud prevention, while Hacash.com introduces a multi-layered architecture that supports high transaction throughput and utilizes rollup for real-time settlement.

%-------------------------------------
\subsection{SideChain} \label{subsec-sidechain}
Sidechains~\cite{back2014enabling} allow assets to be transferred between Bitcoin (acting as the mainchain) and secondary blockchains (sidechains) through a \textit{two-way peg} mechanism~\cite{ren2023interoperability}. This enables Bitcoin to be transferred to a sidechain, where it can benefit from the sidechain's unique features and capabilities, and then back to the mainchain.

% \begin{center}
% \fbox{%
%     \begin{minipage}{0.95\linewidth}

%     \textbf{(RQ2) Finding 4:} Sidechains allow assets to be transferred between the Bitcoin mainchain and other blockchains through a two-way peg mechanism. This process involves locking Bitcoin on the mainchain and using certain proofs (usually SPV) to unlock it on the sidechain.
    
%     \end{minipage}
% }
% \end{center}

\smallskip
\noindent \textbf{Simplified payment verification.} SPV~\cite{nakamoto2008bitcoin} is an essential component of the Bitcoin network that allows users to verify transactions without maintaining a full copy of the blockchain. The SPV node first calculates the hash \(H(T_x)\) of the transaction \(T_x\) that needs to be verified and then fetches all block headers \(H_b\) of the longest branch. It uses the Merkle path \(P(T_x)\) to calculate the Merkle root hash \(R_m\) and compares it with the Merkle root stored in the local block header to identify the block containing the transaction.

\smallskip
\noindent\textbf{Working mechanism.} The core component of sidechain architecture is the two-way peg mechanism, enabling asset transfer between Bitcoin and a sidechain, typically through SPV proof. The working principle of the sidechain is as follows.

% \begin{circitemize}
%     \item \textit{Locking Bitcoin on the mainchain:} The user creates a transaction \( \mathsf{T_{lock}} \) on the mainchain, sending \( \mathsf{A_{BTC}} \) to a locking script, recorded in block \( \mathsf{B_{main}} \) with Merkle root \( \mathsf{R_{main}} \).

%     \item \textit{Generating the SPV proof:} An SPV proof \( \mathsf{\pi_{SPV}} \) is generated, including the Merkle path \( \mathsf{P_{main}} \) from \( \mathsf{T_{lock}} \) to \( \mathsf{R_{main}} \):
%     \(
%     \mathsf{\pi_{SPV}} = \left\{\mathsf{P_{main}}, \mathsf{R_{main}}, \mathsf{T_{lock}}\right\}.
%     \)

%     \item \textit{Verification on the sidechain:} The sidechain verifies \( \mathsf{T_{lock}} \) using \( \mathsf{\pi_{SPV}} \) by checking whether
%     \( \mathsf{MerkleRoot}( \mathsf{P_{main}} \cup \mathsf{T_{lock}} ) \) equals to \( \mathsf{R_{main}} \),
%     and creates \( \mathsf{A_{side}} \) sidechain tokens, ensuring \( \mathsf{A_{side}} = \mathsf{A_{BTC}} \).

%     \item \textit{Returning Bitcoin to the mainchain:} To transfer back, the user destroys \( \mathsf{A_{side}} \) and provides proof \( \mathsf{\pi_{side}} \). The main chain verifies this and releases the Bitcoin.
% \end{circitemize}

\begin{circitemize}
    \item \textit{Locking Bitcoin on the mainchain:} The user initiates a transaction \( T_{\text{lock}} \) on the mainchain, sending \( \mathsf{A_{\text{BTC}}} \) to a specific locking script, which is recorded with Merkle root \( R_{\text{main}} \).

    \item \textit{Generating the SPV proof:} An SPV proof \( \mathsf{\pi_{\text{SPV}}} \) is generated, including the Merkle path \( \mathsf{P_{\text{main}}} \) from \( T_{\text{lock}} \) to \( R_{\text{main}} \), as well as the block header \( \mathsf{H_{\text{main}}} \):
    \(
    \mathsf{\pi_{\text{SPV}}} = \left\{\mathsf{P_{\text{main}}}, \mathsf{H_{\text{main}}}, R_{\text{main}}, T_{\text{lock}}\right\}.
    \)

    \item \textit{Verification on the sidechain:} The sidechain verifies \(T_{\text{lock}} \) using \( \mathsf{\pi_{\text{SPV}}} \) by ensuring that
    \(
    \mathsf{MerkleRoot}( \mathsf{P_{\text{main}}} \cup T_{\text{lock}} ) = R_{\text{main}}.
    \)
    Upon successful verification, the sidechain creates \( \mathsf{A_{\text{side}}} \) tokens, ensuring \( \mathsf{A_{\text{side}}} = \mathsf{A_{\text{BTC}}} \).

    \item \textit{Returning Bitcoin to the mainchain:} To transfer Bitcoin back to the mainchain, the user must destroy \( \mathsf{A_{\text{side}}} \) tokens and provide proof \( \mathsf{\pi_{\text{side}}} \). The mainchain verifies the proof and, if valid, releases the Bitcoin to the user.
\end{circitemize}

\smallskip
\noindent\textbf{Types of sidechains.} Sidechains can be classified into three main categories based on their consensus mechanisms.% Federated consensus sidechains, Merged-mined sidechains, and PoX-based sidechains.

\begin{packeditemize}

\item \textit{Federated consensus sidechains:} Federated sidechains rely on a group of trusted entities, known as a federation, to manage the locking and unlocking of assets between the mainchain and the sidechain~\cite{dilley2016strong}. This is achieved through federated multi-signature to lock the bitcoins released in the sidechain. The security of federated sidechains depends on the trustworthiness and honesty of the federation members.

\item \textit{Merged-mined sidechains:} A Merged-mined sidechains utilize Bitcoin’s miner network and the same PoW consensus to manage asset transfers.~\cite{scaffino2023glimpse}. Miners validate blocks on both the mainchain and sidechain simultaneously, using a two-way peg mechanism for asset movement. Similarly, Drivechain~\cite{drivechain2018}, employs Bitcoin miners as custodians to manage transfers. It uses miners' collective hash power to secure assets across different sidechains.

% \smallskip
% \item \textit{PoX-based sidechains:} In a PoX-based sidechain, synchronization with Bitcoin is achieved through a consensus mechanism other than PoW. Miners commit BTC transactions, creating anchor points that link the sidechain’s blocks to Bitcoin. These BTC transactions serve as a basis for producing new blocks on the Stacks sidechain, allowing miners to earn rewards in the sidechain’s native token.

\item \textit{PoX-based sidechains:} In a PoX-based sidechain, PoX miners commit BTC to the Bitcoin mainchain through a consensus mechanism other than PoW. These BTC transactions serve as “anchor points” that create a link between Bitcoin and the sidechain, synchronizing both blockchains. The miners are rewarded in the sidechain’s native token, like how miners are rewarded in PoW-based system.

\end{packeditemize}

\smallskip
\noindent \textbf{Limitations.}
Sidechains enhance the scalability of Bitcoin, but they come with several limitations. Security is a prime concern due to reduced decentralization in sidechains, which often rely on a smaller set of validators for consensus, potentially concentrating power. Interoperability is another challenge, as each sidechain may operate under different rules and protocols, complicating asset transfers between chains. The design of sidechains also involves trade-offs, particularly between decentralization and the desired performance characteristics such as transaction speed and cost. 

% \hi{limit: security issues in pegs}

% \hi{limit: interoperability issue due to fragmentation}

% \begin{tikzpicture}
% \draw[dashed] (0,0) -- (8,0);
% \end{tikzpicture}

% \mfcomment{not sure where to place it, leaving it for further discussion}
% Glimpse: On-Demand PoW Light Client with Constant-Size Storage for DeFi
% \url{https://www.usenix.org/system/files/usenixsecurity23-scaffino.pdf}

% Glimpse leverages a novel construction for light clients that enables efficient cross-chain verification. This mechanism works by creating a bridge that does not require extensive on-chain data storage. Instead, it uses constant-size storage and computational overhead to verify the state of another blockchain on demand. Glimpse is particularly effective because it remains compatible with blockchains that have limited scripting capabilities, such as the Liquid Network, a pegged sidechain of Bitcoin. The security of Glimpse is validated within the Universal Composability (UC) framework, ensuring robust and secure cross-chain interactions.

% \begin{tikzpicture}
% \draw[dashed] (0,0) -- (8,0);
% \end{tikzpicture}

\smallskip
\noindent \textbf{Discussion.} The sidechain-based projects~\cite{liquid2024, rootstock2024, mapprotocol2024, stacks2024, mintlayer2024, bouncebit2024, libre2024, bitrexe2024, bevm2024, rgb2024, rgb2024lightpaper, bihelix2024, bitlightlabs2024, yin2022boolnetwork, mercurylayer2024} investigated in the paper can be grouped into the three categories above. Federated consensus sidechains, such athe Liquid Network, implement a consortium model that employs blocksigners and watchmen, diverging from Bitcoin's PoW mechanism. Merged-mined sidechains such as Rootstock leverage Powpeg mechanism, which involves multi-signature management to manage BTC transfers between the mainchain and the sidechain. In the realm of PoX-based sidechains, projects like MAP Protocol push the boundaries by using ZKP and light clients for cross-chain interoperability, while Stacks ties its security to Bitcoin via PoX, allowing miners to earn rewards for BTC commitments. Mintlayer and BounceBit extend this concept by employing PoS mechanisms, with Mintlayer using Bitcoin block hashes for randomness in block production and BounceBit integrating both CeFi and DeFi elements. Libre, operating on a Delegated Proof-of-Stake (DPoS) mechanism, integrates with Bitcoin via the Lightning Network. BitReXe introduces a multi-VM system using the PREDA programming model, and BEVM utilizes Musig2 for secure multi-signature schemes and native BTC as gas.

\subsection{Client-side Verification Plus UTXO} \label{subsec-csv}
Client-side validation (CSV) allows for an efficient method of transaction verification. Instead of every node handling every transaction, clients validate the parts of the transaction history relevant to them. However, by enabling users to authenticate data validity autonomously, this method bypasses the standard consensus model, potentially resulting in data discrepancies among diverse clients.

The UTXO set, recognized for its immutability consensus, serves as a foundational ledger within the Bitcoin network~\cite{delgado2019analysis}. Establishing a dependable linkage between off-chain assets and the state of UTXOs is essential for extending consensus-driven validation. The CSV+UTXO approach combines off-chain asset issuance with a dynamic on-chain UTXO ledger. The UTXO set acts as a time witness that can verify the precise state of Bitcoin transactions at any given time. This feature allows users to synchronize changes in these transactions with changes in the state of other assets.

% \begin{center}
% \fbox{%
%     \begin{minipage}{0.95\linewidth}

%     \textbf{(RQ2) Finding 5:} CSV allows clients to validate only relevant transactions. It integrates the immutable UTXO set as a foundational ledger to synchronize off-chain asset changes with the on-chain state.
    
%     \end{minipage}
% }
% \end{center}

\smallskip
\noindent\textbf{Single use seal.} 
%https://blackpaper.rgb.tech/consensus-layer/3.-client-side-validation/3.2.-single-use-seals
A single-use seal is a cryptographic primitive that facilitates a two-tiered commitment~\cite{todd2016commitments}. It allows a committer to commit to a message at a future point in time, with the assurance that this commitment can only be made once. Bitcoin Transaction Output-based Single-use-Seals (TxO Seals~\cite{rgbtech2024}) apply this concept to the Bitcoin transaction graph. Parties in the protocol agree on a transaction output with special meaning, designated as a "seal." A future transaction, known as the "witness transaction," is required to contain a deterministic Bitcoin commitment to a specific message. There are two properties of TxO Seals:

\begin{packeditemize}
    \item \textit{Hiding:} Independent parties cannot detect the presence of the commitment in the transaction graph, even if the original message is known.
    \item \textit{Verifiability:} Given information about the specific commitment protocol and access to deterministic Bitcoin proof data, any party can verify the commitment's validity for the intended message only.
\end{packeditemize}

\smallskip
\noindent\textbf{Working mechanism.} The CSV+UTXO technology leverages the hiding and verifiability of TxO seals to enhance the Bitcoin network's scalability. Below is a process overview:

\begin{circitemize}
    \item \textit{Transaction initiation:} A client \( C_i \) initiates a transaction \( T_x \). This transaction is committed to the UTXO set, where each UTXO \( U_i \) represents an unspent transaction output.
    
    \item \textit{Client-side validation:} Each client \( C_i \) validates the transaction \( T_x \) by verifying the relevant UTXOs \( U_i \) associated with the transaction. The client computes the transaction hash \( H(T_x) \) and validates it using the Merkle root \( R_m \).
    
    \item \textit{Single-use seal commitment:} A single-use seal \( S \) is applied to the transaction output \( T_x \). Let \( M \) be the message to be committed, and \( C \) be the commitment. The commitment \( C \) is represented as:
    \(
    C = \mathsf{Commit}(H(T_x), M).
    \)
    
    \item \textit{Deterministic Bitcoin proof:} The witness transaction includes a deterministic Bitcoin commitment to \( M \). The proof \( \pi \) is verified using the UTXO set \( U_i \), where \( H(U_i) \) represents the hash of the relevant UTXO:
    \(
    \pi = \mathsf{Proof}(C, H(U_i)).
    \)
    
    \item \textit{Verification:} The commitment's validity is verified by any party with access to the proof data. The verifiability property ensures that the commitment \( C \) can only be valid for the intended message \( M \) and that it is linked to the specific UTXO \( U_i \).
    \(
    \mathsf{Verify}(\pi, H(U_i)) \rightarrow \mathsf{True}.
    \)
\end{circitemize}

\noindent\textbf{Limitations.} The implementation of this approach is highly complex, primarily due to Bitcoin’s inherent design, which does not natively support such intricate computations. Additionally, When a client receives a payment, it may need to verify a large amount of data all at once, which can be considered a drawback. This issue typically arises when a client first acquires an asset with a lengthy transaction history.

\smallskip
\noindent \textbf{Discussion.} The reviewed projects~\cite{rgb2024, rgb2024lightpaper, bihelix2024, bitlightlabs2024, yin2022boolnetwork, mercurylayer2024} collectively demonstrate a focused attempt to broaden Bitcoin's functionalities through the uses of UTXO and CSV. RGB embeds cryptographic commitments in UTXOs for asset issuance and management. RGB++ expands on RGB by integrating other UTXO-based blockchains like CKB and Cardano as verification layers, trading off some privacy for increased global verifiability. BiHelix further builds on the RGB protocol, integrating with the Lightning Network to enhance Bitcoin’s interoperability. Bitlight Labs focuses on CSV to manage smart contracts, employing the Simplicity language to enable Turing-complete scripting. Bool Network uses Dynamic Hidden Committees and integrates CSV for cross-chain security. Lastly, Mercury Layer emphasizes statechain technology to facilitate rapid Bitcoin UTXO transfers, relying heavily on CSV for statechain verification.

% %https://www.pathbtc.com/
% it also uses UTXO
% \noindent\textit{(6) Path Protocol:}

\subsection{State Channel} \label{subsec-sidechannel}
State channels are designed to facilitate off-chain transactions between parties~\cite{dziembowski2018general}. These channels are particularly useful for frequent transactions, such as micropayments.
% State channel is a L2 scaling solution that enables off-chain transactions between parties. These channels are established to facilitate frequent, low-value transactions, such as micropayments, in a manner that is cost-effective and fast.

% \begin{center}
% \fbox{%
%     \begin{minipage}{0.95\linewidth}

%     \textbf{(RQ2) Finding 6:} State channels create a multi-signature address where participants deposit Bitcoin to open the channel, allowing numerous off-chain transactions through state updates that reflect fund distributions. The channel is closed by broadcasting a final transaction reflecting the last agreed state to the blockchain.
    
%     \end{minipage}
% }
% \end{center}

\smallskip
\noindent
\textbf{Working mechanism.} The working mechanism of state channels can be broken down into the key steps involved in the opening, operation, and closure of a state channel.
 
\begin{circitemize}

\item \textit{Opening the channel:} To establish a state channel, two or more participants, say Alice and Bob, create a multi-signature address \( A_{\text{multi}} \), which requires signatures from both parties. Each participant deposits a certain amount of Bitcoin, \( B_{\text{Alice}} \) and \( B_{\text{Bob}} \), into this address. The transaction \( T_{\text{open}} \) marking the channel’s opening is recorded on the blockchain:
\(
T_{\text{open}} = \mathsf{Tx}(A_{\text{multi}}, B_{\text{Alice}} + B_{\text{Bob}}).
\)
Once the channel is open, participants can conduct numerous transactions off-chain. These transactions are reflected as state updates \( S_i \), where each update represents a new distribution of the funds held in the channel.

\item \textit{Conducting off-chain transactions:} During the operation of the state channel, participants exchange signed transactions \( T_i \) that update the state of the channel without broadcasting these transactions to the Bitcoin network. These state updates are essentially promises of future transactions that both parties agree upon:
\(
T_i = \text{SignedTx}(S_i),
\)
with signatures from Alice and Bob. The most recent state update always supersedes previous ones, ensuring that the latest agreement on the fund distribution is maintained. For example, if Alice and Bob agree on a state update \( S_2 \) after initially agreeing on \( S_1 \), then:
\(
S_2 \Rightarrow \mathsf{(new\ distribution:\ B_{\text{Alice}}^{'},\ B_{\text{Bob}}^{'})}.
\)

\item \textit{Closing the channel:} When the participants decide to close the channel, they create a final transaction \( T_{\text{close}} \) reflecting the last agreed-upon state \( S_{\text{final}} \) and broadcast this transaction to the Bitcoin blockchain:
\(
T_{\text{close}} = \mathsf{Tx}(A_{\text{multi}}, S_{\text{final}}).
\)
This final transaction closes the channel and redistributes the funds according to the last state update. By doing so, the blockchain only records two transactions: \( T_{\text{open}} \) to open the channel and \( T_{\text{close}} \) to close it, regardless of the number of off-chain transactions conducted within the channel.

\end{circitemize}

\smallskip
\noindent\textbf{Limitations.} One challenge is the complexity of managing the multi-sig setup required to open a channel~\cite{poon2015bitcoin}. Each transaction within the channel needs to be signed by all parties involved, which introduces delays and coordination issues, particularly when involving more than two participants. Additionally, the mechanism for closing a channel is vulnerable to disputes~\cite{mccorry2016towards}. If a party tries to broadcast an outdated transaction state, it requires timely intervention from the other party to prevent fraud, which necessitates constant monitoring.
zi'xi

\smallskip
\noindent \textbf{Discussion.} The projects~\cite{poon2016lightningnetwork, omnibolt2024, nostrassets2024, arkprotocol2024} exemplify the technical advancements of state channels. Lightning Network utilizes mechanisms like RSMC for secure updates and HTLC for conditional transactions, enabling off-chain payments through a network of payment channels. OmniBOLT builds on the Lightning Network by facilitating the off-chain circulation of smart assets, adding capabilities such as cross-channel atomic swaps through its Golang-based protocol suite. Lnfi Network extends the state channel concept by integrating Taproot assets with Nostr’s key management. Ark Protocol innovates on state channel by eliminating the need for incoming liquidity, allowing for the seamless conversion of on-chain UTXOs to virtual UTXOs and back.

\smallskip
\noindent\textbf{Others.} In addition to the aforementioned solutions, this category of Bitcoin L2 solutions includes a variety of other methods (Appendix \ref{more}). This includes Tectum~\cite{tectum2024fastest}, which introduces SoftNotes~\cite{softnote2024} for fast, off-chain transactions; Nubit~\cite{nubit2024}, which employs \textit{data availability sampling} (DAS) for efficient block verification; and HyperAGI~\cite{hyperagi2024}, designed to support compute-intensive applications for AI. These projects represent the diversity in Bitcoin’s evolving L2 ecosystem.

\begin{center}
\fbox{%
    \begin{minipage}{0.95\linewidth}
    
    \textbf{(RQ2) Finding 2:} Among the examined Bitcoin L2 projects, \textbf{12\%} embed data into transactions to extend functionality. Furthermore, \textbf{81\%} of the projects utilize off-chain computation to alleviate on-chain congestion, while \textbf{72\%} employ cryptographic proofs, including ZKPs and SPV, for secure state verification.

    \end{minipage}
}
\end{center}

%===============================
\section{Security Evaluation}\label{sec-security}
%===============================

%This section introduces a comprehensive evaluation model designed to rigorously assess the security of Bitcoin L2 projects. By applying 

%ref from\url{https://www.btceden.org/?status=active}

\subsection{Security Framework}

By examining various vulnerability sources~\cite{btceden2024, conti2018survey, slowmist2024}, we identified a series threats across the entire lifecycle of asset flow, categorizing them into: \textit{unauthorized withdrawals}, \textit{incorrect transaction verification}, \textit{manipulation and censorship}, \textit{fraudulent disputes}, \textit{data unavailability}, and \textit{the inability to recover assets during emergencies}. 

We accordingly present our structured security evaluation framework (answering \textbf{RQ3}), comprising six critical criteria (denoted by \textbf{C}) as follows. We provide a detailed analysis of each project’s strengths and vulnerabilities as in Table~\ref{tab-summary}.

%— \textit{withdrawal mechanism} (WM), \textit{state verification} (SV), \textit{anti-censorship measures} (ACM), \textit{dispute resolution} (DR), \textit{data availability} (DA), and \textit{emergency asset recovery} (EAR) — 

%************************************
%************************************
%************************************

\begin{table*}[!bth]
\caption{Assessment for Bitcoin L2 Projects}\label{tab-summary}
\vspace{-0.9em}
\begin{center}
\begin{threeparttable}

\resizebox{0.99\textwidth}{!}{
\begin{tabular}{|l|r c|cccccccccccc|ccccc|}
\toprule

& 
\multicolumn{1}{c}{\rotatebox{0}{\textbf{Project}}} & 
\rotatebox{0}{\textbf{Type}} & 
\multicolumn{2}{c}{\rotatebox{30}{\textbf{Withdrawal}}} &
\multicolumn{2}{c}{\rotatebox{30}{\textbf{State verification}}} &
\multicolumn{2}{c}{\rotatebox{30}{\textbf{Anti-censorship}}} & 
\multicolumn{2}{c}{\rotatebox{30}{\textbf{Dispute resolution}}} &
\multicolumn{2}{c}{\rotatebox{30}{\textbf{Data avaliability}}} &
\multicolumn{2}{c|}{\rotatebox{30}{\textbf{\makecell{Emergency\\ recovery}}}} &
\rotatebox{90}{\textbf{Scalability}} & 
\rotatebox{90}{\textbf{Efficiency}}  & 
\rotatebox{90}{\textbf{Functionality}} &
\rotatebox{90}{\textbf{Complexity}} & 
\rotatebox{90}{\textbf{Interoperability}} 
\\   
\midrule %******************************** 
\multirow{5}{*}{\rotatebox{90}{\textbf{BitVM \& BitVM2}}} & \cellcolor{blue!10} Bitlayer~\cite{bitlayer2024}  &  \multirow{5}{*}{BitVM}  & MSA, DV, TLC  & \tikz\pic{sema=magenta/1/white}; & ZKP, PSC & \tikz\pic{sema=teal/1/white}; &  DS, DoP & \tikz\pic{sema=electricpurple/90/white}; & IFP, EDRP & \tikz\pic{sema=limegreen/90/white}; & POC & \tikz\pic{sema=skyblue/90/white};  & FCT, TET & \tikz\pic{sema=goldenrod/90/white}; & \cellcolor{teal!12} H. & \cellcolor{yellow!25} M. & \cellcolor{teal!12} H. & \cellcolor{teal!12} H. & \cellcolor{yellow!25} M. \\

& \cellcolor{blue!10} ZKByte\cite{zbyte2024}   & & MSA, CP, TLC & \tikz\pic{sema=magenta/1/white}; & ZKP, CC & \tikz\pic{sema=teal/1/white}; & DS, GM & \tikz\pic{sema=electricpurple/90/white}; & IFP, RoV & \tikz\pic{sema=limegreen/90/white}; & DDA, POC & \tikz\pic{sema=skyblue/90/white};  & FCT, MV & \tikz\pic{sema=goldenrod/90/white}; & \cellcolor{teal!12} H. & \cellcolor{yellow!25} M. & \cellcolor{yellow!25} M. & \cellcolor{yellow!25} M. & \cellcolor{yellow!25} M. \\

& \cellcolor{blue!10} \text{Bitstake}\cite{bitstake2024} & & \text{MSA, CP} & \tikz\pic{sema=magenta/90/white}; & \text{ZKP} & \tikz\pic{sema=teal/180/white}; & \text{DS} & \tikz\pic{sema=electricpurple/180/white}; & \text{IFP} & \tikz\pic{sema=limegreen/180/white}; & \text{DDA} & \tikz\pic{sema=skyblue/90/white}; & \text{FCT} & \tikz\pic{sema=goldenrod/180/white}; & \cellcolor{yellow!25} M. & \cellcolor{magenta!10} L. & \cellcolor{yellow!25} M. & \cellcolor{yellow!25} M. & \cellcolor{magenta!10} L. \\

& \cellcolor{blue!10} Citrea\cite{citrea2024}  & & MSA & \tikz\pic{sema=magenta/180/white}; & ZKP, PSC & \tikz\pic{sema=teal/1/white}; & DS & \tikz\pic{sema=electricpurple/180/white}; & IFP & \tikz\pic{sema=limegreen/180/white}; & DDA, POC & \tikz\pic{sema=skyblue/90/white}; & TLC & \tikz\pic{sema=goldenrod/180/white}; & \cellcolor{yellow!25} H. & \cellcolor{yellow!25} M. & \cellcolor{teal!12} H. & \cellcolor{teal!12} H. & \cellcolor{yellow!25} M. \\

& \cellcolor{blue!10}  SataoshiVM\cite{satoshivm2024}  &   & MSA, CP & \tikz\pic{sema=magenta/90/white}; & ZKP & \tikz\pic{sema=teal/180/white}; & DS & \tikz\pic{sema=electricpurple/180/white}; & IFP & \tikz\pic{sema=limegreen/180/white}; & DDA & \tikz\pic{sema=skyblue/180/white}; & FCT, MV & \tikz\pic{sema=goldenrod/90/white}; & \cellcolor{yellow!25} M. & \cellcolor{yellow!25} M. & \cellcolor{teal!12} H. & \cellcolor{yellow!25} M. & \cellcolor{yellow!25} M. \\

\midrule

\multirow{14}{*}{\rotatebox{90}{\textbf{Rollups}}} & \cellcolor{blue!10}  B2 Network\cite{bsquared2024}  &  \multirow{7}{*}{ZK rollups} & MSA, CP & \tikz\pic{sema=magenta/90/white}; & ZKP, PSC & \tikz\pic{sema=teal/1/white}; & DS, GM & \tikz\pic{sema=electricpurple/90/white}; & IFP & \tikz\pic{sema=limegreen/180/white}; & DDA & \tikz\pic{sema=skyblue/180/white}; & FCT & \tikz\pic{sema=goldenrod/180/white}; & \cellcolor{yellow!25} M. & \cellcolor{yellow!25} M. & \cellcolor{yellow!25} M. & \cellcolor{yellow!25} M. & \cellcolor{yellow!25} M. \\

& \cellcolor{blue!10} Bison\cite{bisonlabs2024}  & & MSA, CP & \tikz\pic{sema=magenta/90/white}; & ZKP, PSC & \tikz\pic{sema=teal/1/white}; & DS & \tikz\pic{sema=electricpurple/180/white}; & IFP, RoV & \tikz\pic{sema=limegreen/90/white}; & DDA, POC & \tikz\pic{sema=skyblue/90/white}; & FCT & \tikz\pic{sema=goldenrod/180/white}; & \cellcolor{yellow!25} M. & \cellcolor{yellow!25} M. & \cellcolor{teal!12} H. & \cellcolor{yellow!25} M. & \cellcolor{teal!12} H. \\

& \cellcolor{blue!10} Tuna Chain\cite{tunachain2024}  & & MSA, TLC & \tikz\pic{sema=magenta/90/white}; & ZKP, PSC & \tikz\pic{sema=teal/1/white}; & DS & \tikz\pic{sema=electricpurple/180/white}; & IFP, EDRP & \tikz\pic{sema=limegreen/90/white}; & DAS & \tikz\pic{sema=skyblue/180/white}; & FCT, EW& \tikz\pic{sema=goldenrod/90/white}; & \cellcolor{teal!12} H. & \cellcolor{yellow!25} M. & \cellcolor{yellow!25} M. & \cellcolor{teal!12} H. & \cellcolor{yellow!25} M. \\

& \cellcolor{blue!10} BL2\cite{bl22024}  & & MSA, CP & \tikz\pic{sema=magenta/90/white}; & ZKP & \tikz\pic{sema=teal/180/white}; & DS, GM & \tikz\pic{sema=electricpurple/90/white}; & IFP, EDRP & \tikz\pic{sema=limegreen/90/white}; & DDA & \tikz\pic{sema=skyblue/180/white};  & FCT & \tikz\pic{sema=goldenrod/180/white}; & \cellcolor{teal!12} H. & \cellcolor{yellow!25} M. & \cellcolor{yellow!25} M. & \cellcolor{yellow!25} M. & \cellcolor{yellow!25} M. \\

& \cellcolor{blue!10} Sovryn\cite{sovryn2024}  & & MSA, TLC & \tikz\pic{sema=magenta/90/white}; & ZKP, PSC & \tikz\pic{sema=teal/1/white}; & DS, GM & \tikz\pic{sema=electricpurple/90/white}; & IFP, RoV & \tikz\pic{sema=limegreen/90/white}; & DDA, POC & \tikz\pic{sema=skyblue/90/white};  & FCT & \tikz\pic{sema=goldenrod/180/white}; & \cellcolor{yellow!25} M. & \cellcolor{yellow!25} M. & \cellcolor{yellow!25} M. & \cellcolor{yellow!25} M. & \cellcolor{teal!12} H. \\

& \cellcolor{blue!10} Melin Chain\cite{merlinchain2024}  & & MSA, CP & \tikz\pic{sema=magenta/90/white}; & ZKP, CC & \tikz\pic{sema=teal/1/white}; & DS, GM & \tikz\pic{sema=electricpurple/90/white}; & IFP, RoV & \tikz\pic{sema=limegreen/90/white}; & DDA, POC & \tikz\pic{sema=skyblue/90/white};  & TET & \tikz\pic{sema=goldenrod/180/white}; & \cellcolor{teal!12} H. & \cellcolor{yellow!25} M. & \cellcolor{teal!12} H. & \cellcolor{teal!12} H. & \cellcolor{yellow!25} M. \\

& \cellcolor{blue!10} LumiBit\cite{lumibit2024}  & & MSA, CP & \tikz\pic{sema=magenta/90/white}; & ZKP, CC & \tikz\pic{sema=teal/90/white}; & DS, GM & \tikz\pic{sema=electricpurple/90/white}; & IFP & \tikz\pic{sema=limegreen/180/white}; & DDA, POC & \tikz\pic{sema=skyblue/90/white};  & FCT, EW& \tikz\pic{sema=goldenrod/90/white}; & \cellcolor{yellow!25} M. & \cellcolor{yellow!25} M. & \cellcolor{teal!12} H. & \cellcolor{yellow!25} M. & \cellcolor{yellow!25} M. \\

\cmidrule{2-3}
& \cellcolor{blue!10} Biop\cite{bioplabs2024}  & \multirow{6}{*}{Optimistic rollups}& MSA, CP, TLC & \tikz\pic{sema=magenta/1/white}; & ORP, PSC & \tikz\pic{sema=teal/1/white}; & DS, GM & \tikz\pic{sema=electricpurple/90/white}; & IFP, RoV & \tikz\pic{sema=limegreen/90/white}; & DDA, POC & \tikz\pic{sema=skyblue/90/white}; & FCT & \tikz\pic{sema=goldenrod/180/white}; & \cellcolor{teal!12} H. & \cellcolor{yellow!25} M. & \cellcolor{teal!12} H. & \cellcolor{teal!12} H. & \cellcolor{yellow!25} M. \\

& \cellcolor{blue!10} Rollux\cite{rollux2024}  & & MSA, DV & \tikz\pic{sema=magenta/90/white}; & ORP, CC & \tikz\pic{sema=teal/1/white}; & DS & \tikz\pic{sema=electricpurple/180/white}; & IFP & \tikz\pic{sema=limegreen/180/white}; & - & \tikz\pic{sema=skyblue/359/white}; & FCT & \tikz\pic{sema=goldenrod/180/white}; & \cellcolor{teal!12} H. & \cellcolor{yellow!25} M. & \cellcolor{yellow!25} M. & \cellcolor{yellow!25} M. & \cellcolor{magenta!10} L. \\

& \cellcolor{blue!10} BOB\cite{gobob2024}  & & MSA, CP & \tikz\pic{sema=magenta/90/white}; & ORP, PSC & \tikz\pic{sema=teal/1/white}; & DS & \tikz\pic{sema=electricpurple/180/white}; & IFP, IP & \tikz\pic{sema=limegreen/90/white}; & DDA, POC & \tikz\pic{sema=skyblue/90/white};  & TET, MV & \tikz\pic{sema=goldenrod/90/white}; & \cellcolor{teal!12} H. & \cellcolor{yellow!25} M. & \cellcolor{teal!12} H. & \cellcolor{teal!12} H. & \cellcolor{yellow!25} M. \\

& \cellcolor{blue!10} BeL2\cite{bel22024}  &
& MSA, CP & \tikz\pic{sema=magenta/90/white}; 
& ORP, CC & \tikz\pic{sema=teal/90/white}; 
& DS, GM & \tikz\pic{sema=electricpurple/90/white}; 
& IFP, RoV & \tikz\pic{sema=limegreen/90/white}; 
& DDA, POC & \tikz\pic{sema=skyblue/90/white}; 
& FCT, TET & \tikz\pic{sema=goldenrod/90/white}; & \cellcolor{teal!12} H. & \cellcolor{yellow!25} M. & \cellcolor{teal!12} H. & \cellcolor{teal!12} H. & \cellcolor{yellow!25} M. \\

& \cellcolor{blue!10} Hacash.com\cite{hacash2024}  & & MSA, CP & \tikz\pic{sema=magenta/90/white}; & ORP, PSC & \tikz\pic{sema=teal/90/white}; & DS & \tikz\pic{sema=electricpurple/180/white}; & IFP & \tikz\pic{sema=limegreen/180/white}; & - & \tikz\pic{sema=skyblue/359/white}; & FCT & \tikz\pic{sema=goldenrod/180/white}; & \cellcolor{yellow!25} M. & \cellcolor{yellow!25} M. & \cellcolor{teal!12} H. & \cellcolor{yellow!25} M. & \cellcolor{magenta!10} L. \\

& \cellcolor{blue!10} Rollkit\cite{rollkit2024}  & & MSA & \tikz\pic{sema=magenta/180/white}; & ORP, CC & \tikz\pic{sema=teal/1/white}; & DS, DoP & \tikz\pic{sema=electricpurple/90/white}; & IFP, RoV & \tikz\pic{sema=limegreen/90/white}; & DDA, POC & \tikz\pic{sema=skyblue/90/white}; & FCT, TET & \tikz\pic{sema=goldenrod/90/white}; & \cellcolor{teal!12} H. & \cellcolor{yellow!25} M. & \cellcolor{teal!12} H. & \cellcolor{teal!12} H. & \cellcolor{yellow!25} M. \\

\midrule

\multirow{10}{*}{\rotatebox{90}{\textbf{Sidechains}}} & \cellcolor{blue!10}  Liquid Network\cite{liquid2024}  &  \multirow{1}{*}{Federated consensus} & MSA & \tikz\pic{sema=magenta/180/white}; & PSC & \tikz\pic{sema=teal/180/white}; & DS & \tikz\pic{sema=electricpurple/180/white}; & IFP, EDRP & \tikz\pic{sema=limegreen/90/white}; & DDA, POC & \tikz\pic{sema=skyblue/90/white}; & TET & \tikz\pic{sema=goldenrod/180/white}; & \cellcolor{yellow!25} M. & \cellcolor{yellow!25} M. & \cellcolor{teal!12} H. & \cellcolor{yellow!25} M. & \cellcolor{yellow!25} M. \\

\cmidrule{2-3}
&  \cellcolor{blue!10}  Rootstock\cite{rootstock2024}  &  \multirow{1}{*}{Merged mining} & MSA, CP & \tikz\pic{sema=magenta/90/white}; & ZKP, PSC & \tikz\pic{sema=teal/1/white}; & DS & \tikz\pic{sema=electricpurple/180/white}; & IFP & \tikz\pic{sema=limegreen/180/white}; & DDA & \tikz\pic{sema=skyblue/180/white}; & FCT & \tikz\pic{sema=goldenrod/180/white}; & \cellcolor{teal!12} H. & \cellcolor{yellow!25} M. & \cellcolor{teal!12} H. & \cellcolor{teal!12} H. & \cellcolor{yellow!25} M. \\

\cmidrule{2-3}
&  \cellcolor{blue!10}  MAP Protocol\cite{mapprotocol2024} &  \multirow{7}{*}{PoX consensus}& MSA & \tikz\pic{sema=magenta/90/white}; & ZKP & \tikz\pic{sema=teal/180/white}; & DS & \tikz\pic{sema=electricpurple/180/white}; & - & \tikz\pic{sema=limegreen/359/white}; & DDA & \tikz\pic{sema=skyblue/180/white}; & FCT & \tikz\pic{sema=goldenrod/180/white}; & \cellcolor{yellow!25} M. & \cellcolor{yellow!25} M. & \cellcolor{teal!12} H. & \cellcolor{yellow!25} M. & \cellcolor{yellow!25} M. \\

&  \cellcolor{blue!10} Stacks\cite{stacks2024}  &  & MSA & \tikz\pic{sema=magenta/180/white}; & ZKP, CC & \tikz\pic{sema=teal/1/white}; & DS & \tikz\pic{sema=electricpurple/180/white}; & IFP & \tikz\pic{sema=limegreen/180/white}; & - & \tikz\pic{sema=skyblue/359/white}; & FCT & \tikz\pic{sema=goldenrod/180/white}; & \cellcolor{teal!12} H. & \cellcolor{yellow!25} M. & \cellcolor{yellow!25} M. & \cellcolor{teal!12} H. & \cellcolor{yellow!25} M. \\

&  \cellcolor{blue!10} Mintlayer\cite{mintlayer2024} & & MSA, CP & \tikz\pic{sema=magenta/90/white}; & ZKP, PSC & \tikz\pic{sema=teal/90/white}; & DS, GM & \tikz\pic{sema=electricpurple/90/white}; & IFP, EDRP & \tikz\pic{sema=limegreen/90/white}; & DDA, POC & \tikz\pic{sema=skyblue/90/white}; & FCT, EW& \tikz\pic{sema=goldenrod/90/white}; & \cellcolor{teal!12} H. & \cellcolor{yellow!25} M. & \cellcolor{teal!12} H. & \cellcolor{teal!12} H. & \cellcolor{yellow!25} M. \\

&  \cellcolor{blue!10} BounceBit\cite{bouncebit2024}  & & MSA, CP & \tikz\pic{sema=magenta/90/white}; & ZKP & \tikz\pic{sema=teal/180/white}; & GM, DoP & \tikz\pic{sema=electricpurple/90/white}; & IFP, RoV & \tikz\pic{sema=limegreen/90/white}; & DDA & \tikz\pic{sema=skyblue/180/white}; & TET & \tikz\pic{sema=goldenrod/180/white}; & \cellcolor{teal!12} H. & \cellcolor{yellow!25} M. & \cellcolor{teal!12} H. & \cellcolor{teal!12} H. & \cellcolor{yellow!25} M. \\

&  \cellcolor{blue!10} Libre\cite{libre2024} & & MSA, CP & \tikz\pic{sema=magenta/90/white}; & ZKP & \tikz\pic{sema=teal/180/white}; & GM & \tikz\pic{sema=electricpurple/180/white}; & IFP, RoV & \tikz\pic{sema=limegreen/90/white}; & DDA, POC & \tikz\pic{sema=skyblue/90/white}; & FCT & \tikz\pic{sema=goldenrod/180/white}; & \cellcolor{teal!12} H. & \cellcolor{yellow!25} M. & \cellcolor{yellow!25} M. & \cellcolor{yellow!25} M. & \cellcolor{yellow!25} M. \\

&  \cellcolor{blue!10} BitReXe\cite{bitrexe2024} & & MSA & \tikz\pic{sema=magenta/180/white}; & ZKP & \tikz\pic{sema=teal/180/white}; & GM & \tikz\pic{sema=electricpurple/180/white}; & IFP & \tikz\pic{sema=limegreen/180/white}; & DDA, POC & \tikz\pic{sema=skyblue/90/white}; & FCT & \tikz\pic{sema=goldenrod/180/white}; & \cellcolor{yellow!25} M. & \cellcolor{yellow!25} M. & \cellcolor{magenta!10} L. & \cellcolor{yellow!25} M. & \cellcolor{magenta!10} L. \\

&  \cellcolor{blue!10} BEVM\cite{bevm2024} & & MSA, CP & \tikz\pic{sema=magenta/90/white}; & ZKP & \tikz\pic{sema=teal/180/white}; & DS, GM & \tikz\pic{sema=electricpurple/90/white}; & IFP, RoV & \tikz\pic{sema=limegreen/90/white}; & DDA & \tikz\pic{sema=skyblue/180/white}; & TET, MV & \tikz\pic{sema=goldenrod/90/white}; & \cellcolor{yellow!25} M. & \cellcolor{yellow!25} M. & \cellcolor{teal!12} H. & \cellcolor{yellow!25} M. & \cellcolor{teal!12} H. \\

\midrule

\multirow{6}{*}{\rotatebox{90}{\textbf{CSV Plus UTXO}}} & \cellcolor{blue!10}  RGB\cite{rgb2024}  & \multirow{6}{*}{\makecell{Client side\\ verification}}& MSA, CP & \tikz\pic{sema=magenta/90/white}; & ZKP, CC, PSC & \tikz\pic{sema=teal/1/white}; & DS & \tikz\pic{sema=electricpurple/180/white}; & IFP, EDRP & \tikz\pic{sema=limegreen/90/white}; & DDA, POC & \tikz\pic{sema=skyblue/90/white}; & FCT, MV & \tikz\pic{sema=goldenrod/90/white}; & \cellcolor{yellow!25} M. & \cellcolor{yellow!25} M. & \cellcolor{teal!12} H. & \cellcolor{teal!12} H. & \cellcolor{yellow!25} M. \\

&  \cellcolor{blue!10} RGB++\cite{rgb2024lightpaper}  &  & CP, TLC & \tikz\pic{sema=magenta/90/white}; & CC, PSC & \tikz\pic{sema=teal/90/white}; & DS, DoP & \tikz\pic{sema=electricpurple/90/white}; & IFP, EDRP & \tikz\pic{sema=limegreen/90/white}; & POC & \tikz\pic{sema=skyblue/180/white}; & - & \tikz\pic{sema=goldenrod/359/white}; & \cellcolor{teal!12} H. & \cellcolor{yellow!25} M. & \cellcolor{yellow!25} M. & \cellcolor{yellow!25} M. & \cellcolor{yellow!25} M. \\

&  \cellcolor{blue!10} BiHelix\cite{bihelix2024} & & MSA, CP & \tikz\pic{sema=magenta/90/white}; & ZKP & \tikz\pic{sema=teal/180/white}; & DS & \tikz\pic{sema=electricpurple/180/white}; & IFP & \tikz\pic{sema=limegreen/180/white}; & - & \tikz\pic{sema=skyblue/359/white}; & FCT & \tikz\pic{sema=goldenrod/180/white}; & \cellcolor{yellow!25} M. & \cellcolor{magenta!10} L. & \cellcolor{yellow!25} M. & \cellcolor{yellow!25} M. & \cellcolor{magenta!10} L. \\

&  \cellcolor{blue!10} Bitlight Labs\cite{bitlightlabs2024}  & & MSA & \tikz\pic{sema=magenta/180/white}; & ZKP & \tikz\pic{sema=teal/180/white}; & DS & \tikz\pic{sema=electricpurple/180/white}; & - & \tikz\pic{sema=limegreen/359/white}; & DDA & \tikz\pic{sema=skyblue/180/white}; & FCT & \tikz\pic{sema=goldenrod/180/white}; & \cellcolor{yellow!25} M. & \cellcolor{yellow!25}  M. & \cellcolor{yellow!25} M. & \cellcolor{yellow!25} M. & \cellcolor{magenta!10} L. \\

&  \cellcolor{blue!10} Bool Network\cite{yin2022boolnetwork} & & MSA, CP & \tikz\pic{sema=magenta/90/white}; & ZKP & \tikz\pic{sema=teal/180/white}; & DS & \tikz\pic{sema=electricpurple/180/white}; & IFP & \tikz\pic{sema=limegreen/180/white}; & - & \tikz\pic{sema=skyblue/359/white}; & FCT & \tikz\pic{sema=goldenrod/180/white}; & \cellcolor{yellow!25} M. & \cellcolor{teal!12} H. & \cellcolor{teal!12} H. & \cellcolor{teal!12} H. & \cellcolor{yellow!25} M. \\

&  \cellcolor{blue!10} Mercury Layer\cite{mercurylayer2024} & & MSA & \tikz\pic{sema=magenta/180/white}; & ZKP & \tikz\pic{sema=teal/180/white}; & - & \tikz\pic{sema=electricpurple/359/white}; & IFP & \tikz\pic{sema=limegreen/180/white}; & DDA & \tikz\pic{sema=skyblue/180/white}; & TET & \tikz\pic{sema=goldenrod/180/white}; & \cellcolor{yellow!25} M. & \cellcolor{yellow!25}  M. & \cellcolor{yellow!25} M. & \cellcolor{yellow!25} M. & \cellcolor{magenta!10} L. \\

\midrule

\multirow{4}{*}{\rotatebox{90}{\textbf{State Channel}}} & \cellcolor{blue!10}  Lightning Net.\cite{poon2016lightningnetwork}  &  \multirow{4}{*}{State channel} & MSA, CP & \tikz\pic{sema=magenta/90/white}; & PSC & \tikz\pic{sema=teal/180/white}; & DS & \tikz\pic{sema=electricpurple/180/white}; & IFP & \tikz\pic{sema=limegreen/180/white}; & DDA & \tikz\pic{sema=skyblue/180/white}; & FCT & \tikz\pic{sema=goldenrod/180/white}; & \cellcolor{teal!12} H. & \cellcolor{teal!12} H. & \cellcolor{yellow!25} M. & \cellcolor{yellow!25} M. & \cellcolor{yellow!25} M. \\

& \cellcolor{blue!10}  OmniBOLT\cite{omnibolt2024}  & & MSA, TLC & \tikz\pic{sema=magenta/90/white}; & ZKP, PSC & \tikz\pic{sema=teal/90/white}; & DS & \tikz\pic{sema=electricpurple/180/white}; & IFP, RoV & \tikz\pic{sema=limegreen/90/white}; & DDA & \tikz\pic{sema=skyblue/180/white}; & FCT & \tikz\pic{sema=goldenrod/180/white}; & \cellcolor{teal!12} H. & \cellcolor{yellow!25} M. & \cellcolor{yellow!25} M. & \cellcolor{yellow!25} M. & \cellcolor{yellow!25} M. \\

& \cellcolor{blue!10}  Nostr Assets\cite{nostrassets2024}  & & MSA, TLC & \tikz\pic{sema=magenta/90/white}; & ZKP & \tikz\pic{sema=teal/180/white}; & DS, GM & \tikz\pic{sema=electricpurple/90/white}; & IFP & \tikz\pic{sema=limegreen/180/white}; & DDA & \tikz\pic{sema=skyblue/180/white}; & FCT & \tikz\pic{sema=goldenrod/180/white}; & \cellcolor{teal!12} H. & \cellcolor{yellow!25} M. & \cellcolor{teal!12} H. & \cellcolor{yellow!25} M. & \cellcolor{yellow!25} M. \\

& \cellcolor{blue!10}  Ark Protocol\cite{arkprotocol2024}  & & MSA, DV & \tikz\pic{sema=magenta/90/white}; & - & \tikz\pic{sema=teal/359/white}; & DS & \tikz\pic{sema=electricpurple/180/white}; & IFP & \tikz\pic{sema=limegreen/180/white}; & DDA & \tikz\pic{sema=skyblue/180/white}; & FCT & \tikz\pic{sema=goldenrod/180/white}; & \cellcolor{teal!12} H. & \cellcolor{yellow!25} M. & \cellcolor{teal!12} H. & \cellcolor{yellow!25} M. & \cellcolor{magenta!10} L. \\

\midrule

\multirow{4}{*}{\rotatebox{90}{\textbf{Others}}} & \cellcolor{blue!10}  Tectum\cite{tectum2024fastest}  &  \multirow{1}{*}{SoftNote} & TLC & \tikz\pic{sema=magenta/180/white}; & ZKP & \tikz\pic{sema=teal/180/white}; & GM & \tikz\pic{sema=electricpurple/180/white}; & IFP & \tikz\pic{sema=limegreen/180/white}; & DDA & \tikz\pic{sema=skyblue/180/white}; & FCT & \tikz\pic{sema=goldenrod/180/white}; & \cellcolor{teal!12} H. & \cellcolor{yellow!25} M. & \cellcolor{yellow!25} M. & \cellcolor{yellow!25} M. & \cellcolor{yellow!25} M. \\

\cmidrule{2-3}

& \cellcolor{blue!10}  Nubit\cite{nubit2024}  &  \multirow{1}{*}{Data availability } & CP, DV & \tikz\pic{sema=magenta/90/white}; & - & \tikz\pic{sema=teal/359/white}; & DS, GM & \tikz\pic{sema=electricpurple/90/white}; & IFP, RoV & \tikz\pic{sema=limegreen/90/white}; & DDA, POC, DAS & \tikz\pic{sema=skyblue/1/white}; & FCT & \tikz\pic{sema=goldenrod/180/white}; & \cellcolor{teal!12} H. & \cellcolor{yellow!25} M. & \cellcolor{teal!12} M. & \cellcolor{teal!12} H. & \cellcolor{yellow!25} M. \\

\cmidrule{2-3}

& \cellcolor{blue!10}  HyperAGI\cite{hyperagi2024}  &  \multirow{1}{*}{AGI} & MSA & \tikz\pic{sema=magenta/180/white}; & ZKP & \tikz\pic{sema=teal/180/white}; & DS, GM & \tikz\pic{sema=electricpurple/90/white}; & IFP, RoV & \tikz\pic{sema=limegreen/90/white}; & DDA, DAS & \tikz\pic{sema=skyblue/90/white}; & FCT, TET & \tikz\pic{sema=goldenrod/90/white}; & \cellcolor{teal!12}  H. & \cellcolor{yellow!25} M. &  \cellcolor{teal!12} H. & \cellcolor{teal!12} H. & \cellcolor{yellow!25} M. \\

\midrule %************************************

\multicolumn{1}{|c|}{} 
& \multicolumn{1}{c}{} 
& \multicolumn{1}{c}{} 
& \multicolumn{12}{|c|}{\textbf{Security Features}} 
& \multicolumn{5}{c|}{\textbf{Properties}} 
\\
%\cmidrule(lr){4-15}
%\cmidrule(lr){16-20}

%&  &  &  &  &   &  &  &  &   &   &  &   & &   &  \\

\end{tabular} 
}

\begin{tablenotes}
      \scriptsize
     % \footnotesize
      \item[] \textbf{Security Level:} the greater the proportion of color, the higher the safety in a criterion.
      \item[]  \tikz\pic{sema=magenta/180/white}; in magenta represents WM series;   
      \tikz\pic{sema=teal/180/white}; in teal indicates SV series; 

      \item[]  \tikz\pic{sema=electricpurple/180/white}; in electricpurple represents ACM series;   \tikz\pic{sema=limegreen/180/white}; in limegreen indicates DR series;  
    
      \item[]   \tikz\pic{sema=skyblue/180/white}; in skyblue represents DA series;  \tikz\pic{sema=goldenrod/180/white}; in goldenrod indicates EAR series.

      \item[] \textbf{Abbr.}: (C1) MSA = Multi-Signature Authentication; CP = Cryptographic Proofs; DV = Decentralized Validators; TLC = Time-Locked Contracts;
      \item[] (C2) ZKP = Zero-Knowledge Proofs; ORP = Optimistic Rollups Proofs; CC = Cryptographic Commitments; PSC = Periodic State Commitments;
      \item[] (C3) DS = Decentralized Sequencers; GM = Governance Models; DoP = Distribution of Power; MMR = Mitigation of Manipulation Risks;
      \item[] (C4) IFP = Implementation of Fraud Proofs; EDRP = Efficiency of Dispute Resolution Process; RoV = Role of Validators; IP = Incentives and Penalties;
      \item[] (C5) DDA = Decentralized Data Availability Solutions; DAS = Data Availability Sampling; EC = Erasure Coding; POC = Periodic On-Chain Commitments;
      \item[] (C6) FCT = Force-Close Transactions; TET = Time-Locked Exit Transactions; EW = Exit Windows; MV = Monitoring by Validators.
      \item[] \textbf{Property}:  $-$ = Does NOT provide property/feature;  H./M./L. with background colors = High/Medium/Low.
     \end{tablenotes}
     
   \end{threeparttable}
   %\vspace{-0.2in}
\end{center}

\end{table*}

\begin{comment}
    \item 

\begin{center}
\fbox{%
    \begin{minipage}{0.95\linewidth}

    \textbf{(RQ3) Finding 2:} 

    \end{minipage}
}
\end{center}

\end{comment}

\smallskip
\noindent\textbf{C1. Withdrawal mechanism} (WM) evaluates whether the L2 solution employs robust security measures to safeguard the withdrawal process, ensuring that funds can be securely and correctly transferred back to the Bitcoin mainchain. The primary security threat associated with this criterion is the risk of \textit{unauthorized withdrawals}, where malicious actors could potentially gain access to funds without proper authorization. Insecure or flawed withdrawal mechanisms could lead to the loss of assets, as attackers might exploit vulnerabilities to initiate fraudulent withdrawals or intercept the process. To mitigate the threat, WM considers security features as below:

\begin{packeditemize}
    \item[\textcolor{magenta}{$\triangleright$}] \textit{Multi-signature authentication:} This involves requiring multiple parties to sign off on a transaction before it can be executed. It adds an additional layer of security, ensuring that no single entity can unilaterally withdraw funds.

    \item[\textcolor{magenta}{$\triangleright$}] \textit{Cryptographic proofs:} These are mathematical techniques that allow one party to prove to another that a statement is true without revealing any information beyond the validity of the statement itself. In the context of withdrawals, the proofs are typically used to guarantee that the transaction data is valid without revealing sensitive information.

    \item[\textcolor{magenta}{$\triangleright$}] \textit{Decentralized validators:} Using a group of validators to approve transactions can distribute the power to authorize withdrawals across multiple independent entities, reducing the risk of collusion or single points of failure (SPoF).

    \item[\textcolor{magenta}{$\triangleright$}] \textit{Time-locked contracts:} They are, in most cases, a pair of smart contracts on both legs that delay the execution of a transaction until a specified amount of time has passed. The contracts can be used to create a buffer period during which any unauthorized withdrawal can be detected and canceled.

\end{packeditemize}

\noindent\textbf{C2. State verification} (SV) assesses the robustness of the L2 solution’s validation techniques to ensure the correctness of off-chain transactions and their accurate reflection on the mainchain. The primary security threat associated with this criterion is the risk of \textit{incorrect transaction verification}, where invalid or manipulated off-chain transactions could be improperly validated and recorded on the mainchain. This could lead to incorrect balances, loss of funds, or unauthorized asset transfers. To mitigate this threat, it is crucial that the L2 solution employs stringent state verification mechanisms that can effectively detect and prevent invalid state transitions:

\begin{packeditemize}
\item[\textcolor{teal}{$\triangleright$}] \textit{Zero-knowledge proofs:} These proofs allow validators to confirm the validity of transactions without needing to see the transaction details. This preserves privacy while ensuring that only legitimate transactions are processed.

\item[\textcolor{teal}{$\triangleright$}] \textit{Fraud proofs for Optimistic rollups:} These are mechanisms that allow any network participant to challenge and prove a fraudulent transaction. If a transaction is deemed fraudulent, it can be reverted, and the dishonest party penalized.

\item[\textcolor{teal}{$\triangleright$}] \textit{Cryptographic commitments:} This technique involves creating a cryptographic hash of transaction data, which can later be used to verify the integrity of the data. It ensures that the data has not been tampered with.

\item[\textcolor{teal}{$\triangleright$}] \textit{Periodic state commitments to mainchain:} Regularly committing the state of off-chain transactions to the mainchain creates a verifiable and immutable record. This helps maintain the integrity of the Bitcoin blockchain.

\end{packeditemize}

\noindent\textbf{C3. Anti-censorship measures} (ACM) evaluate the L2 solution’s ability to prevent any single entity from controlling or censoring transactions within the network. The primary security threat associated with this criterion is the potential for \textit{manipulation and censorship}, where a powerful entity or group could selectively block or reorder transactions to their advantage, undermining the fairness and neutrality of the network. This threat can lead to scenarios where certain transactions are unfairly delayed or never processed. To mitigate the threat, ACM should consider the following aspects:

\begin{packeditemize}
\item[\textcolor{electricpurple}{$\triangleright$}] \textit{Decentralized sequencers:} Sequencers are responsible for ordering transactions in a decentralized manner, preventing any single party from manipulating the transaction order.

\item[\textcolor{electricpurple}{$\triangleright$}] \textit{Governance models:} These models distribute decision-making power across a broad range of stakeholders, preventing any single entity from having undue influence.

\item[\textcolor{electricpurple}{$\triangleright$}] \textit{Distribution of power:} Ensuring that power is distributed among multiple participants reduces the risk of censorship and promotes a more democratic system.

\item[\textcolor{electricpurple}{$\triangleright$}] \textit{Mitigation of manipulation risks:} Measures to detect and prevent manipulation of the transaction process help ensure that all transactions are processed fairly.

\end{packeditemize}

\noindent\textbf{C4. Dispute resolution} (DR) examines the mechanisms that an L2 solution employs to resolve conflicts, particularly those involving \textit{fraudulent activities or disputes}. If disputes are not resolved efficiently, it could lead to manipulation of the network, where malicious actors might take advantage of unresolved conflicts, further leading to network instability, loss of user funds, and reduced trust in the system.  %To mitigate risks, L2 solutions should incorporate the following aspects.

%a combination of fraud proofs, governance roles, and incentives or penalties.

\begin{packeditemize}

\item[\textcolor{limegreen}{$\triangleright$}] \textit{Fraud proofs:} These proofs allow network participants to challenge and prove fraudulent transactions, ensuring that any dishonest activities are promptly addressed.

\item[\textcolor{limegreen}{$\triangleright$}]\textit{Efficiency of dispute resolution process:} The speed and effectiveness with which disputes are resolved are critical to maintaining trust in the system. Efficient processes ensure that issues are addressed quickly and fairly.

\item[\textcolor{limegreen}{$\triangleright$}]\textit{Role of validators/arbitrators:} Validators or arbitrators play a crucial role in resolving disputes. Their impartiality is essential for a fair dispute-resolution process.

\item[\textcolor{limegreen}{$\triangleright$}]\textit{Incentives and penalties:} A system of rewards and penalties encourages honest behavior and discourages fraud. This includes rewarding participants who detect and report fraud and penalizing those who attempt to commit fraud.

\end{packeditemize}

\noindent\textbf{C5. Data availability} (DA) assesses the strategies implemented by an L2 solution to ensure that all transaction data is available for verification at all times. The primary concern here is to prevent scenarios where crucial transaction data becomes inaccessible, leading to potential validation failures or network manipulation. \textit{Data unavailability} poses a severe security threat, as it could allow malicious actors to hide or manipulate transaction details, undermining the integrity of the verification process. Without guaranteed data availability, the network risks losing the ability to detect fraudulent activities, making it vulnerable to exploitation. 

%The L2 solution should integrate decentralized data availability solutions, data availability sampling, erasure coding, and periodic on-chain commitments to mitigate these risks.

\begin{packeditemize}
\item[\textcolor{skyblue}{$\triangleright$}] \textit{Decentralized data availability:} The metric involves using distributed nodes within the network to serve transaction data, ensuring no SPoF can compromise data availability.

\item[\textcolor{skyblue}{$\triangleright$}] \textit{Data availability sampling:} This technique involves checking random samples of data to ensure that the entire dataset is available and has not been tampered with.

\item[\textcolor{skyblue}{$\triangleright$}] \textit{Erasure coding:} This is a data protection method that breaks data into fragments, encodes them, and distributes them across multiple locations. It ensures that data can be reconstructed even if some fragments are lost or corrupted.

\item[\textcolor{skyblue}{$\triangleright$}] \textit{Off-chain storage with periodic on-chain commitments:} Storing data off-chain reduces the burden on the mainchain, while periodic on-chain commitments provide a verifiable record that the data has not been tampered with.

\end{packeditemize}

\noindent\textbf{C6. Emergency asset recovery} (EAR) focuses on evaluating the effectiveness of the mechanisms to allow users to reclaim their assets during emergencies, such as network failures, breaches, or unexpected shutdowns. The corresponding security threat in this context is the potential loss of assets due to system failures or malicious attacks that prevent users from \textit{recovering assets}. This criterion examines whether the L2 solution provides robust methods to initiate and complete asset recovery, ensuring that users have a fallback plan. %to protect their investments. 

%The detailed criteria include implementing force-close transactions, time-locked exit transactions, exit windows, and continuous monitoring by validators. 

\begin{packeditemize}

\item[\textcolor{goldenrod}{$\triangleright$}]\textit{Force-close transactions:} These transactions allow users to close their positions and recover their assets even if the network is experiencing issues (e.g., asynchrony).

\item[\textcolor{goldenrod}{$\triangleright$}]\textit{Time-locked exit transactions:} These transactions provide a delayed period during which users can exit the system securely. Assets can be protected against immediate threats.

\item[\textcolor{goldenrod}{$\triangleright$}]\textit{Exit windows:} These are specified periods during which users can exit the system, providing regular opportunities for asset recovery.

\item[\textcolor{goldenrod}{$\triangleright$}]\textit{Monitoring by validators and watchtowers:} Validators and watchtowers continuously monitor the network for issues and ensure that exit transactions are processed correctly.

\end{packeditemize}

\subsection{Case Study}

We select a representative project (Bitlayer~\cite{bitlayer2024}) as a case study. We examine each criterion with a scale of \textit{low}, \textit{medium}, and \textit{high} to show the security performance.

%that demonstrates how we apply our security framework for an assessment.
%The proposed six criteria will be examined to determine how BitLayer meets the security requirements for a secure L2 solution.

% Furthermore, we designed a scoring mechanism (Table~\ref{tab:evaluation_scoring}) that assigns scores to each criterion based on a predefined set of evaluation points. Each criterion will be scored on a scale of 0 to 5, and the scores will be weighted according to their importance. The final score will provide a comprehensive assessment of the project’s security.

%https://www.aicoin.com/en/article/404243
%https://www.aicoin.com/en/article/398004

%------------ may need it ---------------------
% Bitlayer~\cite{bitlayer2024} addresses the trade-off between security and Turing completeness in Bitcoin L2 technologies by leveraging inspirations from BitVM, DLC, and Lightning Network protocols. Its working mechanism relies on interactive off-chain communications and pre-signed transactions only broadcasted when certain conditions are met.
%------------ ------------- ---------------------

% DLC~\cite{mitdci2024} is a smart contract technology introduced by MIT’s Digital Currency Initiative, designed to facilitate conditional payment on Bitcoin. BitLayer leverages DLC to facilitate complex contract logic off-chain (shown in Figure~\ref{fig:dlc}). 

\smallskip
\noindent\textbf{Multi-sig authentication.} The process begins with participants like Alice and Bob creating an off-chain fund transaction on Bitlayer. Each locks a certain amount of Bitcoin into a 2-of-2 multi-sig address, ensuring that funds can only be moved with both parties’ consent. This setup directly addresses the WM criterion by preventing unauthorized access to the locked funds through multi-sig authentication. The pre-signed nature of these transactions ensures that funds are securely committed without immediate blockchain involvement. Consequently, BitLayer’s multi-sig authentication is effective in securing funds, thus rating this aspect as \textit{high}.

\smallskip
\noindent\textbf{Creation of CETs.} Next, Alice and Bob pre-sign various CETs representing all potential outcomes of their contract. These CETs include conditions to distribute the funds based on an oracle’s outcome announcement. This process aligns with state verification by employing cryptographic commitments to ensure data integrity and tamper-proof records of off-chain states. The CETs are designed to execute based on ZKPs, ensuring that the transaction details remain private while validating the outcome. Therefore, the robust creation and use of CETs in BitLayer provide strong state verification capabilities, making this criterion rate \textit{high}.

\smallskip
\noindent\textbf{Time-locked contracts.} Oracles play a crucial role in determining the outcome by signing the corresponding CET based on real-world events. To ensure security, time-locked contracts are used so that if the oracle does not provide a signature within a specified period, the locked funds are returned to the participants. This ensures that funds are not indefinitely locked, satisfying the criteria of EAR by providing a secure exit strategy. Consequently, rating this aspect as \textit{high}.

\smallskip
\noindent\textbf{Broadcasting the fund transaction.} After all CETs are pre-signed, the fund transaction is broadcasted, locking the funds in the multi-signature address. Upon receiving the oracle’s signed outcome, the appropriate CET is executed to redistribute the funds. If an incorrect CET is broadcasted, BitLayer’s fraud-proof mechanism enables participants to challenge the transaction, ensuring disputes are resolved fairly. This dispute resolution process effectively meets the DR criterion, resulting in a \textit{high} rating for BitLayer.

% Once all CETs are pre-signed, the fund transaction is broadcasted, locking the funds in the multi-signature address. Upon receiving the oracle’s signed outcome, the correct CET is executed to redistribute the funds. If an incorrect CET is broadcasted, BitLayer’s fraud-proof mechanism allows participants to challenge the transaction, ensuring that disputes are resolved fairly. This dispute resolution process meets the DR criterion by enabling secure resolution of fraudulent activities. As a result, the effectiveness of BitLayer’s dispute resolution (DR) mechanism rates \textit{high}.

\smallskip
\noindent\textbf{Decentralized validators.} BitLayer employs decentralized validators to validate transactions, preventing any single entity from manipulating the process. This measure fulfills the ACM criterion by distributing power and ensuring unbiased transaction sequencing. Furthermore, the CETs and fund transactions are stored off-chain, with only necessary data being committed on-chain when required. This hybrid storage approach ensures that data is always accessible for validation purposes while maintaining transparency. Thus, BitLayer’s use of decentralized validators and a hybrid storage approach provides strong ACM and DA, rating these two criteria \textit{high}.

\begin{center}
\fbox{%
    \begin{minipage}{0.95\linewidth}

    \textbf{(RQ3) Finding 3:} Most projects (\textbf{85\%}) exhibit strong withdrawal mechanisms, particularly through MSA and TLC. State verification is robustly supported by cryptographic proofs like ZKPs among projects (\textbf{72\%}). Anti-censorship measures (\textbf{55\%}) vary, with decentralized sequencers and governance models ensuring fair transaction processing in some projects. Dispute resolution mechanisms (\textbf{78\%}) are generally effective, with fraud-proof implementations and the role of validators playing key roles. However, data availability (\textbf{42\%}) and emergency asset recovery (\textbf{37\%}) are less consistently implemented, with only a subset of projects offering comprehensive solutions.

    \end{minipage}
}
\end{center}

% Our security evaluation reveals a varied performance among Bitcoin L2 solutions across key security criteria.

% to conclude the result of all projects in the aspect of security

% %=================================================   
% \section{Progress}
% \mfcomment{Existing L2 Project Summary Progress}
% \url{https://gitmind.com/app/docs/mogscf56}
% %=================================================  

%=================================================  
\section{Further Discussion} 
\label{sec-conclu}
%=================================================   
%\mfcomment{TODO: review}

% https://adiabat.github.io/dlc.pdf
%https://github.com/discreetlogcontracts/dlcspecs/blob/master/Introduction.md
%https://arxiv.org/abs/1904.05234
%https://github.com/discreetlogcontracts/dlcspecs/
% https://medium.com/@Bitlayer/bitlayer-core-technology-dlc-and-its-optimization-considerations-6fc5ebaae92c

\smallskip
\noindent\textbf{More evaluated properties} (last \textit{five} columns in Table~\ref{tab-summary}). We also assess the selected projects using a broader set of criteria (i.e., properties from existing studies) at three levels (high, medium, and low) to provide extended references.

\smallskip
\noindent\textbf{Limitations of DLC.} One issue is the contract’s funding guarantee. For example, Alice and Bob can use a DLC to hedge against Bitcoin price volatility by agreeing on a future price for Bitcoin, settled in USD. Since funds are pre-committed, the contract can only handle limited price fluctuations. If the price of Bitcoin falls below a certain threshold, the pre-committed funds may not cover the contract’s value, locking the funds inefficiently. Increasing the amount of committed funds could mitigate this, but it also means locking up more capital. Another challenge is the extensive number of signatures required for every possible contract outcome. This can become unmanageable in scenarios with many potential outcomes, like forward contracts with varied price points. A proposed solution~\cite{daian2019flash} involves breaking down values into fractional parts, reducing the number of signatures needed for a wide range of values, but this solution has yet to be verified. 

% \textit{c. Trust in oracles also poses a challenge.} The contract relies on the oracle to provide accurate data. If an oracle fails to sign or provide necessary data, the contract could become unresolvable unless multiple oracles are used to mitigate this risk. 

% Adding multiple oracle signatures~\cite{daian2019flash} can ensure that the contract can still be settled even if one oracle fails. 

% Lastly, the current DLC setup does not support the transfer of contracts between participants, which could be useful in scenarios where one party wants to settle early. 

% DLCs enable parties to enter into a contract based on predefined conditions, with the settlement of funds occurring privately off-chain~\cite{discreetlogcontracts2024}. This is achieved through the use of oracles, which provide signed messages indicating the outcome of the contract conditions.

% Implementing a mechanism for transferring contract participation could enhance the flexibility and utility of DLCs.

\smallskip
\noindent\textbf{Centralization of indexers.}
% e.g., \url{https://www.hiro.so/blog/the-challenges-of-building-an-indexer-for-bitcoin-ordinals}
Bitcoin indexers are crucial for enabling off-chain computations. For example, B²nodes~\cite{bsquared2024} are equipped with indexers to parse and index inscriptions. These indexers work by scanning the blockchain, extracting necessary data, and maintaining states. However, the reliance on centralized indexers (i.e. UniSat~\cite{unisat2024}) introduces several security risks. Centralized indexers can become SPoF, susceptible to attacks that could manipulate data. For instance, a malicious indexer could provide incorrect transaction histories or balances. Additionally, the concentration of indexing power raises concerns about the potential for Sybil attacks. To mitigate risks, some works~\cite{nubit2024, hiro2024ordhook, yu2023bridging,rooch2024btc} utilize decentralized indexing solutions, but they are still in infancy.

% -------------------may not need solutions?
% To mitigate these risks, some projects are exploring decentralized indexing solutions. One approach involves creating a network of decentralized indexers~\cite{li2021bringing} that verify each other’s data, thereby reducing the risk of manipulation. Additionally, to reduce the burden on individual indexers, stateless computation~\cite{wen2024state} is explored, allowing for the verification of only essential state data, which enables the validation of indexer outputs without requiring the complete state data.

% % https://eprint.iacr.org/2024/408.pdf
% In industry, the Nubit project~\cite{nubit2024} proposes a modular indexing solution where a committee of indexers publishes checkpoints to a data availability layer, ensuring data can be verified even if most indexers are compromised. Similarly, Hiro’s Ordhook project~\cite{hiro2024ordhook} for Ordinals adopts a backward traversal approach to track satoshi transfers, improving performance in the face of blockchain reorganizations and other challenges. Additionally, the Rooch project~\cite{rooch2024btc} leverages on-chain indexers to expand the Bitcoin ecosystem. RoochBTC enables smart contracts to verify Bitcoin block headers and UTXO states as objects within its MoveVM, allowing cross-verification and indexing by a lightweight client. This setup ensures that UTXO states are accurately reflected without locking assets on the Bitcoin network.

\smallskip
\noindent\textbf{Congested mempool.} The popularity of Bitcoin inscriptions has introduced significant congestion to mempool. Additionally, the proposal of BitVM facilitates more complex smart contracts on Bitcoin, which further increases network congestion. The mempool serves as a staging area for valid but unconfirmed transactions, and its size reflects the network’s congestion level~\cite{mikhaylov2023bitcoin}. A congested mempool indicates a high volume of transactions waiting to be processed, which inherently leads to longer confirmation times and increased transaction fees. For miners, this congestion can be beneficial as it increases the fees they earn per block mined. However, for regular users, this translates into higher costs and delays, making Bitcoin less efficient for everyday transactions. Therefore, optimizing Bitcoin’s network efficiency while preserving its programming functionality is a critical challenge. Researchers could investigate more efficient transaction validation methods or develop new smart contract frameworks that minimize on-chain data storage to manage mempool congestion.

% https://mempool.space/graphs/mempool#3y
% https://arxiv.org/pdf/2404.11189

% When the mempool becomes congested, transactions are delayed and incur higher fees as they wait for inclusion in a block, which can hold up to a million vbytes.
% With the introduction of Ordinals, there has been a noticeable impact on the mempool size. When Ordinals launched in January 2023, the mempool size was approximately 7.1 million vbytes. However, as the popularity of Ordinals grew, the mempool expanded significantly, peaking at 247.9 million vbytes by March 24, the highest level in nearly two years~\cite{mempool2024}.

% - Unreliable offchain Indexers/validators -> inconsistency

% - Liquidity Concerns: require users to lock up funds in channels, liquidity issues, where funds are not readily available for other uses.

% - integration and Interoperability: 

% - Complexity:

% - the lack of (homogeneous) finality guarantees pose challenge 

% the introduction of temporary data storage on Ethereum — blobs —
% and rollups’ dependence on them — puts the security and
% longevity of blob- and rollup-based inscription into question.
% \qw{result observed in}
% \url{https://arxiv.org/pdf/2404.11189}

%=================================================  
\section{Conclusion}
%\label{sec-conclu}
%=================================================   
%\smallskip
%\noindent\textbf{Conclusion.} 
We present the first systematic study of Bitcoin L2 solutions by collecting and evaluating existing projects. We further propose and apply a security reference framework to assess the identified design patterns within each selected project. We conclude that L2 brings Bitcoin functionalities but also introduces increased complexity and additional attack vectors.

% see this 
% https://www.usenix.org/conference/usenixsecurity25/ethics-guidelines
%\smallskip
%\noindent\underline{\textbf{Ethic considerations.}} This SoK paper does not involve any ethical risks. All the data used is publicly accessible. There are no concerns related to animals, human subjects, the environment, healthcare, or military factors. We confirm that we have have adhered to all ethical principles  outlined in CFPs.

% reviewed the ethics considerations outlined in the conference CFPs and

%================================================
%================================================
\bibliographystyle{unsrt}
\bibliography{bib(short).bib}
%================================================
%================================================

\appendix

\section{Technical Components}\label{sec-technical}

\subsection{UTXO Model}

A Unspent Transaction Output (UTXO) represents a specific amount of Bitcoin that has been received but not yet spent by the owner. Essentially, it is the cryptocurrency equivalent of cash in hand, where each UTXO can be thought of as an individual bill or coin.

UTXO represents a discrete amount of Bitcoin that remains constant until it is spent. The association with a Bitcoin address denotes ownership and the right to spend, while the transaction ID provides a traceable history, linking the UTXO back to its origin in the blockchain.

\smallskip
\noindent\textbf{Working mechanism.} The process of creating and spending UTXOs is fundamental to how Bitcoin transactions are structured. The model allows the miners to verify transactions without tracking account balances. Instead, miners check whether UTXOs being spent are valid and unspent. 

\begin{circitemize}
\item \textit{Creation of UTXOs:} UTXOs are generated through Bitcoin transactions. When a sender \( A \) sends an amount \( x \) BTC to a recipient \( B \), this amount \( x \) becomes a UTXO associated with \( B \)’s address. Mathematically, if the transaction \( T \) is made, then:
\(
\mathsf{U_B} = T_{\mathsf{A \rightarrow B}}(x),
\)
where \( \mathsf{U_B} \) is the new UTXO for recipient \( B \).

\item \textit{Spending UTXOs:} To create a new transaction, a sender's wallet selects (via certain coin selection algorithms~\cite{ramezan2023survey}) one or more UTXOs as inputs to cover the transaction amount \( y \). If the total value of the selected UTXOs \( \left(\sum \mathsf{U_i}\right) \) exceeds the amount \( y \), the excess amount \( z = \sum \mathsf{U_i} - y \) is sent back to the sender as a new UTXO, often referred to as "change". This can be expressed as:
\(
\mathsf{New\ U_{change}} = T_{\mathsf{B \rightarrow B}}(z),
\)
where \( T_\mathsf{{B \rightarrow B}} \) is the transaction from the sender back to themselves.

\item \textit{Transaction inputs \& outputs:} A Bitcoin transaction consists of inputs and outputs, where inputs \( I \) are UTXOs being spent, and outputs \( O \) create new UTXOs for the recipients. If a transaction \( T \) consumes \( n \) inputs to produce \( m \) outputs, it can be represented as:
\(
T(I_1, I_2, \dots, I_n) \rightarrow (O_1, O_2, \dots, O_m),
\)
where \( I_n \) are the UTXOs used as inputs and \( O_m \) are the new UTXOs created. This cycle of consuming old UTXOs to create new ones continues with each transaction.

\end{circitemize}

\begin{figure}[!hbt]
    \centering
    \includegraphics[width=0.8\linewidth]{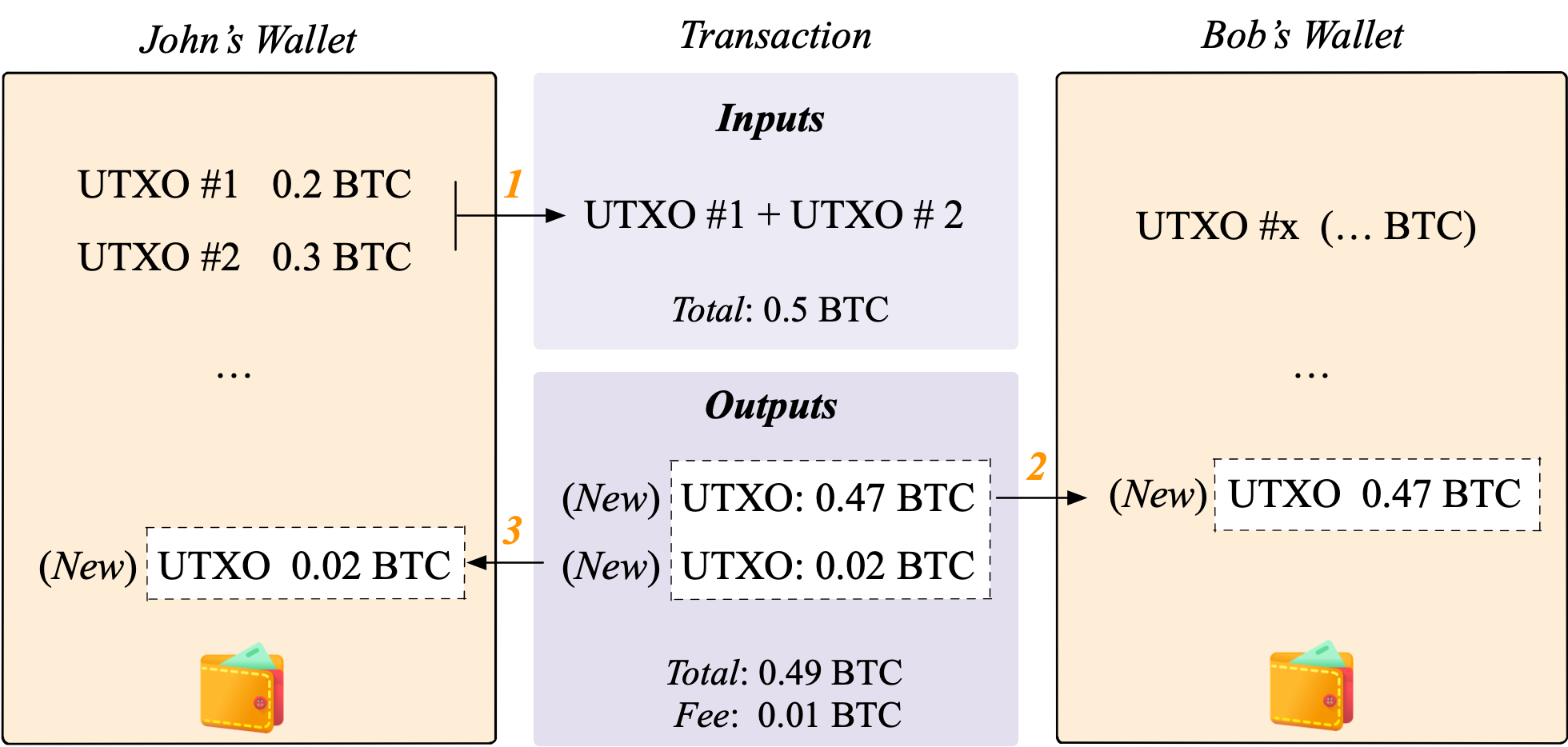}
   \caption{Bitcoin UTXO Model}
   \label{fig-utxo}
   \vspace{-0.1in}
\end{figure}

% a Bitcoin transaction consists of inputs and outputs, where inputs are UTXOs being spent, and outputs create new UTXOs for the recipients (and potentially change for the sender). This cycle of consuming old UTXOs to create new ones continues with each transaction.

% \smallskip
% \noindent\textbf{UTXO model Vs accounts model.} The UTXO and Accounts models represent two distinct approaches to transaction and balance management within blockchain networks, each with its unique advantages and trade-offs. The UTXO model, utilized by Bitcoin, tracks individual transaction outputs that have not been spent, offering enhanced privacy and security through its requirement for specific UTXO references in transactions, thereby complicating double-spending attacks. This model, however, introduces complexity in managing individual UTXOs, potentially leading to higher transaction fees and the need for meticulous UTXO management. On the other hand, the Accounts model, as seen in Ethereum, operates similarly to traditional bank accounts, directly adjusting account balances with each transaction, which simplifies the user experience but may offer less privacy and poses challenges in auditability and transparency. While the UTXO model excels in security and auditability, making it ideal for a decentralized ledger emphasizing transparency and security, the Accounts model provides a straightforward and intuitive approach, favoring ease of use and efficiency in transactions.

\subsection{Merkle Tree}
A Merkle tree, also known as a hash tree, is a cryptographic concept employed in Bitcoin to ensure data integrity. It organizes data into a binary tree structure, where each leaf node represents the hash of transaction data, and each non-leaf node is the hash of its immediate child nodes.

% \noindent\textbf{Merkle root.}
% The Merkle root is a singular hash that represents the entirety of transactions within a blockchain block, serving as a comprehensive digital fingerprint. It is derived from the process of hashing pairs of transactions together in a hierarchical manner until a single hash remains. This hash, the Merkle root, is then stored in the block's header, alongside other critical block information such as the software version, the previous block's hash, the timestamp, the difficulty target, and the nonce. The Merkle root is crucial for efficiently verifying the integrity and completeness of the transactions in a block without needing to examine each transaction individually. It enables users to confirm the inclusion of a specific transaction within a block through a minimal amount of data, making it a key component in the security of Bitcoin.

\smallskip
\noindent\textbf{Cryptographic hash functions.} SHA-256 (Secure Hash Algorithm 256-bit) \cite{sha256} is the hash function used to generate transaction hashes and block hashes in Bitcoin. It produces a 256-bit (32-byte) hash value, typically rendered as a hexadecimal number, 64 digits long. In addition to SHA-256, Bitcoin uses the RIPEMD-160~\cite{ripemd160} hash function as part of the process to create Bitcoin addresses. After a public key is generated from a private key, it is hashed using SHA-256, and then the result is hashed again using RIPEMD-160. This produces a shorter hash used, along with a network byte and a checksum, to form the Bitcoin address. 
% Cryptographic hash functions play a pivotal role in the functioning and security of Bitcoin. These functions are designed to take an input (or 'message') and return a fixed-size string of bytes, typically a digest that appears to be random. The output hash is unique to each unique input, making it virtually impossible for two different inputs to produce the same output hash (a property known as collision resistance). 

\smallskip
\noindent\textbf{Binding and hiding.} The binding property of a Merkle tree is rooted in the cryptographic hash functions used to generate the tree. Specifically, altering any single transaction \( T_i \) in the block will change its corresponding hash \( \mathsf{H}(T_i) \), thereby altering the hashes of all parent nodes up to the Merkle root \( R \). This property can be expressed as:
\(
R = \mathsf{H(H(\dots H(H}(T_1) \| \mathsf{H}(T_2)) \| \mathsf{H(H}(T_3) \| \mathsf{H}(T_4))) \dots),
\)
where \( \mathsf{H(\cdot)} \) represents the cryptographic hash function, and \( \| \) denotes concatenation. This recursive hashing ensures that the Merkle root \( R \) is a binding commitment to the exact set and order of transactions in the block.

Although the Merkle root and the associated Merkle path can be used to verify the inclusion of a transaction \( T_i \), they do not reveal the specifics of the transaction. For example, to prove the inclusion of a transaction \( T_1 \) in the block, we provide the hash of \( T_1 \) and the necessary hashes in the Merkle path, such as \( \mathsf{H}(T_2) \), \( \mathsf{H(H}(T_3) \| \mathsf{H}(T_4)) \), and \( \mathsf{H(H(H}(T_5) \| \mathsf{H}(T_6)) \| \mathsf{H(H}(T_7) \| \mathsf{H}(T_8))) \), which allows us to reconstruct the Merkle root. If the reconstructed Merkle root matches the known Merkle root \( R \), we can confirm that \( T_1 \) is part of the block without knowing all transactions.

\subsection{Segregated Witness (SegWit)}
The Bitcoin network consistently verifies a new block approximately every 10 to 15 minutes, with each block encompassing a specific number of transactions. Consequently, the size of these blocks directly influences the number of transactions that can be confirmed within each block. SegWit, representing one of the key protocol upgrades \cite{bip141}, was proposed to address the scalability issue by changing the way transaction data is structured and stored in the blocks.

\begin{lstlisting}[caption={Transaction Details}\label{listing-tx}, basicstyle=\ttfamily\scriptsize, breaklines=true]
# Input:
- Previous tx: fd9...b33a
- Index: 0
- scriptSig: 304...8932
# Output:
- Value: 5000000000
- scriptPubKey: ... OP_CHECKSIG

-------after Segwit-------

# Input:
- Previous tx: fd9...b33a
- Index: 0
- scriptSig: (empty)
# Output:
- Value: 5000000000
- scriptPubKey: ... OP_CHECKSIG
# Witness Data:
- Input 0
- ScriptSig: 304...8932

\end{lstlisting}

\lstset{
    language=,
    basicstyle=\ttfamily\scriptsize,
    keywordstyle=\color{blue}, % Color for general keywords
    morekeywords={scriptSig}, % These will be in blue
    morekeywords=[2]{scriptPubKey}, % Use a different set for another color
    keywordstyle=[2]\color{red}, % Set color for the second keyword set
}

SegWit separates signatures  from the transaction data. A SegWit transaction (above listing) consists of two main components: the original transaction structure without the signature (i.e., $\mathsf{ScriptSig}$) and a separate \textit{witness} section containing the signatures and scripts. The witness information is still transmitted and stored in the blockchain but is no longer a part of the transaction's txid calculation. The txid is now calculated without including the witness data. The change means that the txid remains constant even if the signature data is altered (i.e., fixing transaction malleability~\cite{decker2014bitcoin}). Secondly, SegWit introduces a new concept called \textit{block weight}, which is a blend of the block's size with and without the witness data. The maximum block weight is set to 4MB, while the size of the non-witness data is still capped at 1MB.

\subsection{Taproot Upgrade}

The Taproot upgrade cover three BIPs: Schnorr signatures (BIP340), Taproot (BIP341) and Tapscript (BIP342). The core idea of upgrade centers around BIP341~\cite{bip341}, which is to combine the strengths of \textit{Merkelized abstract syntax trees} (MAST) and \textit{Schnorr signatures} by committing a single Schnorr public key in the output that can represent both a single public key spend and a complex script spend. It introduces a new Taproot output (SegWit version 1) that includes a signature, a control block, and a script path. Moreover, it also specifies the rules for spending the Taproot output, which can be either a key-path spend (using a single signature) or a script-path spend (using the scripts committed to in the MAST structure).

To further illustrate the working principle of Taproot, we consider a situation where a Bitcoin address is controlled by three parties: A, B, and C. We created a Taproot output address that allows for spending conditions. The spending conditions are as follows: (i)
\textit{any two of the three parties (A, B, C) can jointly spend the funds}; and (ii) \textit{if the funds are not moved for a year, party A can unilaterally spend them (a time-lock condition).} 

Initially, the parties aggregate their individual public keys ($pk_A, pk_B, pk_C$) to form a single Schnorr aggregated public key $P$. This aggregate key, alongside the Merkle root derived from two hashed scripts: Script1 adds up the valid signatures (ensuring that at least two signatures are provided); Script2 uses $\mathsf{OP\_CheckSequenceVerify}$ to enforce the time-lock (guaranteeing that the script can only be executed after a year). Then, it checks the signature from A, and forms the Taproot output. When it's time to spend, the parties can opt for a key-path spend using $P$ for a private transaction, or a script-path spend, revealing the chosen script and its corresponding Merkle proof to fulfill specific conditions.

\section{Compendium of Projects}\label{sec-project}
\subsection{Statistical Websites and Our Results}\label{sec-web}

We source the projects from four major statistical websites that extensively catalog a large number of registered projects. 

\begin{packeditemize}

\item \textit{Rootdata}~\cite{rootdata2023} provides a comprehensive ecosystem map that categorizes various projects within the Bitcoin Layer 2 space. It offers detailed listings, including projects in different stages of development, such as testnet and upcoming. 

\item \textit{BTCEden}~\cite{btceden2023} compiles active and upcoming Bitcoin L2 projects. It presents data in a tabular format, highlighting the project's name, the type of L2 solution (e.g., ZK Rollup, Side Chain), its purpose, and the total capitalization. 

\item \textit{L2.watch}~\cite{l2watch2023} offers a detailed overview of L2 projects, including those in pre-testnet, testnet, and mainnet stages. 

\item \textit{EdgeIn.io}\footnote{To ensure a comprehensive search on EdgeIn.io, we employed a set of targeted keywords that are pivotal to our research. We utilized the search functionality with the keyword "Bitcoin" combined with various terms related to Layer 2 technologies, such as "{layer2, l2, sidechain, bitvm, client-side, rollup, state channel}". This strategic approach yielded 139 companies that align with our research focus, providing a broad spectrum of L2-related projects within the Bitcoin ecosystem.}~\cite{edgein2023} is a platform that provides business data and knowledge for the Web3 industry, connecting users with a vast network of Web3 organizations. 

\end{packeditemize}

\begin{table}[!htbp]
  \centering
  \caption{Bitcoin L2 Projects In The Wild}  \label{tab-distribution}%
  \vspace{-0.1in}
\resizebox{\linewidth}{!}{
    \begin{tabular}{c|c|c}
  % \toprule
    \textbf{Source} &  \textbf{Initial collection}  &   \textbf{Final collection}  \\
    \midrule
    
   \textit{Rootdata} & \cellcolor{blue!8} \textcolor{magenta}{\textbf{60}}: 42 (mainnet), 4 (testnet), 14 (initiated)  & {\cellcolor{yellow!20}\multirow{4}{*}{  \makecell{\cellcolor{yellow!20}\textcolor{magenta}{\textbf{40}} in toal:  \\ \cellcolor{yellow!20}19 (mainnet), 14 (testnet), \\ \cellcolor{yellow!20}7 (pre-testnet)}}} \\
    
    \textit{BTCEden} & \cellcolor{blue!8} \textcolor{magenta}{\textbf{51}}: 18 (active), 33 (initiated)  &  \\
    
    \textit{L2.watch} &\cellcolor{blue!8} \textcolor{magenta}{\textbf{85}}: 28 (pre-testnet) 17 (testnet), 39 (mainnet), 1 (n/a) &  \\
     
    \textit{EdgeIn.io} & \cellcolor{blue!8} \textcolor{magenta}{\textbf{139}} companies  & \cellcolor{yellow!20} \\
   % \bottomrule
    \end{tabular}%
    }

\end{table}%

Following the data collection method, we obtained an initial list of 335 projects (the \textit{second column} in Table~\ref{tab-distribution}). We then conducted a deduplication process, refining our dataset to a list of 126 unique Bitcoin L2 projects. This dataset was divided according to the projects’ developmental and fundraising stages (illustrated in Fig.~\ref{fig:projects}). 

In the Mainnet Stage, there were 33 projects, of which 18 received funding and 25 had complete documentation. The Testnet Stage included 25 projects, with 11 securing funding and 14 having comprehensive documentation. In the Pre-testnet Stage, 48 projects were identified, 8 of which obtained funding, while 10 had complete documentation. Lastly, the Initiated Stage consisted of 20 projects, with 2 receiving funding and only 1 having complete documentation.

% https://crypto-fundraising.info/
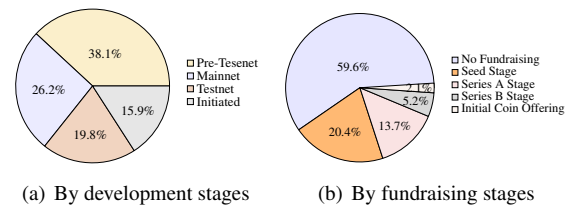
\begin{figure}[!h]
\centering
\subfigure[By development stages]{
\resizebox{0.40\linewidth}{!}{
        \begin{tikzpicture}
            \pie[
                text=legend,
                radius=3,
                color={bananamania!70, blue!8, desertsand!80, gray!20},font=\Large % Increase font size  
            ]{
                38.1/Pre-Tesenet, 
                26.2/Mainnet, 
                19.8/Testnet, 
                15.9/Initiated
            }
        \end{tikzpicture}
        }
        }
\subfigure[By fundraising stages]{
\resizebox{0.45\linewidth}{!}{
        \begin{tikzpicture}
            \pie[
                text=legend,
                radius=3,
                color={blue!10, orange!55, babypink!45, gray!30, desertsand!20},font=\Large
            ]{
                59.6/No Fundraising, 
                20.4/Seed Stage, 
                13.7/Series A Stage, 
                5.2/Series B Stage, 
                2.1/Initial Coin Offering
            }
        \end{tikzpicture}
        }
        }
\caption{Project statistical distribution}
\label{fig:projects}
\end{figure}

While applying our second round of filtering process, we finally reported 40 projects by removing projects that lacked sufficient documentation, community engagement, or a clear path to Mainnet deployment.

\begin{comment}
\begin{center}
\fbox{%
    \begin{minipage}{0.95\linewidth}
\textbf{(RQ1) Finding 1 (State-of-the-art projects):} Mainnet Stage (33 projects): Among these, 18 projects received funding, and 25 had complete documentation. Testnet Stage (25 projects): Within this group, 11 projects were funded, and 14 had comprehensive documentation. Pre-testnet Stage (48 projects): 8 projects obtained funding, and 10 had complete documentation. Initiated Stage (20 projects): 2 projects obtained funding, and 1 project had complete documentation.
    
    \end{minipage}
}
\end{center}
\end{comment}

\subsection{BitVM-based Solutions}\label{subsec-data2}

%https://static.bitlayer.org/Bitlayer-Technical-Whitepaper.pdf
\noindent\textit{(1) Bitlayer:} Building on the foundation of BitVM (also using NAND gates, see Fig.~\ref{fig:nand}), BitLayer \cite{bitlayer2024} extends the Turing completeness of smart contracts on Bitcoin by introducing enhanced architectural components. Specifically, BitLayer incorporates a more sophisticated state management system that allows for the execution of complex logic without altering the core Bitcoin protocol. This is achieved through an implementation of Taproot and Schnorr signatures, which enables more efficient execution of multi-party contracts and off-chain computations.

% https://whitepaper.zbyte.io/
\noindent\textit{(2) ZKByte} \cite{zbyte2024} builds upon the principles of BitVM by integrating ZKPs into the Bitcoin L2. It introduces a ZKP-enabled validator that verifies the state transitions of the blockchain without exposing sensitive data. This validator leverages the UTXO model to track and manage states across the main network and L2 network. Additionally, ZKByte incorporates a trusted oracle system, which validates the correctness of UTXO inputs and outputs and ensures that script execution adheres to the L2 protocol.

% https://bitstakeplatform.com/whitepaper
\noindent\textit{(3) Bitstake} \cite{bitstake2024} leverages the BitVM for implementing PoS protocols. Embedding Permissioned Optimistic Bridge, Bitstake enables dynamic participation through staking, where stakeholders can actively engage in verifying and managing transactions over the bridge. Bitstake also enhances BitVM's security model by implementing a challengeable operator environment. In this setup, validators can dispute potentially fraudulent transactions within predefined challenge periods.

% https://docs.citrea.xyz/technical-specs/introduction
\noindent\textit{(4) Citrea} \cite{citrea2024} extends the functionality of BitVM by implementing a ZK Rollup architecture on the Bitcoin network. Technically, Citrea has introduced the concept of execution slices, where thousands of transactions are batch-processed and validated on the Bitcoin mainchain using a compact ZKP. Additionally, Citrea introduces a bi-directional anchoring mechanism that allows seamless interaction between Bitcoin as a settlement layer and its L2 environment, where more complex smart contracts can be executed.

% https://docs.satoshivm.io/
%https://github.com/SatoshiVM/whitepaper/blob/main/SatoshiVM%20Overview.pdf
\noindent\textit{(5) SataoshiVM} \cite{satoshivm2024} advances the BitVM framework by adopting the Bristol format for its logical gate circuit structure, which simplifies the circuit’s complexity. While BitVM requires multiple rounds of interaction to identify specific data discrepancies, SatoshiVM introduces a more streamlined approach with a single round of non-interactive on-chain verification. This method allows validators to observe commitment values and verify the corresponding off-chain data without continuous interaction. If any discrepancies are detected, validators can immediately identify these errors and penalize the proposer by transferring their Bitcoin UTXO.

\begin{figure}[!hbt]
    \centering
    \includegraphics[width=0.9\linewidth]{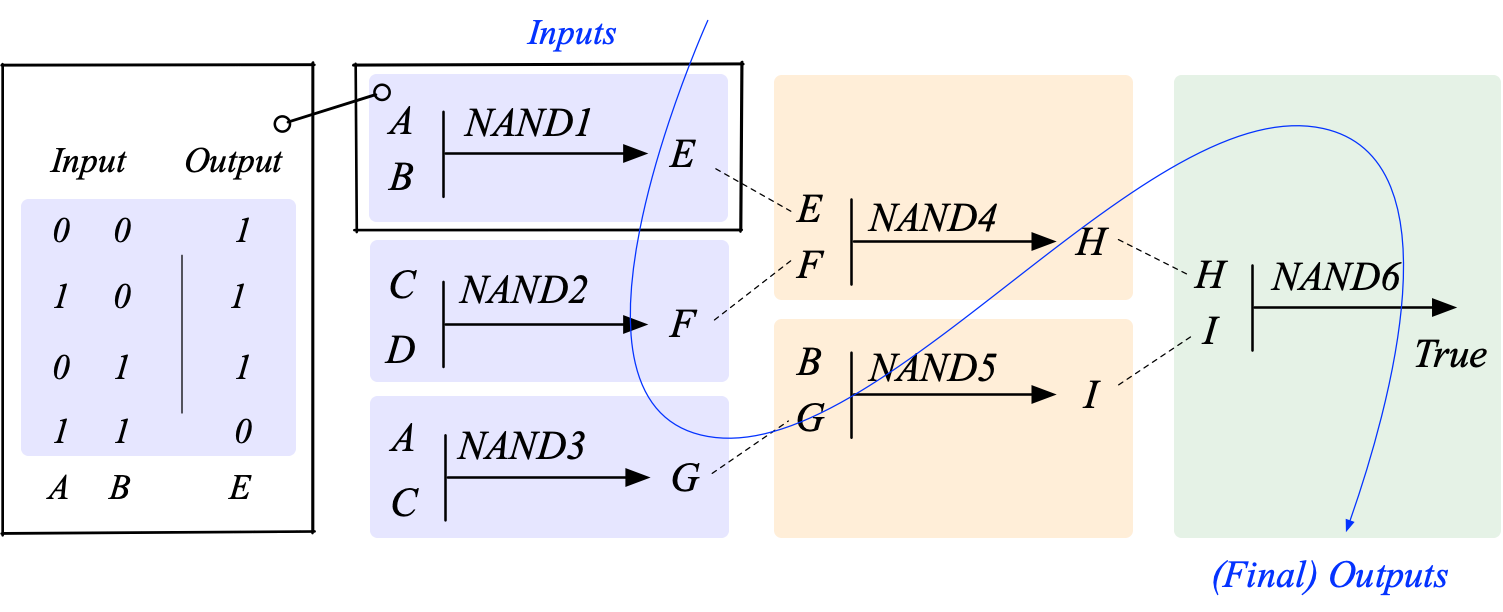}
        \vspace{-0.1in}
    \caption{Binary Circuit Commitment in BitVM}
    \label{fig:nand}
    \vspace{-0.2in}
\end{figure}

\subsection{Rollups}

\noindent{$\bullet$ \textit{Zero-knowledge rollups}
\smallskip

% https://bitvm.gitbook.io/doc/btc-layer2/rollup/b-network
% https://www.bsquared.network/
\noindent\textit{(1) B² Network} \cite{bsquared2024} has implemented a hybrid model resembling a blend of zk and optimistic features. This network architecture is structured into two distinct layers: the rollup layer and the data availability layer. The Rollup layer employs the zkEVM system, responsible for sorting, packaging transactions, and generating zero-knowledge proofs. Meanwhile, the DA layer offers a decentralized storage solution, where off-chain storage nodes conduct zk validations and inscribe Rollup data into Bitcoin’s Ordinals script. Given that Bitcoin L1 lacks native smart contract validation, B² Network has introduced a fraud-proof challenge mechanism akin to Optimistic Rollups. This mechanism allows challengers to challenge the commitment to zkp verification within a specified period. If a challenge is upheld, the Rollup reverts the transactions, and the challenger is rewarded with assets locked by the node. Conversely, if no challenge occurs or the challenge fails, the Rollup receives final confirmation.

% https://dmaster-1.gitbook.io/bisonlabs/bison-network-overview
\noindent\textit{(2) Bison} \cite{bisonlabs2024} employs the ZK-STARK rollup on Bitcoin, conceptualized to operate as a sovereign rollup. Bison introduces a Bison OS system composed of the sequencer and the prover. The Sequencer is responsible for collecting and ordering user transactions, while the Prover leverages STARK technology to generate ZKPs. A key improvement in Bison’s approach is its support for client-side verification, where users can directly download and verify ZK proofs independently.

%https://tunachain.io/
%https://tunachain.gitbook.io/tunachain
\noindent\textit{(3) Tuna Chain} \cite{tunachain2024} leverages zk-Rollups to bundle multiple transactions into single batches that are validated on the Bitcoin main network. It employs ZK proofs for off-chain computations and uses Taproot and Bitcoin Script to verify contracts on-chain without altering Bitcoin's consensus rules. Additionally, it uses native BTC as gas for EVM transactions, integrating Bitcoin's value anchoring with Ethereum's programmability to enable application development.

% https://bL2.live/
\noindent\textit{(4) BL2} \cite{bl22024} builds on zk rollup technology to enhance Bitcoin’s scalability, implementing a dual-layer architecture. The Rollup Layer processes transactions off-chain using a zkEVM system, which aggregates these transactions and generates ZPKs for validation. Unlike traditional rollups, BL2 introduces the Celestia Data Availability Layer, where batch data and zk proofs are stored across decentralized nodes. This setup allows zk-proof verifiers and Bitcoin committers within the BL2 network to validate transactions before committing them to the Bitcoin blockchain.

%https://wiki.sovryn.com/en/home
\noindent\textit{(5) Sovryn:} \cite{sovryn2024} introduces an architecture that combines BOB (Bitcoin Optimistic Bridge) with zk-rollup technology through BitcoinOS. BOB enables the seamless interaction between Bitcoin and Sovryn’s sidechain, operating on Rootstock by locking BTC on the Bitcoin main chain and releasing RBTC on the sidechain. Sovryn leverages zk-rollups to aggregate and process transactions off-chain. These transactions are then compressed into zk-proofs and periodically settled on the Bitcoin mainnet.

% https://docs.merlinchain.io/merlin-docs
\noindent\textit{(6) Melin Chain:} Building on the zk-rollup foundation, Merlin Chain \cite{merlinchain2024} compresses multiple transaction proofs into compact batches that are submitted to the Bitcoin network as a single transaction. Beyond the standard zk-rollup process, Merlin Chain integrates a decentralized oracle network, which serves to validate off-chain data and ensure its accuracy before inclusion in the rollup.

% https://lumibit.xyz/
\noindent\textit{(7) LumiBit} \cite{lumibit2024} utilizes the ZK scheme of Halo2 to explore a L2 scaling solution for Bitcoin. As a ZK-rollup, LumiBit leverages Bitcoin as its Data Availability layer. ZK proofs are inscribed on the Bitcoin network and verified by open-source clients, allowing users to access and verify the latest off-chain states. Unlike other ZK-rollups, LumiBit's prime feature lies in the distinctive use of Halo2 scheme and KZG commitments to reduce verification costs. Additionally, LumiBit is building a Type 2 ZK-EVM with its universal circuit design, ensuring compatibility with the Ethereum ecosystem. This PoS-based security model requires participants to stake BIOP tokens, which incentivizes proper block generation and validation.

\smallskip
\noindent{$\bullet$  \textit{Optimistic rollups}
\smallskip

% https://zhuanlan.zhihu.com/p/644288767
\noindent\textit{(1) Biop} \cite{bioplabs2024} integrates the Optimistic Rollup protocol with a PoS consensus algorithm, creating a system known as BiopOS. In this setup, Sequencers are tasked with aggregating and packaging transactions into blocks, which are then submitted to the Bitcoin mainnet for verification. Validators are responsible for monitoring the Sequencers’ actions and can submit fault proofs if discrepancies are detected.

%https://docs.rollux.com/docs/protocol/protocol-2.0/#verification
\noindent\textit{(2) Rollux} \cite{rollux2024} is a L2 scaling protocol built on Bitcoin and Syscoin, leveraging Optimism Bedrock to provide EVM equivalence. It uses optimistic rollups, batching transactions off-chain, and then submitting them to Syscoin for finality. Key components include the Canonical Transaction Chain (CTC) for transaction ordering, the State Commitment Chain (SCC) for state root proposals, and Bridge contracts for L1-L2 communication. The current system relies on a centralized sequencer but plans to decentralize this role.

%https://docs.gobob.xyz/docs/learn/bob-stack/stack-overview
\noindent\textit{(3) BOB} \cite{gobob2024} employs EVM to facilitate smart contract execution, leveraging Optimistic Rollup to batch process transactions off-chain before submitting them to Ethereum for final settlement. BOB’s integration of Bitcoin’s Rust libraries via the RISC Zero zkVM allows for off-chain execution of Rust programs. These off-chain computations are then verified through zk-proofs, which are validated within EVM contracts.

% https://www.docdroid.net/Xqk17l9/be-L2-whitepaper-english-pdf#page=2
\noindent\textit{(4) BeL2} \cite{bel22024} integrates Bitcoin with Elastos SmartWeb technology to enhance scalability and programmability. Utilizing optimistic rollup technology, BeL2 batches transactions off-chain and periodically submits them to the Bitcoin mainnet, leveraging Elastos’ merged mining for security. Additionally, a relayer mechanism is used for fraud prevention, staking deposits as collateral.

%https://hacash.org/whitepaper.pdf
\noindent\textit{(5) Hacash.com} \cite{hacash2024} is designed to support real-time settlement using a multi-layered architecture for Bitcoin. It includes three layers: Layer 1 features ASIC-resistant mining and support for multi-signature transactions; Layer 2 operates as a channel chain settlement network enabling high transaction throughput by allowing multiple off-chain transactions with only final balances submitted on-chain; and Layer 3 serves as an application ecosystem scaling layer, supporting rollup technology and multi-chain protocols.

%https://rollkit.dev/learn/stack
\noindent\textit{(6) Rollkit}  \cite{rollkit2024} is a modular framework for sovereign rollups that utilizes Bitcoin for data availability. Rollup sequencer nodes aggregate transactions into blocks and post them to a data availability layer like Celestia for ordering and finalization. Full nodes execute and verify these blocks, and propagate fraud proofs in optimistic rollup setups, while light clients verify proofs and authenticate state queries. Rollkit integrates with Cosmos SDK via an enhanced Application BlockChain Interface and utilizes Taproot transactions on Bitcoin for DA.

\subsection{Sidechains}

\noindent{$\bullet$ \textit{Federated consensus sidechains}
% https://docs.liquid.net/docs/technical-overview
\smallskip

\noindent\textit{(1) Liquid Network} \cite{liquid2024} introduces a Federated Sidechain Consensus (FSC) model, which uses a consortium of blocksigners and watchmen instead of Bitcoin’s traditional PoW. The network employs a two-way peg system, allowing the transfer of Bitcoin to and from the sidechain, where Bitcoin is converted into Liquid Bitcoin. Additionally, it uses Confidential Transactions to encrypt transaction details and a one-minute block time to reduce confirmation delays.

\smallskip
\noindent{$\bullet$ \textit{Merged-Mined sidechains}
\smallskip

% https://rootstock.io/rsk-white-paper-updated.pdf
% https://eprint.iacr.org/2022/684.pdf
\noindent\textit{(1) Rootstock} (RSK) \cite{rootstock2024} is designed to bring Ethereum-compatible smart contracts to the Bitcoin ecosystem. It uses RBTC, a token pegged 1:1 to BTC, generated through merged mining. RSK employs a two-way peg system facilitated by Powpeg, an autonomous multi-signature management bridge that involves reputable companies like Xapo, Bitpay, and BitGo. This bridge allows seamless transfer of BTC between the main chain and the RSK sidechain.

\smallskip
\noindent{$\bullet$ \textit{PoX-based sidechains}
\smallskip

%https://files.mapprotocol.io/pdf/mapprotocol_whitepaper_en.pdf
\noindent\textit{(1) MAP Protocol} \cite{mapprotocol2024} operates similarly to PoX-based sidechains as it integrates a mechanism for verifying cross-chain transactions. However, instead of relying on traditional PoX, it utilizes a combination of ZK-proof technology and light clients to achieve decentralized interoperability across multiple blockchain networks.

%https://gaia.blockstack.org/hub/1AxyPunHHAHiEffXWESKfbvmBpGQv138Fp/stacks.pdf
\noindent\textit{(2) Stacks} \cite{stacks2024} extends Bitcoin’s functionality by enabling Proof of Transfer while leveraging Bitcoin’s security. It ties the security of Stacks directly to Bitcoin by anchoring each Stack block to a Bitcoin block. Miners are rewarded in Stacks’ native token for their BTC commitments.

% https://docs.mintlayer.org/
\noindent\textit{(3) Mintlayer} \cite{mintlayer2024} utilizes a PoS consensus mechanism, where participants stake ML tokens to become validators and block producers. Each Mintlayer block references a Bitcoin block, using the Bitcoin block hashes as a source of randomness to select block producers. The network also employs a Dynamic Slot Allotment system to randomly select block signers who validate and sign new blocks.

% https://docs.bouncebit.io/
\noindent\textit{(4) BounceBit} \cite{bouncebit2024}’s PoS-based sidechain mechanism involves validators staking both BTC and BounceBit’s native token BB. This staking process secures the network and enables validators to produce new blocks. Users can convert BTC to BBTC and stake it on the BounceBit platform to earn rewards. The system integrates both CeFi and DeFi elements, allowing for funding rate arbitrage and the creation of on-chain certificates for restaking and mining. 

% https://docs.libre.org/libre-docs/readme/libre-level-1-beginner
\noindent\textit{(5) Libre} \cite{libre2024} operates using a Delegated proof-of-stake (DPoS) consensus mechanism. It integrates with Bitcoin via non-custodial wrapping (pegged assets) and the Lightning Network, providing seamless on-and-off ramps for Bitcoin users. This design allows Libre to process over 4,000 transactions per second with zero transaction fees.

% https://bitrexe.gitbook.io/docs
\noindent\textit{(6) BitReXe} \cite{bitrexe2024} operates as a PoS-based sidechain. It uses a unique approach by incorporating the PREDA programming model to scale out general smart contracts within multi-VM blockchain systems. Validators on the BitReXe network stake RXBTC (a token pegged to Bitcoin) to validate transactions and produce new blocks.

%https://www.bevm.io/
\noindent\textit{(7) BEVM} \cite{bevm2024} is an EVM-compatible sidechain solution that integrates the Taproot consensus mechanism. The Taproot consensus is a cornerstone of BEVM's architecture, combining Musig2 for secure multi-signature schemes, a BFT PoS network of Bitcoin SPVs for decentralized verification, and the Signal Protocol for encrypted, secure communication between nodes. Additionally, by using native BTC as gas, BEVM lowers barriers for Bitcoin holders to participate in L2 transactions, while its full EVM compatibility invites developers to leverage existing tools and languages.

\subsection{Client Side + UTXO}

% https://docs.rgb.info/distributed-computing-concepts/client-side-validation
\noindent\textit{(1) RGB} \cite{rgb2024} leverages CSV and the UTXO model to create a scalable method for issuing and managing assets on the Bitcoin network. It initiates with an asset issuer creating a new asset and generating a unique one-time seal along with a cryptographic commitment on their client. This commitment is then embedded within a transaction's UTXO, anchoring the asset to the blockchain. Recipients verify the asset's legitimacy by examining the commitment against the one-time seal. Upon transfer, the old seal is invalidated, and a new set of seal, commitment, and transaction data are recorded on the Bitcoin network, ensuring an immutable transaction history.

%https://talk.nervos.org/t/rgb-protocol-light-paper-translation/7790
\noindent\textit{(2) RGB++} \cite{rgb2024lightpaper} combines the RGB protocol with UTXO-based public blockchains such as CKB, Cardano, and Fuel, utilizing these platforms as verification and data storage layers for RGB assets. This approach shifts data validation tasks from the user to the third-party platforms, replacing client-side validation with decentralized platform verification, provided users trust these blockchains. The protocol achieves compatibility with the original RGB through the concept of “isomorphic binding” where the extended UTXOs on chains like CKB or Cardano act as containers for RGB asset data, directly reflecting asset parameters on the blockchain. For instance, if Alice wants to transfer 30 out of her 100 RGB tokens to Bob, she generates a commitment and spends her UTXO on Bitcoin, while consuming the corresponding UTXO container on CKB, creating new containers for both Alice and Bob. This transaction is validated by the consensus mechanisms of CKB/Cardano, without Bob’s direct involvement. While this setup improves global verifiability and DeFi applications, it sacrifices privacy, requiring a trade-off between security and usability. If users prioritize privacy, they can revert to the traditional RGB mode. RGB++ also assumes trust in the reliability of the CKB/Cardano networks. Users can operate their RGB assets on these UTXO chains directly using their Bitcoin accounts, linking UTXO conditions to Bitcoin addresses, and can use “transaction folding” to reduce costs by batching multiple transfers into a single commitment.

%https://bihelix.net/
\noindent\textit{(3) BiHelix} \cite{bihelix2024} functions as a Bitcoin-native infrastructure that amplifies the capabilities of the Bitcoin network through several key mechanisms. It integrates the RGB protocol, which facilitates private and efficient transactions by allowing parties to reach consensus without the need for on-chain contract recording, thus keeping individual transaction histories and status data off-chain. The SLR (Security-Lighting-RGB) Protocol enhances interoperability by building upon the migration of Core Lightning's functionality to rust-lightning, thereby reinforcing the synergy between the RGB protocol and the Lightning Network. Additionally, BiHelix employs a set of socially consensus-driven nodes on the sidechain, using native economic incentives to draw in node service providers.

%https://bitlightlabs.com/
\noindent\textit{(4) Bitlight Labs} \cite{bitlightlabs2024} develops infrastructure based on the RGB protocol and deploys a suite of applications on Lightning Network, including Bitswap (asset exchange) and the Bitlight (wallet). The project is featured for managing smart contracts with privacy and security through client-side validation, where data is controlled by ``state owner'' rather than being publicly accessible. It operates on Bitcoin transactions, potentially using the Lightning Network's off-chain capabilities, and allows for scripting with Blockstream's formally verified Simplicity language (claimed to be Turing complete).

% %https://bitmask.app/
% \noindent\textit{(5) BitMask:} BitMask Wallet is a  browser extension acting as a versatile tool for Bitcoin users, offering a gateway to decentralized applications and financial sovereignty through the innovative RGB protocol. By leveraging Bitcoin's Taproot upgrade and integrating the Lightning Network, BitMask ensures efficient, low-cost transactions, and full control over assets. 

%https://docs.xrgb.xyz/
% \noindent\textit{(6) XRGB:}
% \noindent\textit{(7) HueHub:}
% \noindent\textit{(8) Nervape:}

% https://eprint.iacr.org/2022/1290.pdf
\noindent\textit{(5) Bool Network} \cite{yin2022boolnetwork} is a Bitcoin Verification Layer that integrates client-side verification within its L2 solution. The protocol designs a client-side verification process where transaction validation is performed collectively by distributed key management systems without relying on a central authority. Its Dynamic Hidden Committees (DHC) act as security guardians, ensuring cross-chain message integrity, while its public blockchain, Bool Chain, serves as an EVM-compatible ledger for recording committee activities and supporting future application development. Bool Network's client-side verification is secured by Ring VRF~\cite{burdges2023ring} (where a user generates a unique pseudorandom output and a proof of its correctness using their private key and a set of public keys, i.e., the ``ring'').

%https://mercurylayer.com/
\noindent\textit{(6) Mercury Layer} \cite{mercurylayer2024} represents an implementation of client-side verification within the Bitcoin ecosystem, focusing on statechain technology to enable efficient asset transfers. At its core, Mercury Layer operates through a system of blind co-signing and key updates, allowing for the instantaneous and cost-free transfer of Bitcoin UTXOs. The protocol emphasizes client-side verification by leveraging client software to perform all transaction operations and statechain validations, ensuring that the server remains unaware of transaction details and UTXO identities. This approach requires clients to actively verify partial signatures, manage backup transactions, and oversee key updates, thereby upholding the security of the statechain without relying on server-side control.

\subsection{State Channel (Fig.\ref{fig:statechannel})} 

%https://lightning.network/lightning-network-paper.pdf
\noindent\textit{(1) Lightning Network} \cite{poon2016lightningnetwork} is a L2 scaling solution for Bitcoin that addresses the blockchain's limitations in transaction throughput and confirmation times by moving the majority of transactions off-chain. It operates through a network of payment channels, utilizing two primary smart contract mechanisms: the Recoverable Sequence Maturity Contract (RSMC), which facilitates secure allocation updates and dispute resolution, and HTLC, which enables the creation of timed, conditional transactions across multiple parties.

        \vspace{-0.1in}
\begin{figure}[!hbt]
    \centering
    \includegraphics[width=0.9\linewidth]{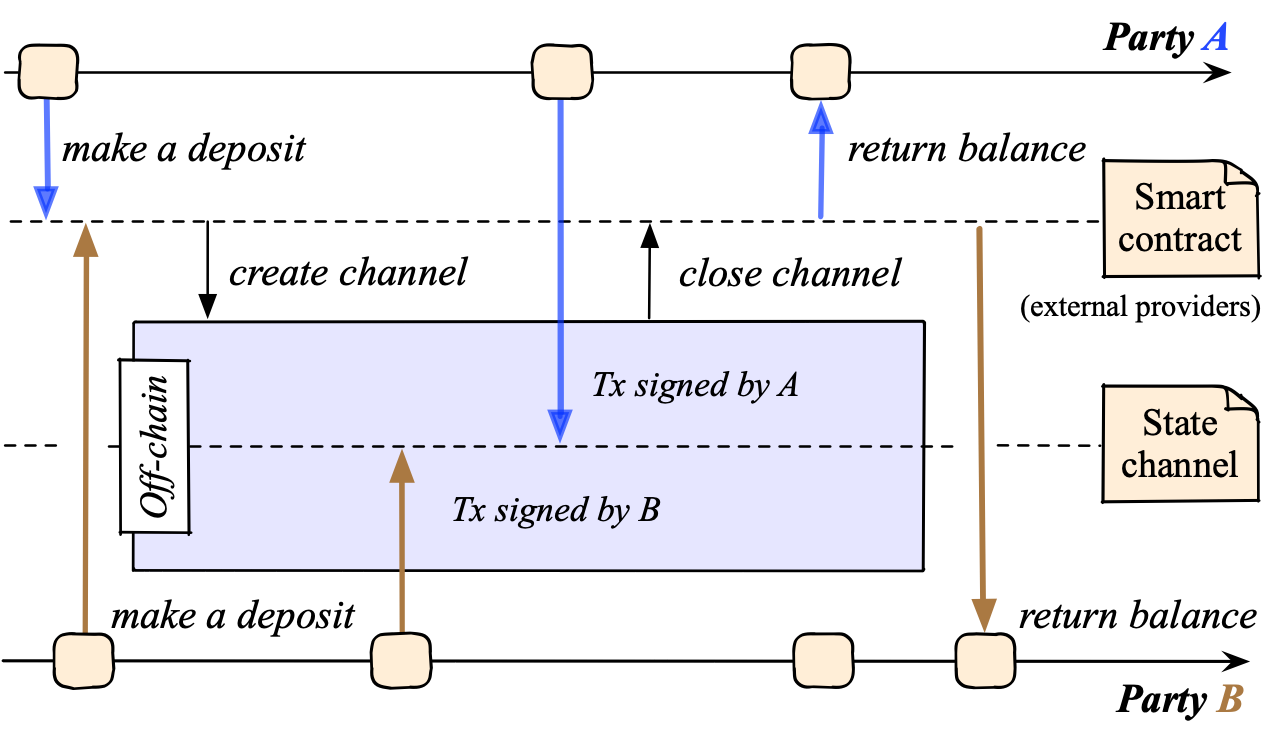}
            \vspace{-0.1in}
    \caption{State channel}
    \label{fig:statechannel}
        \vspace{-0.15in}
\end{figure}

%https://omnilaboratory.github.io/obd/#/?id=omnibolt-facilitates-smart-assets-lightning-transactions
\noindent\textit{(2) OmniBOLT} \cite{omnibolt2024} extends Lightning Network's capabilities to facilitate smart asset transactions. Built on the BTC/OmniLayer network, OmniBOLT enables the circulation of OmniLayer assets through lightning channels, providing instant payments, cross-channel atomic swaps, and decentralized exchange functionalities. Its protocol suite, written in Golang, includes implementations for basic HTLC payments, atomic swaps of multiple currencies, and an automatic market maker with a liquidity pool for DEX.

%https://doc.nostrassets.com/
\noindent\textit{(3) Lnfi Network} (formerly NostrAssets) \cite{nostrassets2024} integrates Taproot assets and Satoshis, allowing users to send and receive these assets using Nostr’s public and private keys. Leveraging Lightning Network for settlement and security, NAP ensures high transaction throughput, minimal latency, and low fees through off-chain transactions. This integration enables near-instantaneous asset transfers, crucial for applications like micro-payments and real-time trading.

%https://ark-protocol.org/
\noindent\textit{(4) Ark Protocol} \cite{arkprotocol2024} is designed to address the scalability limitations of Lightning Network. It allows users to send and receive payments without the need for incoming liquidity. Ark does not require the recipient to be online or interactive for the transaction to be completed. The protocol's lifting mechanism allows users to convert on-chain UTXOs into virtual UTXOs without reliance on third parties, and the redemption function allows to move funds back to the base layer.

\subsection{More}
\label{more}

% https://tectum.gitbook.io/tectum-blockchain-and-softnote-whitepaper/v.-softnote-tectum-components
\noindent\textit{(1) Tectum} \cite{tectum2024fastest} introduces SoftNotes, a L2 solution that enhances Bitcoin's scalability. SoftNotes are digital banknotes representing Bitcoin wallet ownership, facilitated by Tectum blockchain. Technically, SoftNotes operates on a bearer instrument model, allowing for off-chain transactions that are untraceable and anonymous, with cryptographic finality. Tectum's also includes a hybrid mode for transactions, a smart contract-based security model for handling BTC private keys, and the capability to represent and transfer NFTs.

%https://drive.google.com/file/d/1HEcLB8fjW7Sabn_AJap9syEhWDCJs1-R/view
%https://docs.nubit.org/overview/architecture
\noindent\textit{(2) Nubit} \cite{nubit2024}, a Bitcoin data availability layer, introduces DAS solution to address the scalability challenge by allowing nodes to verify block data availability through random sampling rather than downloading entire blocks. DAS employs two protocols: the sampling protocol for verifying data availability through random chunk requests, and the decoding protocol for reconstructing the full block from verified chunks. The system involves three types of participants: validators who run the sampling protocol and sign block headers upon successful verification; full storage nodes that maintain the entire block and respond to chunk requests; and light clients who initiate sampling to ensure data availability.

%https://drive.google.com/file/d/1wKeOc59OSpUuK_YhRgkZbipWloTdx09b/view
%https://www.hyperagi.network/
\noindent\textit{(3) HyperAGI} \cite{hyperagi2024}, a Bitcoin L2 solution, focuses on decentralized, compute-intensive applications, particularly for Artificial General Intelligence (AGI) and metaverse integration. HyperAGI claims to support complex, high-demand tasks such as 3D rendering and AI computations. It achieves this by integrating a 3D pipeline-based Pinpoint protocol, which achieves computation verification in a decentralized manner. Additionally, HyperAGI employs a HyperTrust consensus algorithm that operates with the Graph Virtual Machine.

\section{Bitcoin Improvement Proposals (BIPs)}\label{sec-bip}

We reviewed all existing BIPs (cf. Table~\ref{tab:bip}) and captured the major ones (focusing on the \textit{final} stage, as of Aug. 2024) for Bitcoin's basic design and development lineage.

This work references the following BIPs: 174/370/371 (PSBT), 141/144 (SegWit), and 340-343 (Taproot).

\begin{table*}[!hbt]
\caption{BIPs At A Glance (Aug. 2024, only collecting BIPs in the \textit{final} category)~\cite{bips}}\label{tab:bip}
\renewcommand\arraystretch{1.1}
\vspace{-0.7em}
\begin{center}
\resizebox{\linewidth}{!}{
\begin{tabular}{c|ccc|cc} 
\toprule 
    \multicolumn{1}{c}{} & \textbf{BIP-} &  \textbf{Year} & \textbf{Type} & \textbf{Title} &         \textbf{Key notes}  \\
        \midrule

        \multirow{24}{*}{\rotatebox{90}{\textbf{Application}}} & \hlhref{https://github.com/bitcoin/bips/blob/master/bip-0009.mediawiki}{9} & 2015 & info. & Version bits with timeout and delay   &  Modifies the 'version' field to track multiple soft forks.  \\

        & \hlhref{https://github.com/bitcoin/bips/blob/master/bip-0011.mediawiki}{11} & 2011 & stan. & M-of-N Standard Transactions &   Proposes M-of-N-signatures (\textcolor{teal}{$\mathsf{scriptPubKey}$}).   \\

        & \hlhref{https://github.com/bitcoin/bips/blob/master/bip-0013.mediawiki}{13} & 2011 &  stan. & Address Format for pay-to-script-hash  & Introduces P2SH address format. \\
        
        & \hlhref{https://github.com/bitcoin/bips/blob/master/bip-0014.mediawiki}{14} &  2011 &  stan. & Protocol Version and User Agent   & Defines user agent and protocol version fields. \\
        
        & \hlhref{https://github.com/bitcoin/bips/blob/master/bip-0016.mediawiki}{16}  & 2012 &  stan. & Pay to Script Hash   & Standardizes pay-to-script-hash (P2SH) transactions. \\
        
        & \hlhref{https://github.com/bitcoin/bips/blob/master/bip-0021.mediawiki}{21} & 2012 & stan. & URI Scheme & Specifies URI format for Bitcoin payments. \\
        
        & \hlhref{https://github.com/bitcoin/bips/blob/master/bip-0032.mediawiki}{32} & 2012 & info. & Hierarchical Deterministic Wallets & Introduces HD wallets with tree structure. \\ 
        
        & \hlhref{https://github.com/bitcoin/bips/blob/master/bip-0043.mediawiki}{43} & 2014  & stan. & Purpose Field for Deterministic Wallets & Adds purpose field to HD wallets. \\ 
        
        & \hlhref{https://github.com/bitcoin/bips/blob/master/bip-0044.mediawiki}{44} & 2014 & stan. & Multi-Account Hierarchy for Deterministic Wallets & Defines multi-account hierarchy in HD wallets. \\ 
        
        & \hlhref{https://github.com/bitcoin/bips/blob/master/bip-0047.mediawiki}{47} & 2021 &  info. & Reusable Payment Codes for Hierarchical Deterministic Wallets & Enables reusable payment codes. \\ 
        
        & \hlhref{https://github.com/bitcoin/bips/blob/master/bip-0049.mediawiki}{49} & 2016  & stan. & Derivation scheme for P2WPKH-nested-in-P2SH based accounts & P2WPKH nested in P2SH account derivation. \\ 
        
        & \hlhref{https://github.com/bitcoin/bips/blob/master/bip-0070.mediawiki}{70} & 2013  & stan. & Payment Protocol & Introduces a protocol for payment requests. \\ 
        
        & \hlhref{https://github.com/bitcoin/bips/blob/master/bip-0071.mediawiki}{71} & 2013  & stan. & Payment Protocol MIME types & Specifies MIME types for payment protocol. \\ 
        
        & \hlhref{https://github.com/bitcoin/bips/blob/master/bip-0072.mediawiki}{72} & 2013 & stan. & bitcoin: uri extensions for Payment Protocol & Extends URI scheme for payment protocol. \\ 
        
        & \hlhref{https://github.com/bitcoin/bips/blob/master/bip-0073.mediawiki}{73} & 2013 & stan. & Use "Accept" header for response type negotiation with Payment Request URLs & Uses "Accept" header for payment request responses. \\ 
        
        & \hlhref{https://github.com/bitcoin/bips/blob/master/bip-0075.mediawiki}{75} & 2015 & stan. & Out of Band Address Exchange using Payment Protocol Encryption & Secure address exchange out of band. \\ 
        
        & \hlhref{https://github.com/bitcoin/bips/blob/master/bip-0084.mediawiki}{84} & 2017 & stan. & Derivation scheme for P2WPKH based accounts & P2WPKH account derivation scheme. \\ 
        
        & \hlhref{https://github.com/bitcoin/bips/blob/master/bip-0086.mediawiki}{86} & 2021 & stan. & Key Derivation for Single Key P2TR Outputs & Defines key derivation for P2TR outputs. \\ 
        
        & \hlhref{https://github.com/bitcoin/bips/blob/master/bip-0137.mediawiki}{137} & 2019 & stan. & Signatures of Messages using Private Keys & Defines signing messages with private keys. \\
        
        & \hlhref{https://github.com/bitcoin/bips/blob/master/bip-0173.mediawiki}{173} & 2017 & info. & Base32 address format for native v0-16 witness outputs & Introduces SegWit native address format. \\ 
       
        & \cellcolor{cambridgeblue!40}  \hlhref{https://github.com/bitcoin/bips/blob/master/bip-0174.mediawiki}{174} & \cellcolor{cambridgeblue!40}  2017  & \cellcolor{cambridgeblue!40}  stan. & Partially Signed Bitcoin Transaction Format & Standardizes partially signed transactions. \\    
        
        & \hlhref{https://github.com/bitcoin/bips/blob/master/bip-00350.mediawiki}{350} & 2020 & stan. & Bech32m format for v1+ witness addresses & Defines Bech32m address format for Taproot. \\   
        
        & \cellcolor{cambridgeblue!40} \hlhref{https://github.com/bitcoin/bips/blob/master/bip-0370.mediawiki}{370} & \cellcolor{cambridgeblue!40} 2021 & \cellcolor{cambridgeblue!40} stan. & PSBT Version 2 & Updates PSBT format to version 2. \\   
        
        & \cellcolor{cambridgeblue!40} \hlhref{https://github.com/bitcoin/bips/blob/master/bip-0371.mediawiki}{371} &\cellcolor{cambridgeblue!40} 2021  & \cellcolor{cambridgeblue!40} stan. & Taproot Fields for PSBT  & Adds Taproot support to PSBT. \\ 
       
        & \hlhref{https://github.com/bitcoin/bips/blob/master/bip-0380.mediawiki}{380}+ & 2021 & info.  & Output Script Descriptors - Related  & \makecell{General (\hlhref{https://github.com/bitcoin/bips/blob/master/bip-0380.mediawiki}{380}), Non-Segwit (\hlhref{https://github.com/bitcoin/bips/blob/master/bip-0381.mediawiki}{381}), Segwit (\hlhref{https://github.com/bitcoin/bips/blob/master/bip-00380.mediawiki}{382}), Multisig (\hlhref{https://github.com/bitcoin/bips/blob/master/bip-0383.mediawiki}{383}), \\ combo() (\hlhref{https://github.com/bitcoin/bips/blob/master/bip-00380.mediawiki}{384}),
        raw()/addr() (\hlhref{https://github.com/bitcoin/bips/blob/master/bip-0385.mediawiki}{385}), tr() (\hlhref{https://github.com/bitcoin/bips/blob/master/bip-0386.mediawiki}{386}), Tapscript Multisig (\hlhref{https://github.com/bitcoin/bips/blob/master/bip-00380.mediawiki}{387}) } \\ 
        
        \cmidrule{1-5}

        \multirow{3}{*}{\rotatebox{90}{\textbf{API/RPC}}}   & \hlhref{https://github.com/bitcoin/bips/blob/master/bip-0022.mediawiki}{22} & 2012 & stan. & getblocktemplate - Fundamentals & Introduces a new way to create block templates. \\ 
        
        & \hlhref{https://github.com/bitcoin/bips/blob/master/bip-0023.mediawiki}{23} & 2012 & stan. & getblocktemplate - Pooled Mining & Adds support for pooled mining using getblocktemplate. \\ 
        
        & \hlhref{https://github.com/bitcoin/bips/blob/master/bip-00145.mediawiki}{145} & 2016 & stan. & getblocktemplate Updates for Segregated Witness & Updates getblocktemplate for SegWit compatibility. \\ 

        \cmidrule{1-5}

       \multirow{10}{*}{\rotatebox{90}{\textbf{Peer Services}}}   
       & \hlhref{https://github.com/bitcoin/bips/blob/master/bip-0031.mediawiki}{31} & 2012 & stan. & Pong message & Adds a ping/pong mechanism for P2P connections. \\ 

        & \hlhref{https://github.com/bitcoin/bips/blob/master/bip-0035.mediawiki}{35} & 2012 & stan. & Mempool Message & Allows peers to request mempool contents. \\ 
        
        & \hlhref{https://github.com/bitcoin/bips/blob/master/bip-0037.mediawiki}{37} & 2012 & stan. & Connection Bloom filtering & Enables Bloom filtering for connections. \\ 
        
        & \hlhref{https://github.com/bitcoin/bips/blob/master/bip-0061.mediawiki}{61} & 2014 & stan. & Reject P2P message & Introduces a message to reject other messages. \\ 
        
        & \hlhref{https://github.com/bitcoin/bips/blob/master/bip-0130.mediawiki}{130} & 2015 & stan. & sendheaders message & Requests peer to send block headers. \\ 
        
        & \hlhref{https://github.com/bitcoin/bips/blob/master/bip-0133.mediawiki}{133} & 2016 & stan. & feefilter message & Filters transactions by minimum fee. \\ 
        
        & \cellcolor{cambridgeblue!40} \hlhref{https://github.com/bitcoin/bips/blob/master/bip-0144.mediawiki}{144} & \cellcolor{cambridgeblue!40} 2016 & \cellcolor{cambridgeblue!40} stan. & Segregated Witness (Peer Services) & Adds SegWit support for peer services. \\ 
        
        & \hlhref{https://github.com/bitcoin/bips/blob/master/bip-0152.mediawiki}{152} & 2016 & stan. & Compact Block Relay & Reduces bandwidth by relaying compact blocks. \\ 
        
        & \hlhref{https://github.com/bitcoin/bips/blob/master/bip-0159.mediawiki}{159} & 2017 & stan. & NODE\_NETWORK\_LIMITED service bit & Introduces a new service bit for nodes. \\ 
        
        & \hlhref{https://github.com/bitcoin/bips/blob/master/bip-0324.mediawiki}{324} & 2019 & stan. & Version 2 P2P Encrypted Transport Protocol & Adds encrypted transport for P2P connections. \\ 
        
        & \hlhref{https://github.com/bitcoin/bips/blob/master/bip-0339.mediawiki}{339} & 2020 & stan. & WTXID-based transaction relay & Improves transaction relay with WTXID. \\
       
        \cmidrule{1-5}

        \multirow{16}{*}{\rotatebox{90}{\textbf{Consensus}}}  &  \hlhref{https://github.com/bitcoin/bips/blob/master/bip-0030.mediawiki}{30} & 2012 & stan. & Duplicate transactions & Prevents duplicate transactions in the blockchain. \\ 

        & \hlhref{https://github.com/bitcoin/bips/blob/master/bip-0034.mediawiki}{34} & 2012 & stan. & Block v2, Height in Coinbase & Includes block height in the coinbase transaction. \\ 
        
        & \hlhref{https://github.com/bitcoin/bips/blob/master/bip-0042.mediawiki}{42} & 2014 & stan. & A finite monetary supply for Bitcoin & Caps the total supply of Bitcoin. \\ 
        
        & \hlhref{https://github.com/bitcoin/bips/blob/master/bip-0065.mediawiki}{65} & 2014 & stan. & OP\_CHECKLOCKTIMEVERIFY & Adds a new opcode for transaction lock time. \\ 
        
        & \hlhref{https://github.com/bitcoin/bips/blob/master/bip-0066.mediawiki}{66} & 2015 & stan. & Strict DER signatures & Enforces strict DER encoding for signatures. \\ 
        
        & \hlhref{https://github.com/bitcoin/bips/blob/master/bip-0068.mediawiki}{68} & 2015 & stan. & Relative lock-time using consensus-enforced sequence numbers & Implements relative lock-time with sequence numbers. \\ 
        
        & \hlhref{https://github.com/bitcoin/bips/blob/master/bip-0091.mediawiki}{91} & 2017 & stan. & Reduced threshold Segwit MASF & Lowers the threshold for Segwit activation. \\ 
        
        & \hlhref{https://github.com/bitcoin/bips/blob/master/bip-0112.mediawiki}{112} & 2015 & stan. & CHECKSEQUENCEVERIFY & Adds a new opcode for sequence lock time. \\ 
        
        & \hlhref{https://github.com/bitcoin/bips/blob/master/bip-0113.mediawiki}{113} & 2015 & stan. & Median time-past as endpoint for lock-time calculations & Uses median time-past for lock-time. \\ 
        
        & \cellcolor{cambridgeblue!40} \hlhref{https://github.com/bitcoin/bips/blob/master/bip-0141.mediawiki}{141} & \cellcolor{cambridgeblue!40} 2015 & \cellcolor{cambridgeblue!40} stan. & Segregated Witness (Consensus layer) & Adds SegWit support to the consensus layer. \\ 
        
        & \hlhref{https://github.com/bitcoin/bips/blob/master/bip-0143.mediawiki}{143} & 2016 & stan. & Transaction Signature Verification for Version 0 Witness Program & Defines signature verification for SegWit. \\ 
        
        & \hlhref{https://github.com/bitcoin/bips/blob/master/bip-0147.mediawiki}{147} & 2016 & stan. & Dealing with dummy stack element malleability & Addresses dummy stack element malleability. \\ 
        
        & \hlhref{https://github.com/bitcoin/bips/blob/master/bip-0143.mediawiki}{148} & 2016 & stan. & Mandatory activation of segwit deployment & Forces activation of SegWit. \\ 
        
        & \cellcolor{cambridgeblue!40} \hlhref{https://github.com/bitcoin/bips/blob/master/bip-0341.mediawiki}{341} & \cellcolor{cambridgeblue!40} 2020 & \cellcolor{cambridgeblue!40} stan. & Taproot: SegWit version 1 spending rules & Introduces Taproot spending rules. \\ 
        
        & \cellcolor{cambridgeblue!40} \hlhref{https://github.com/bitcoin/bips/blob/master/bip-0342.mediawiki}{342} & \cellcolor{cambridgeblue!40} 2020 & \cellcolor{cambridgeblue!40} stan. & Validation of Taproot Scripts & Adds validation rules for Taproot scripts. \\ 
        
        & \cellcolor{cambridgeblue!40} \hlhref{https://github.com/bitcoin/bips/blob/master/bip-0343.mediawiki}{343} & \cellcolor{cambridgeblue!40} 2021 & \cellcolor{cambridgeblue!40} stan. & Mandatory activation of taproot deployment & Forces activation of Taproot. \\

        \cmidrule{1-5}

        \multirow{3}{*}{\rotatebox{90}{\textbf{N/A}}}   & \hlhref{https://github.com/bitcoin/bips/blob/master/bip-0050.mediawiki}{50} & 2013 & info. & March 2013 Chain Fork Post-Mortem & Analysis of the March 2013 chain fork incident. \\ 

        & \hlhref{https://github.com/bitcoin/bips/blob/master/bip-0090.mediawiki}{90} & 2016 & info. & Buried Deployments & Describes the process for burying soft fork deployments. \\ 
        
        & \cellcolor{cambridgeblue!40} \hlhref{https://github.com/bitcoin/bips/blob/master/bip-0340.mediawiki}{340} & \cellcolor{cambridgeblue!40} 2020 & \cellcolor{cambridgeblue!40} stan. & Schnorr Signatures for secp256k1 & Introduces Schnorr signatures for improved security and efficiency. \\
   
\bottomrule
                
\end{tabular}
}

\begin{tablenotes}
      \scriptsize
     % \footnotesize
      \item[]  Items highlighted with a \colorbox{cambridgeblue!40}{colored} background are closely related to this paper.  \quad \textbf{Abbr.}: stan. = standard; info. = informational. 
     \end{tablenotes}
\end{center}
\vspace{-0.1in}
\end{table*}

\end{document}